\documentclass{aa}   

\usepackage[varg]{txfonts}

\usepackage{graphicx, natbib}
\bibpunct{(}{)}{;}{a}{}{,}

\usepackage{amssymb}

\begin{document}

\title{Modulation of neutral interstellar He, Ne, O in the heliosphere. Survival probabilities and abundances at IBEX}

\author{M. Bzowski \inst{1}
\and J. M. Sok{\'o}{\l} \inst{1}
\and M. A. Kubiak \inst{1}
\and H. Kucharek \inst{2}}
\offprints{M. Bzowski, \email{bzowski@cbk.waw.pl}}

\institute{Space Research Centre of the Polish Academy of Sciences, Warsaw, Poland, \email{bzowski@cbk.waw.pl}
\and Space Science Center and Department of Physics, University of New Hampshire, Durham NH, USA}

\date{Received  / Accepted }

\abstract {Direct sampling of neutral interstellar (NIS) atoms by the Interstellar Boundary Explorer (IBEX) can potentially provide a complementary method for studying element abundances in the Local Interstellar Cloud and processes in the heliosphere interface.}{We set the stage for abundance-aimed in-depth analysis of measurements of NIS~He, Ne, and O by IBEX and determine systematic differences between abundances derived from various calculation methods and their uncertainties.}{Using a model of ionization rates of the NIS species in the heliosphere, based on independent measurements of the solar wind and solar EUV radiation, we develop a time-dependent method of calculating the survival probabilities of NIS atoms from the termination shock (TS) of the solar wind to IBEX. With them, we calculate densities of these species along the Earth's orbit and simulate the fluxes of NIS species as observed by IBEX. We study pairwise ratios of survival probabilities, densities and fluxes of NIS species at IBEX to calculate correction factors for inferring the abundances at TS.}{The analytic method to calculate the survival probabilities gives acceptable results only for He and Ne during low solar activity. For the remaining portions of the solar cycle, and at all times for O, a fully time dependent model should be used. Electron impact ionization is surprisingly important for NIS~O. Interpreting the IBEX observations using the time dependent model yields the LIC Ne/O abundance of $0.16\pm40\%$. The uncertainty is mostly due to uncertainties in the ionization rates and in the NIS gas flow vector.}{The Ne/He, O/He and Ne/O ratios for survival probabilities, local densities, and fluxes scaled to TS systematically differ and thus an analysis based only on survival probabilities or densities is not recommended, except the Ne/O abundance for observations at low solar activity.} 


\maketitle

\section{Introduction}
Abundance of interstellar Ne relative to O and He has important astrophysical implications \citep{slavin_frisch:07a} since it provides insight into the processes shaping the Local Interstellar Cloud (LIC), in which the Sun is embedded \citep{redfield_linsky:08a, bzowski_etal:12a, mobius_etal:12a}. However, it cannot be directly measured in the LIC using traditional astrophysical methods because Ne is not visible in the absorption lines. Also, determining  other abundances in the LIC is challenging because of the ambiguity of the components of the absorption lines observed in the spectra of nearby stars.  The problem is how to reliably find the component from the LIC among components from other clouds filling the line of sight. Hence a complementary method of inferring the abundances in the interstellar gas in the solar immediate neighborhood is welcome. 

Such a method has recently become available with the launch of the Interstellar Boundary Explorer (IBEX) satellite \citet{mccomas_etal:09a}, which, owing to its time of flight (TOF) mass spectrometer IBEX-Lo \citep{fuselier_etal:09b}, is able to sample neutral interstellar (NIS) H, He, Ne, and O at Earth's orbit \citep{mobius_etal:09b, bochsler_etal:12a}. 

Neutral interstellar species enter freely the heliosphere from the LIC and flow towards the Sun. While direct sampling of NIS atoms at Earth's orbit is ideologically simple, interpretation of the measurements to infer the abundances in the LIC requires taking into account ionization losses and modifications of the flux by solar gravitational accretion. This is usually referred to as the heliospheric filtration. As discussed by a number of authors \citep{izmodenov_etal:04a, mueller_zank:04a}, this filtration is a complex, 2-step, species-dependent process. The first step is the filtration through the heliospheric interface region, the second is extinction due to the ionization inside the termination shock (TS), i.e., within the supersonic solar wind. 

Upon entry into the heliospheric interface region, neutral interstellar gas first passes through the outer heliosheath (OHS) just outside the heliopause, which may be regarded as a bow wave in the interstellar gas. Until recently it was believed that due to the supersonic speed of the heliosphere in the LIC the outer boundary of this region is a bow shock, but recent findings by \citet{bzowski_etal:12a, mobius_etal:12a} suggest that this velocity is significantly lower, which prompted \citet{mccomas_etal:12b} to propose that the bow wave is not a shock and its exact nature depends on a number of parameters whose values are not precisely known.

Regardless, however, of the exact properties of the heliospheric bow wave, the plasma flow decouples from the neutral component flow in the OHS. Consequently, charge exchange reactions between the neutral interstellar atoms and ions in this region lead to the creation of another population of neutral atoms, the so-called secondary population \citep{izmodenov_etal:01a}. This is in significant portion at the expense of the primary population of NIS atoms, which is depleted. This depletion is the first step of the filtration of neutral interstellar atoms on entry to the heliosphere. The filtration factor, defined as the ratio of post- to pre-OHS densities, is species-dependent. Review of the filtration through heliospheric boundary based on models available in the literature was recently presented by \citet{bochsler_etal:12a}, but one has to realize that the filtration of NIS species through the outer heliosheath without a bow shock has not yet been modeled. 

The second step of heliospheric filtration of NIS species is the topic of this paper. Once inside the heliopause, NIS gas is subject to ionization processes due to the interaction with the heliospheric environment. The gas is collisionless and the atoms follow individual hyperbolic trajectories in the solar gravitational potential. Some of them get ionized and are eliminated from the sample. Thus ionization acts as another filtration process for the NIS flux. The magnitude of this filtration depends on the species, location in the heliosphere and time because the intensity of the ionization processes is time- and solar distance-dependent. 

Effectively, the counts number registered by a spacecraft traveling with Earth and observing the neutral species is proportional to the exposure time and the local atom flux. The flux varies along the orbit because of the combination of time- and space-variations of the local NIS flux. This variation must be appropriately taken into account when determining the abundance of NIS species at the inner boundary of the heliospheric interface region.

In the following sections of the paper, we briefly discuss ionization processes relevant for NIS~He, Ne, and O as a function of time and heliocentric distance based on independently measured solar EUV and solar wind output. 
With the ionization rates on hand, we investigate the survival probabilities of NIS~He, Ne, and O atoms from TS to Earth's orbit and comment on the accuracy of various methods used to assess them. Next, we present calculated densities of NIS~He, Ne, and O in the ecliptic plane and their variation in time during IBEX observations. In particular, we dwell on the evolution of these densities at Earth and on the change of the abundances relative to the values at TS. Finally, we discuss the evolution of the NIS~Ne and O fluxes that enter the IBEX-Lo aperture and derive their abundance. We close with comments on the Ne/O abundance in the LIC, mostly repeating the approach exercised by \citet{bochsler_etal:12a}, but using the assessment of the differential filtration of He, Ne, and O derived in this paper.

\section{Ionization rates for neutral He, Ne, and O in the inner heliosphere}

Ionization processes for neutral He, Ne, and O in the heliosphere include photoionization by solar EUV radiation, charge exchange with solar wind ions, and ionization by impact of solar wind electrons. The total ionization rate $\beta\left(\vec{r}\left(t\right), t, \vec{v}\left(t\right)\right)$ for an atom traveling with velocity $\vec{v}\left(\vec{r}\left(t\right)\right)$ at a time $t$ and location $\vec{r}\left(t\right)$ in the heliosphere is given by a sum of the rates of photoionization $\beta_{\mathrm{ph}}$, charge exchange $\beta_{\mathrm{cx}}$, and electron impact $\beta_{\mathrm{el}}$:
\begin{eqnarray}
\label{eq:betaSum}
\beta\left(\vec{r}\left(t\right), t, \vec{v}\left(t\right) \right) &=& 
\beta_{\mathrm{ph}}\left(\vec{r}\left(t\right), t\right)  \nonumber \\
& + &\beta_{\mathrm{cx}}\left(\vec{r}\left(t\right), t, \vec{v}\left(t\right) \right)
 \\ 
& + & \beta_{\mathrm{el}}\left(\vec{r}\left(t\right), t\right) \nonumber	
\end{eqnarray}

In the following, we present ionization rates at 1~AU from the Sun, calculated as Carrington period averages centered at halves of Carrington rotations, which are nodes of the time grid used in the simulations discussed in farther parts of the paper.

	\begin{figure*}
		\begin{tabular}{cc}
			\includegraphics[width=.45\textwidth]{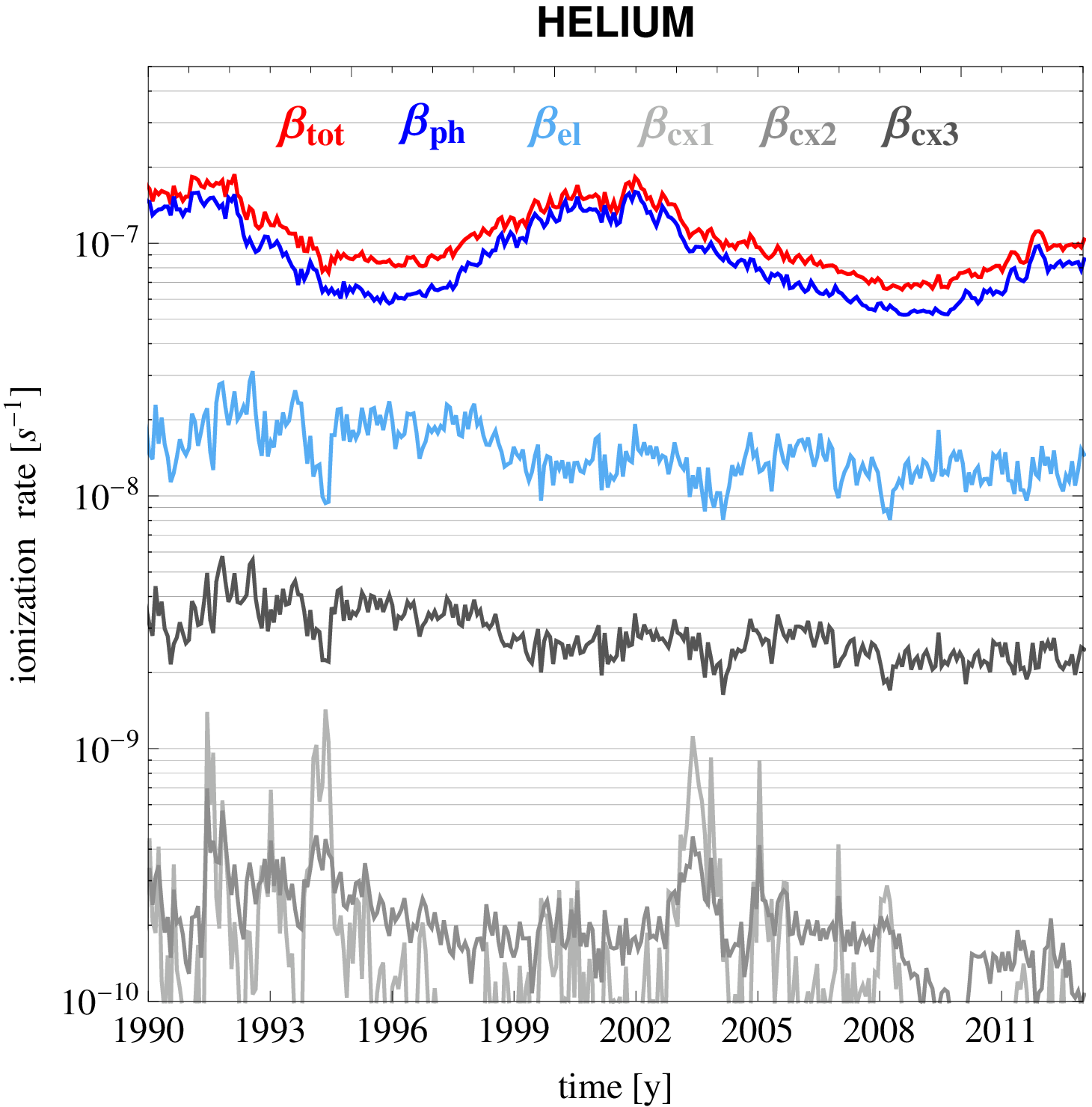} & \includegraphics[width=.45\textwidth]{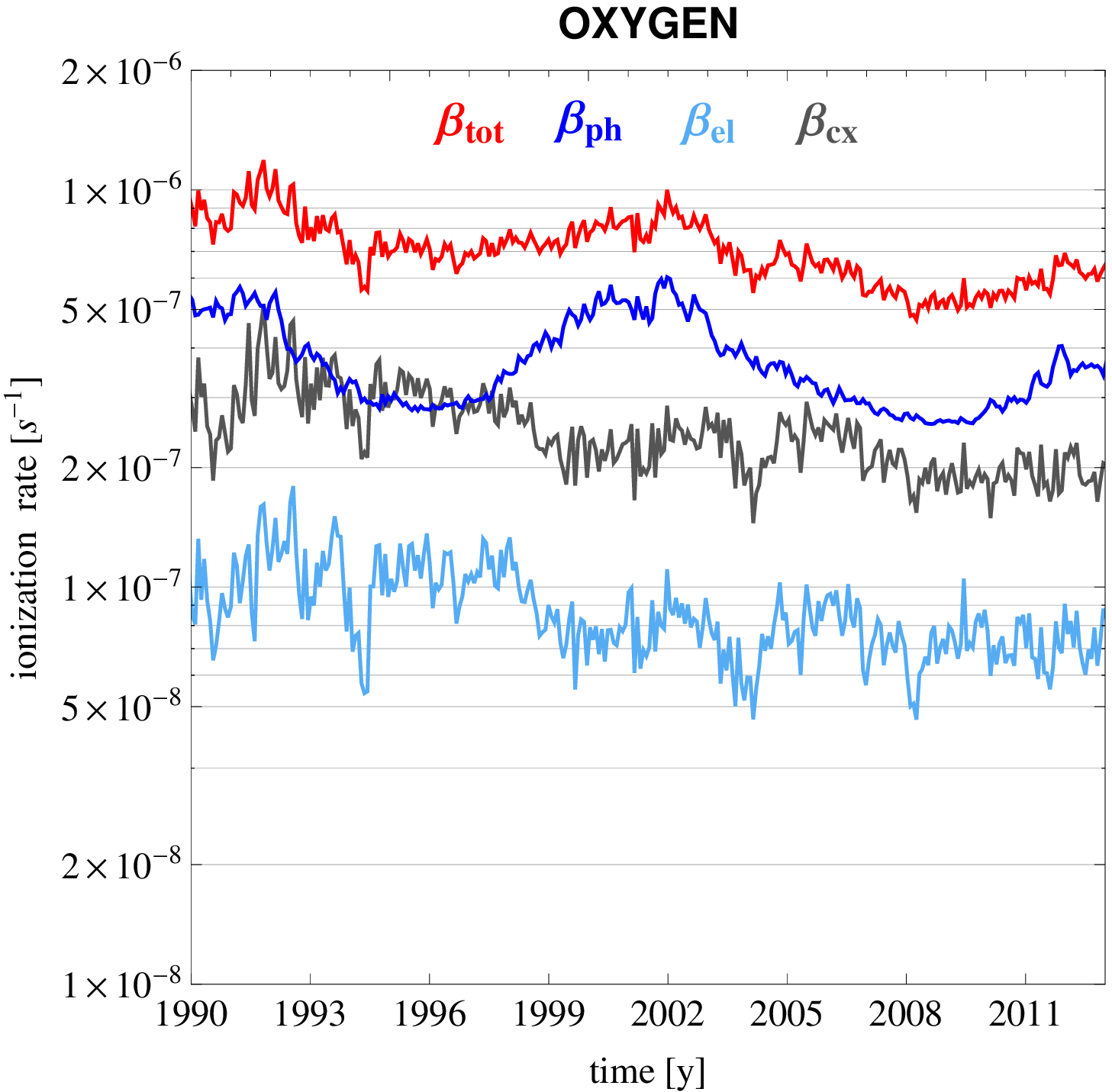}\\
			\includegraphics[width=.45\textwidth]{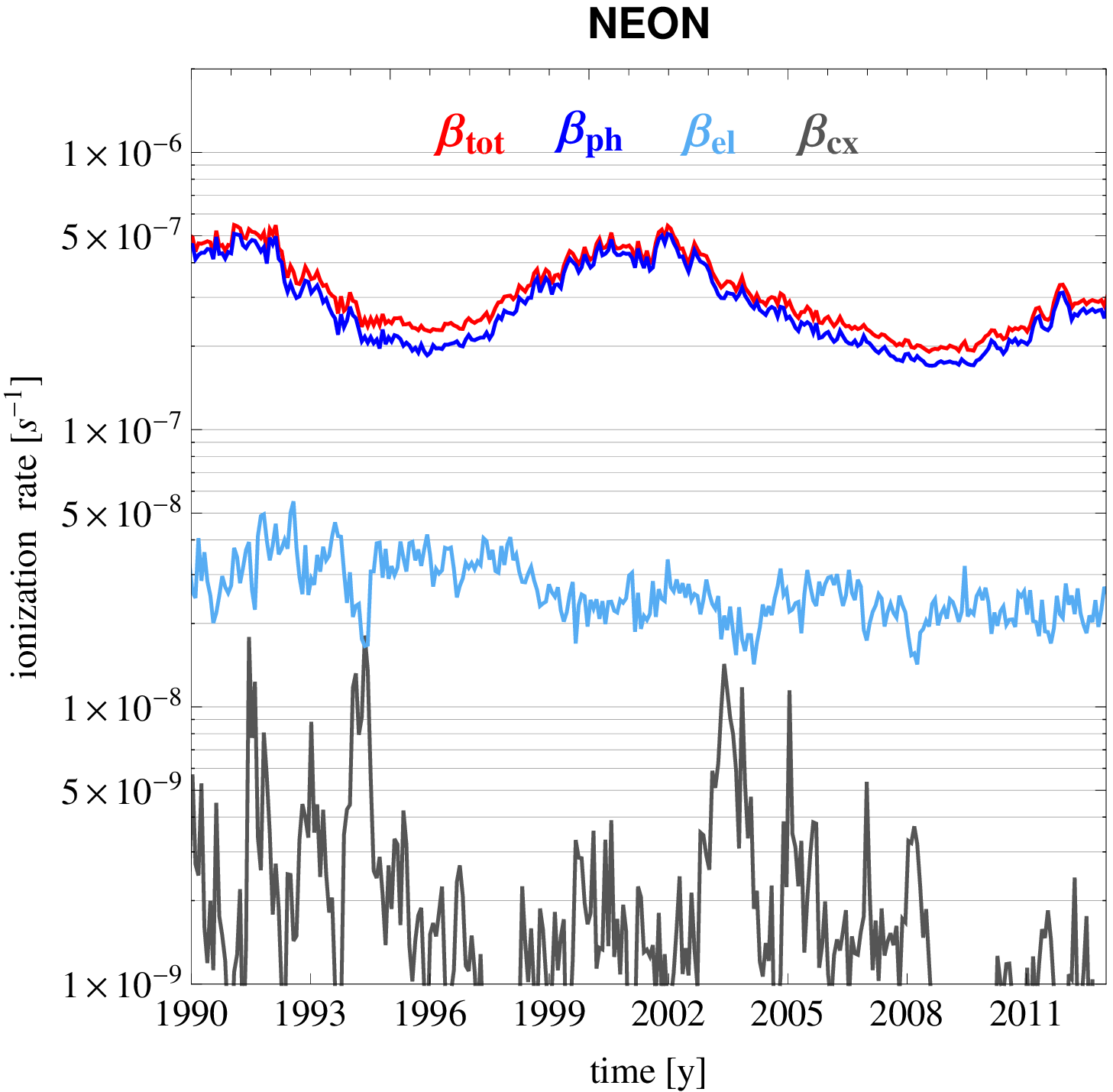} & \includegraphics[width=.45\textwidth]{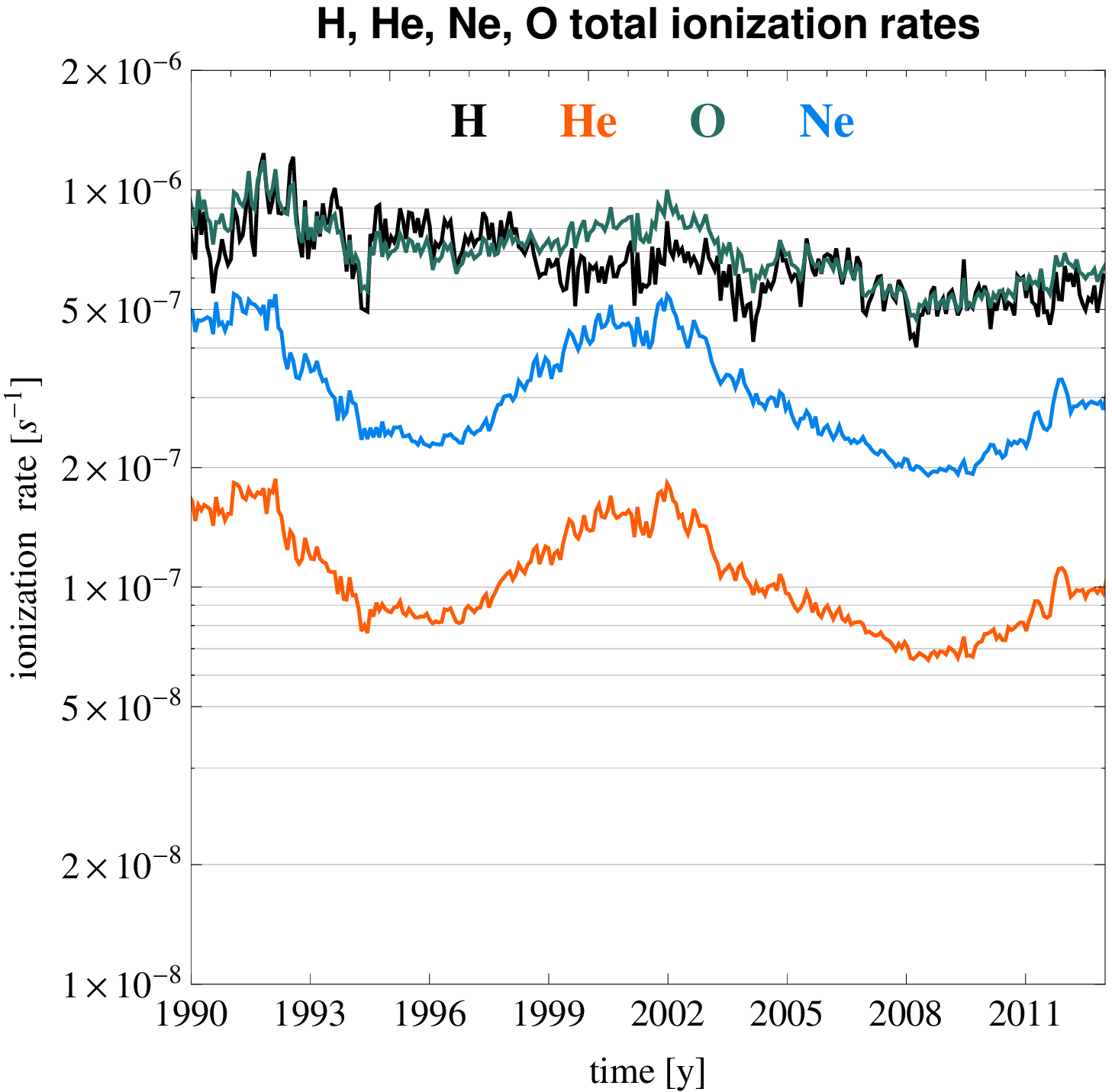}\\
		\end{tabular}
		\caption{Time series of ionization rates of neutral He, Ne, and O used in this study. Shown are total ionization rates at 1 AU from the Sun as well as photoionization, electron-impact, and charge exchange rates for all relevant reactions. The lower-right panels show the total ionization rates of He, Ne, and O in a common scale. For comparison, the ionization rate of H \citep{bzowski_etal:12b} is added.}
		\label{fig:ionRates}
	\end{figure*}

\subsection{Photoionization}
Photoionization is the dominant reaction for all three species in question. It features a modulation with the solar activity cycle as well as variations on shorter time scales. The amplitude of solar cycle modulation of the photoionization rates is approximately 2 (see Fig.~\ref{fig:photoRates}) and the maximum rates occur during the intervals of high solar activity. The intensity of photoionization goes down with the square of solar distance, just as the solar EUV flux does. Photoionization rates are predicted to feature some latitudinal anisotropy \citep{auchere_etal:05a} weakly evolving during the solar cycle and featuring some north-south asymmetry. Since in this paper we focus on NIS gas in the ecliptic plane and mostly out of the focusing cones, we have adopted photoionization rates that evolve as indicated in Fig.~\ref{fig:photoRates} and feature a fixed pole-to-equator flattening of 0.8 (for details, see Eq.~(3) in \citet{bzowski:08a}).


We have developed a time series of photoionization rates of H, He, Ne, and O on a common time grid covering time interval from 1948 until present (Fig.~\ref{eq:photoIonRate}). The construction is based on direct measurements of solar spectrum from TIMED/SEE \citep{woods_etal:05a} and a hierarchy of proxies. Details are presented in the Appendix. 

\subsection{Charge exchange}
\label{sec:ChargeExchange}
Charge exchange is the second important ionization reaction for oxygen, but practically negligible for the noble gases He and Ne (see Fig.~\ref{fig:ionRates}), being at a level of 1\% of the total rate, which is much less than the uncertainty in the photoionization rate. Details of the physics of this reaction and of the reaction products have been extensively discussed in the past and will not be repeated here (see, e.g., \citet{bzowski_etal:12b}). The charge exchange rate also features decrease with the square of solar distance and does not show a clear periodic modulation, but does show a latitudinal modulation with the phase of solar activity.

For O the only relevant charge exchange reaction is   	
	\begin{displaymath}
		O + H^+ \to H_{\mathrm{ENA}} + O^{+}_{\mathrm{PUI}}
	\end{displaymath}
where $H^+$ is a solar wind proton and O is a NIS~O atom. Other ions in the solar wind are not sufficiently abundant to make a non-negligible effect \citep{zurbuchen_etal:02a}. The cross section for this reaction was taken from \citet{lindsay_stebbings:05a}, who conservatively assess its accuracy in the energy range relevant for our applications at 25\%.

Similarly for neon, the only potentially important reaction is
	\begin{displaymath}
		Ne + H^+ \to H_{\mathrm{ENA}} + Ne^{+}_{\mathrm{PUI}}
	\end{displaymath}
	The cross section for this reaction was adopted from \citet{nakai_etal:87a}, with the rms error quoted at 20\%.

For He the situation is different. There are three potentially important reactions, namely: 
	\begin{displaymath}
		\begin{array}{l}
			\mathrm{cx1}: \qquad He + H^+ \to H_{\mathrm{ENA}} + He^{+}_{\mathrm{PUI}}\\
			\mathrm{cx2}: \qquad He + \alpha \to H^+_{\mathrm{sw}} + He^{+}_{\mathrm{PUI}}\\
			\mathrm{cx3}: \qquad He + \alpha \to He_{\mathrm{ENA}} + He^{++}_{\mathrm{PUI}}\\
		\end{array}
	\end{displaymath}
This is because the cross sections for charge exchange between He atoms and alpha particles is orders of magnitude larger than for charge exchange between He atoms and protons (Fig.~\ref{fig:crossSectionsCX}; \citet{barnett_etal:90}) and the abundance of solar wind alphas is quite large -- about $4\%$ \citep{kasper_etal:12a}. The accuracy of the fits of the cross section formula to data is quoted at $\mathrm{rms} = 10\%$. 

Because of the similarity in the first ionization potential of O and H, the charge exchange reaction for oxygen has a quasi-resonant character and is quite intense. During the past solar cycles, it was the dominant loss process for O, nowadays, however, because of the overall drop in the solar wind flux, which begun in 1990-ties, the charge exchange rate is lower than the photoionization rate (see Fig.~\ref{fig:ionRates}).

In the modeling presented in the paper, we calculate the charge exchange rates based on the in-ecliptic time series of solar wind speed and density known as the OMNI-2 time series \citep{king_papitashvili:05} for the equatorial rates and on a compilation of the evolution of latitudinal structure of the solar wind speed and density by \citet{sokol_etal:12a}. This compilation is based on remote-sensing observations of interplanetary scintillation \citep{tokumaru_etal:10a} and in situ Ulysses measurements \citep{mccomas_etal:02b, mccomas_etal:08a}. The densities and speeds are calculated on a uniform grid in time and heliolatitude with nodes at halves of Carrington rotations and at full tens degrees of heliolatitude. It is further assumed that density falls off with $r^2$ and speed is distance independent. 

The charge exchange for a given species, location in space, and time moment is calculated ``on the fly'' during calculations based on the local density of solar wind and the computed relative speed between the local solar wind and the neutral atom in question \citep{bzowski_etal:12b}. The evolution of the charge exchange rates in the ecliptic plane at 1~AU from the Sun in the stationary atom approximation is presented in Fig.~\ref{fig:ionRates}. Details of the calculation scheme are available in the Appendix.

\subsection{Electron impact ionization}
\begin{figure}
		\resizebox{\hsize}{!}{\includegraphics{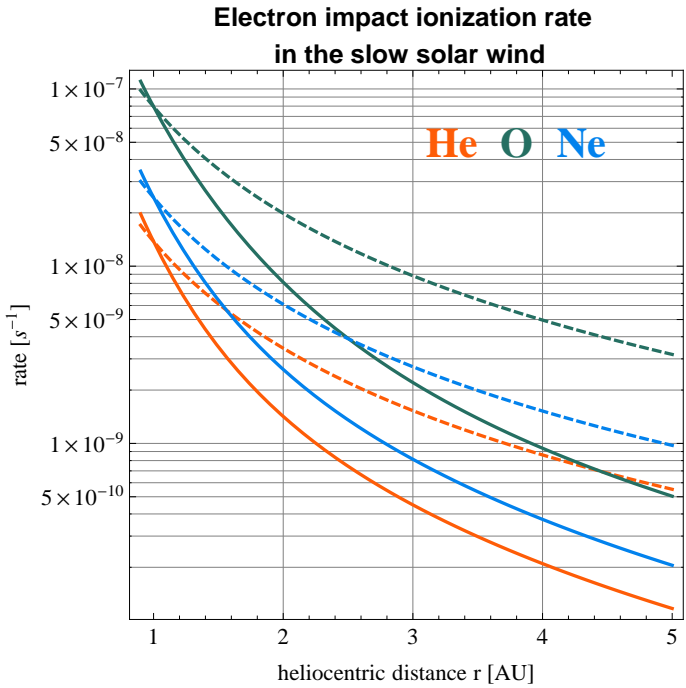}}
		\caption{Comparison of the radial dependence of electron impact ionization adopted in this paper with the $1/r^2$-dependence for He, Ne, and O. Solid lines indicate the radial profiles calculated using $\beta_{\mathrm{el}}^{\mathrm{slow}}\left( r \right)$ from Eq.~(\ref{eq:betaElSlow}), broken lines are for electron ionization calculated from $\beta_{\mathrm{el}}^{\mathrm{slow}}\left( r = 1 \mathrm{AU}\right)/r^2$, both for solar wind proton density at 1~AU $n_{\mathrm{sw}} \left( r = 1 \mathrm{AU}\right) = 6.14 \, \mathrm{cm^{-3}}$, assuming the decrease of proton density like $1/r^2$.}
		\label{fig:electronRateRadial}
	\end{figure}

The electron impact ionization rates are the poorest known among the ionization processes in the heliosphere. Electron ionization is special in the sense that its radial dependence significantly differs from $1/r^2$ (see Fig.~\ref{fig:electronRateRadial}). This is because the distribution function of solar wind electrons is complex, with a Maxwellian-like core and a halo (elevated wings) approximated by the Maxwellian or kappa-like function \citep{pilipp_etal:87c, maksimovic_etal:97a}. The abundance of the halo population relative to the core varies with heliocentric distance. Also the cooling rates of the two populations are different and change with solar distance. Neither of them is adiabatic. Details of processes modifying the electron distribution function are poorly investigated. A recent review of the evolution of solar wind electron distribution function with solar distance and during solar cycle was recently provided by \citet{issautier:09a}. 

It is worth to mention that outside 1~AU only a very small portion of the core population electrons have sufficient energies to ionize He and Ne. A slightly higher portion of the core electrons can ionize O. The peak energies of the two electron populations at 1~AU are marked in the drawing illustrating the energy dependence of the electron ionization cross sections shown in Fig.~\ref{fig:crossSectionsElectrons} in the Appendix. Thus, electron ionization outside 1~AU from the Sun is approximately equally due to the core and halo populations. The formulae used to approximate the electron rate with coefficients specific for the species in question are presented in the Appendix (Eqs.~(\ref{eq:betaElSlow}), (\ref{eq:betaElFastHeNe}), (\ref{eq:betaElFastO})). 

The radial profiles of the electron ionization are steeper than $1/r^2$ (Fig.~\ref{fig:electronRateRadial}), so the relative contribution of this reaction to the total ionization rate rapidly decreases with solar distance. Beyond $\sim5$~AU from the Sun the electron rate for all species in question becomes negligible, as is easily inferred from Fig.~\ref{fig:electronRateRadial}. The evolution of the electron rate in time at 1~AU in the slow solar wind is presented in Fig.~\ref{fig:ionRates}. It also do not show a clear time modulation with solar activity cycle, but does show a secular downward trend due to the drop of solar wind density, arrested in $\sim 2005$ \citep{mccomas_etal:08a}.

The model we adopted is parametrized by time and heliolatitiude variations of solar wind proton density (tied with the local electron density by the quasi-neutrality condition and taking into account the contribution from solar wind alphas). Its heliocentric distance dependence is based on radial profiles of electron temperatures measured by Ulysses during the previous solar cycle. They don't have to be exactly the same nowadays but we assume they are in lack of newer measurements. Consequently, we attribute large error bars to the electron ionization rates, as detailed in the error estimate section. 

\section{Survival probabilities of NIS~He, Ne, and O}
\subsection{Calculation of survival probabilities}
Survival probabilities are inherent element of any approach to obtain abundances of NIS species at TS based on in-situ observations performed in the inner heliosphere. Therefore in this section we will elaborate on this subject in greater detail. 

The theory of survival probabilities of neutral atoms on Keplerian trajectories under time-constant solar conditions was developed by \cite{blum_fahr:70a, axford:72} and for the time-dependent ionization rates by \citet{rucinski_etal:03} (see their Eq.~(4)). In essence, to calculate survival probability of an atom on a given trajectory between two selected points in space one has to compute the exponent of the ionization exposure function $\epsilon$:
	\begin{equation}
		w=\mathrm{exp}\left[ \epsilon \right]
		\label{eq:surPro}
	\end{equation}
where the ionization exposure function is defined as:
	\begin{equation}
		\epsilon=-\int\limits_{t_{\mathrm{TS}}}^{t_{0}}\beta\left(\vec{r}\left(t\right),t\right)\mathrm{d}t
		\label{eq:epsDef}
	\end{equation}
Here $\beta\left(\vec{r}\left(t\right),t\right)$ is the instantaneous total ionization rate at time $t$ in the location $\vec{r}\left(t\right)$ on the trajectory. The trajectory starts at a time $t_{\mathrm{TS}}$ and ends at $t_0$; $t_{\mathrm{TS}} < t_0$. The instantaneous ionization rate varies because of the changes in the heliocentric distance $r\left(t\right)$ (the spatial effect) and due to the evolution in time of the ionizing factors (the time effect). Note that since the probability value is between 0 and 1, the exposure function value must be negative. 

	\begin{figure*}
		\resizebox{\hsize}{!}{\includegraphics{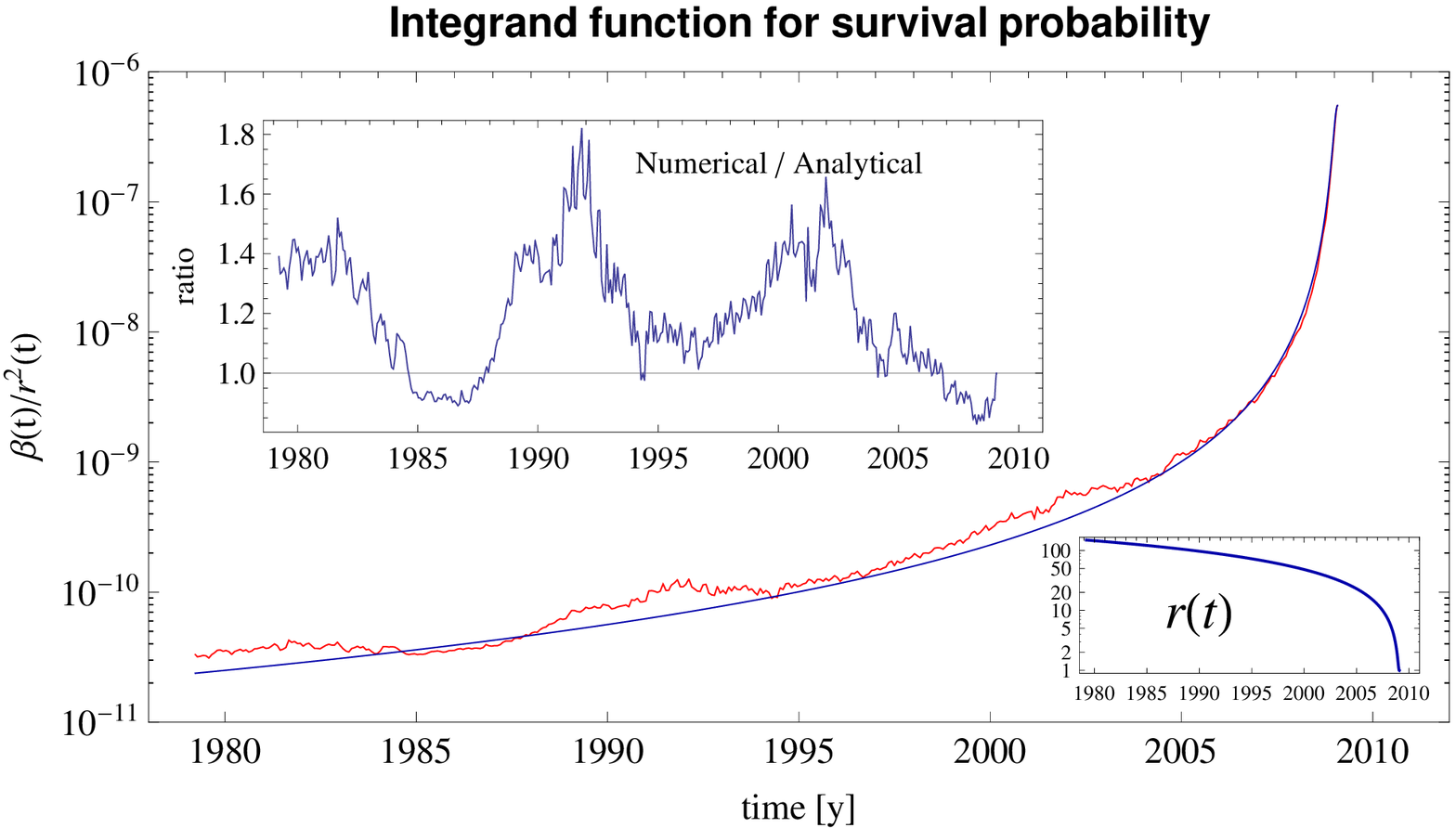}}
		\caption{Integrand function in Eq.~(\ref{eq:epsDef}) for an O atom on the orbit whose perihelion is at the Earth location on IBEX Orbit 16 in 2009 ($\lambda_{\mathrm{E}} = 130\degr$). The red line is the $\beta\left(\vec{r}\left(t\right), \vec{v}, t\right)$ from Eq.~(\ref{eq:betaSum}) taken along the orbit of the atom. The model used in $\beta\left(\vec{r}\left(t\right),t\right)$ is time dependent and has a component of electron impact rate, which does not follow the $1/r^2$ dependence. The dark-blue line is the integrand function calculated for $\beta\left(r_{\mathrm{E}},t_0 \right)\left( \frac{r_0}{r\left( t \right)} \right)^2$, i.e. assuming that the ionization rate is equal to the ionization rate from our model calculated for the instant of detection $t_0$.  The history of the radial distance of the atom from the Sun is shown in the lower-right inset and it is identical for the red and dark-blue lines. The upper-left inset presents the ratio of the actual ionization rate for atoms following the aforementioned trajectory and the rate used in the analytical model $\beta_0\left(t_0\right)\left(r_0/r\right)^2$. To obtain the survival probability one must calculate the exponent of the time integral of the function drawn in red, see Eqs.~(\ref{eq:surPro}) and (\ref{eq:epsDef}). The analytic approximation is equivalent to taking exponent of the integral of the function drawn with the blue line, i.e. to evaluating the exponent of the expression from Eqs.~(\ref{eq:surPro}) and (\ref{eq:epsAnal}).}
		\label{fig:integrandFunction1}
	\end{figure*}

For the special, idealized case of invariable ionization rate falling off with the square of heliocentric distance, this formula can be evaluated analytically \citep{lee_etal:12a}: 
	\begin{equation}
		\epsilon=-\beta_0 r_{\mathrm{E}}^2\Delta \theta / L
		\label{eq:epsAnal}
	\end{equation}
where $\beta_0$ is the total ionization rate at a distance $r_{\mathrm{E}}$ from the Sun, $L$ is the angular momentum of the atom and $\Delta \theta$ is the angle swept by the atom on its orbit between the infinity and the local point at its orbit. 


In reality, however, the simplification of stationary ionization rate falling off with $1/r^2$ works relatively well only for He, and only for epochs of prolonged low solar activity, as experienced recently, but is not reliable for species featuring higher ionization rates, as Ne and O, even for the relatively low activity conditions. 

The reason is illustrated in Fig.~\ref{fig:integrandFunction1}, which shows the ionization rate experienced by an O atom observed in its perihelion by IBEX on Orbit 16. This integrand function $\beta\left(\vec{r}\left(t\right),t\right)$ in Eq.~(\ref{eq:epsDef}) is compared with the ionization rate taken for the same trajectory assuming that the rate is constant in time, falling off strictly as $1/r^2$ and equal to the rate at the moment of detection in the perihelion of the trajectory. Insets in the figure show the evolution of the heliocentric distance along the atom's orbit in time and the ratio of the actual ionization rate to the constant rate taken for the moment of detection.

The differences between the survival probabilities calculated numerically using the time-dependent model and analytically are due to the actual evolution of the solar ionizing factors in time (i.e., due to the time variability of the photoionization rate, solar wind speed and density) and the contribution from electron ionization, not following the $1/r^2$ relation. It is clear that the integral of the function given by the red line in Fig.~\ref{fig:integrandFunction1} will differ from the integral of the function drawn with the blue line. The actual ionization rate differs from the idealized case by a varying factor, up to 80\%. Even the mean value of the variable integrand function differs from the idealized case, which results in different survival probabilities obtained from the analytical and realistic formulations. 

\begin{figure}
		\begin{tabular}{c}
			\resizebox{\hsize}{!}{\includegraphics{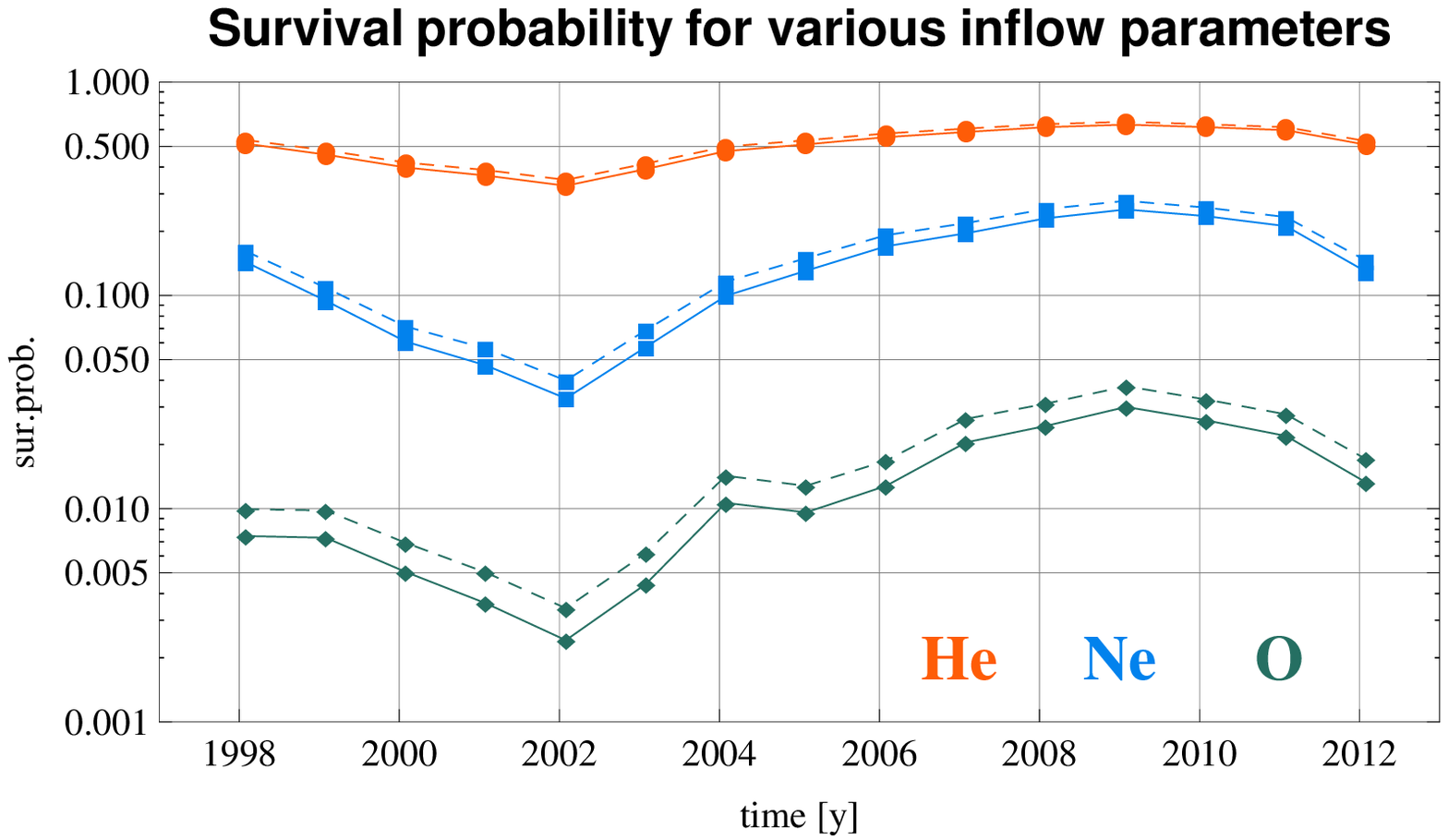}}\\
			\resizebox{\hsize}{!}{\includegraphics{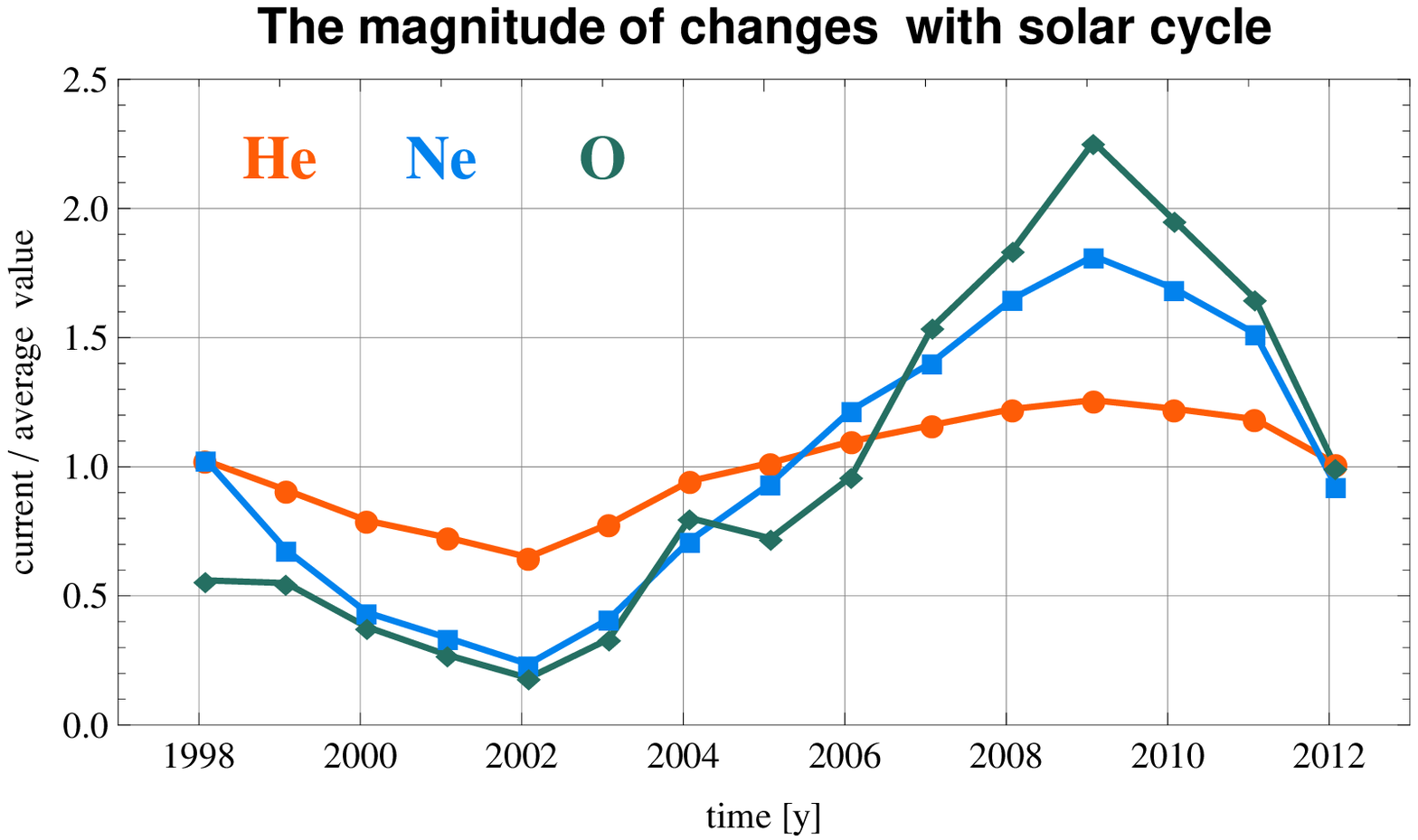}}\\
		\end{tabular}
		\caption{Upper panel: survival probabilities for He, Ne, and O atoms on the trajectory reaching Earth in the perihelion in 2009 at $130\degr$ ecliptic longitude. Solid lines are for the velocity vector at infinity corresponding to the bulk flow of the gas as established by \citet{bzowski_etal:12a} based on IBEX measurements, the broken lines illustrate the sensitivity of survival probabilities to the inflow parameters: they correspond to survival probabilities calculated for the NIS inflow vector as found by \citet{witte:04} based on GAS/Ulysses observations. Lower panel: Ratios of survival probabilities to their mean values calculated from an 11-year interval ending at the detection time.}
		\label{fig:surProHistory}
	\end{figure}

Fig.~\ref{fig:surProHistory} (upper panel) shows time series of survival probabilities of He, Ne, and O atoms reaching perihelia at ecliptic longitude $130\degr$ during a time interval spanning more than one solar cycle. The ratio of maximum to minimum values of survival probabilities during the solar cycle is minimum for He -- 1.9, but distinctly increases for Ne, reaching 7.7, and even stronger for O, up to 12.4. It is also worth to notice that the departures from the mean are by a similar factors for the three species discussed during solar maximum, but definitely different for solar minimum. This is illustrated in the lower panel of Fig.~\ref{fig:surProHistory}.

		\begin{figure}
		\begin{tabular}{c}
			\resizebox{\hsize}{!}{\includegraphics{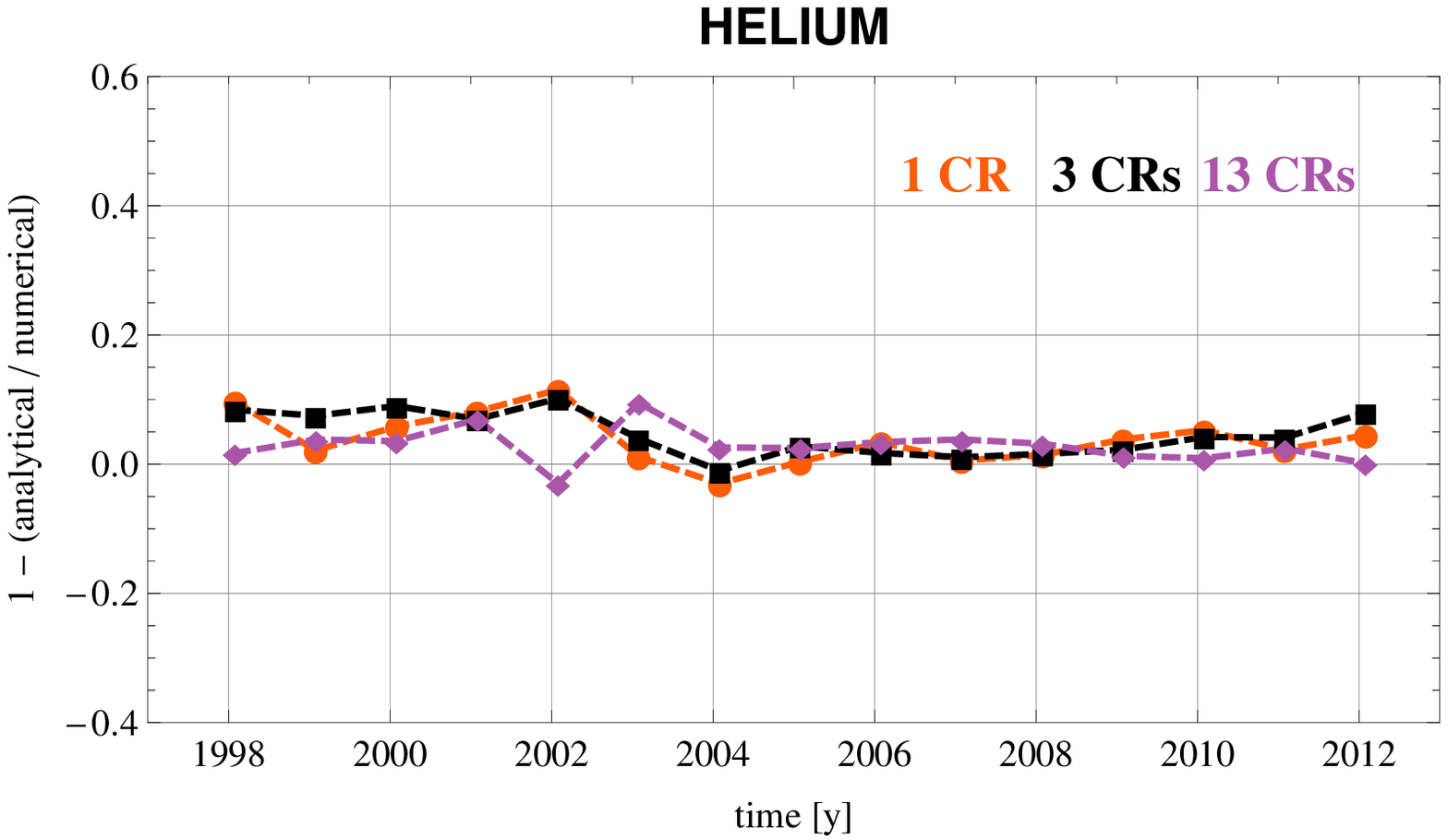}}\\
			\resizebox{\hsize}{!}{\includegraphics{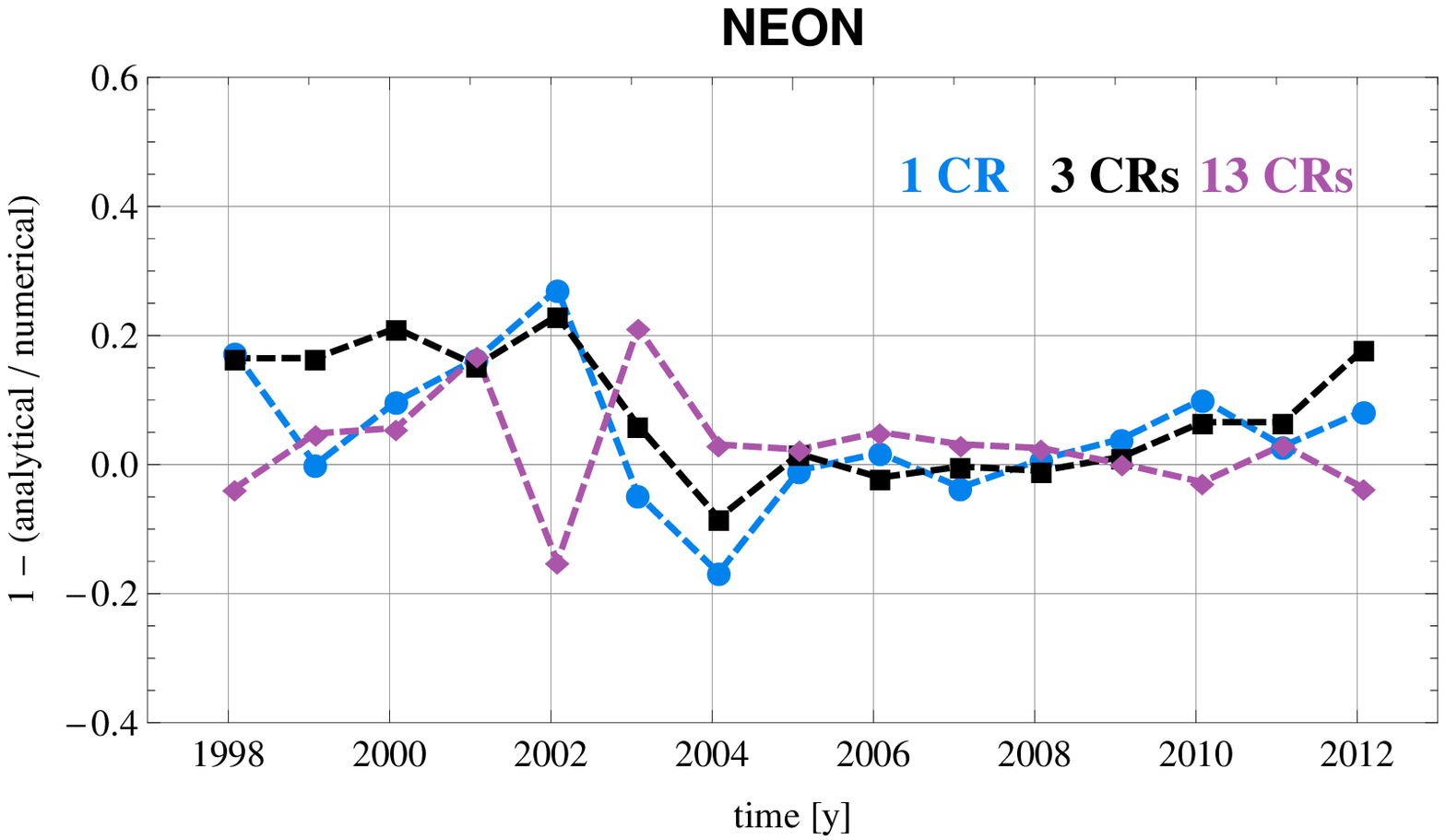}}\\
			\resizebox{\hsize}{!}{\includegraphics{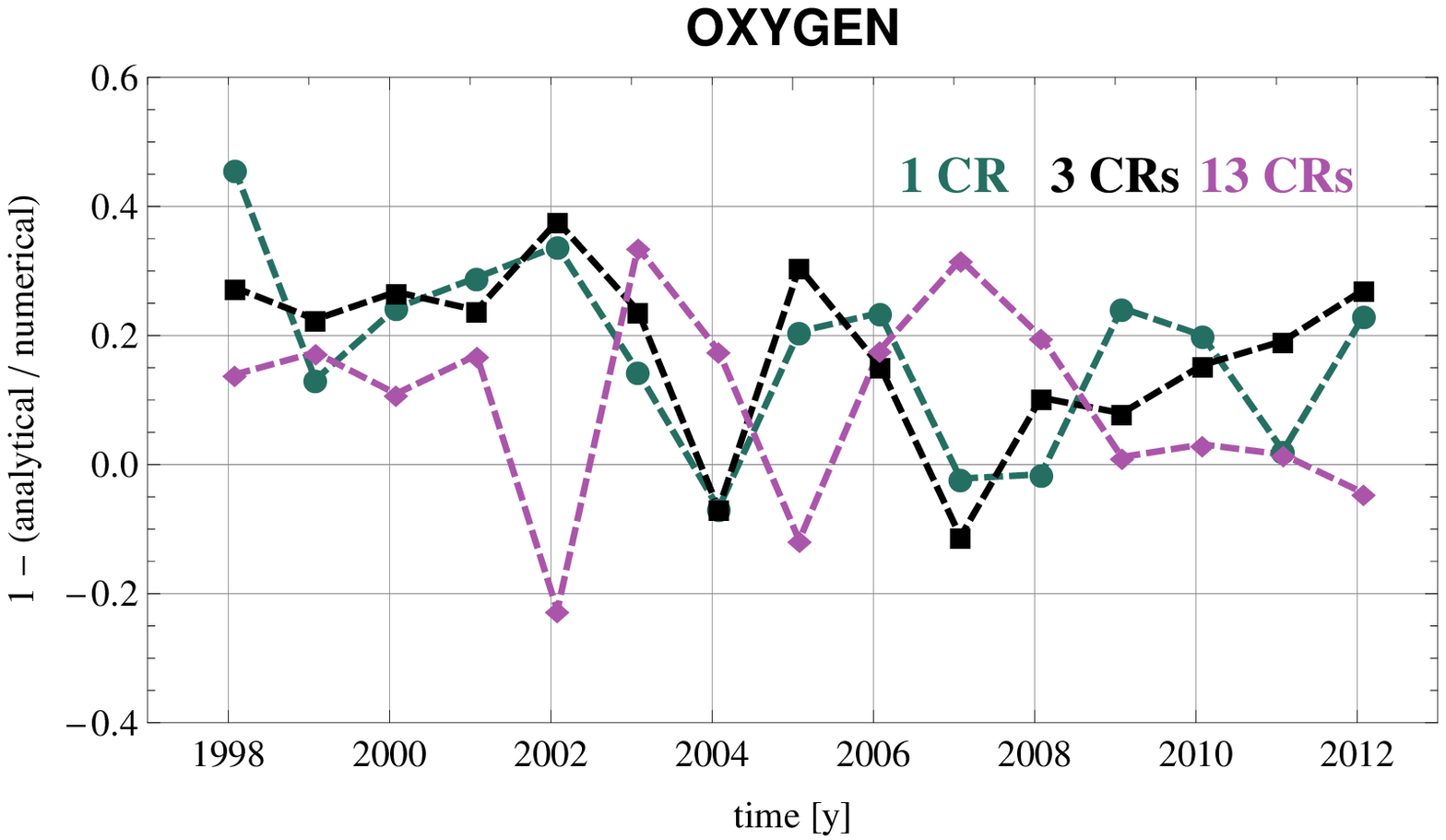}}\\
		\end{tabular}
		\caption{Relative differences between survival probabilities of He (upper panel), Ne (middle panel), and O (lower panel) calculated using the analytical and the full time-dependent models. The time-dependent model is identical as used in Fig.~\ref{fig:surProHistory}, the analytical models were calculated using stationary ionization rates following the $1/r^2$ solar distance dependence, with the rates for 1~AU calculated as average values over intervals ending at the detection moment and spanning backward in time 1, 3, or 13~Carrington rotation periods. Note that the relative differences do not average to 0. This effect is best visible for O, for the entire time interval shown in the figure. The systematic upward departure of the mean values from 0 is due to electron impact.}
		\label{fig:surProAveraging}
	\end{figure}

Since the analytical formulae for survival probabilities under the conventional assumption of stationarity of the ionization rate and its $1/r^2$ dependence on the solar distance are so simple, many researchers use them in the analysis. We tried to find a recipe for the treatment of the actual solar ionizing factors to make such an approach reliable -- and we failed. The idea was to find an interval to average the ionizing factors backward in time and to use the analytical models with the parameters obtained from the averaging. 

The results of this attempt is shown in Fig.~\ref{fig:surProAveraging}. The figure shows relative differences between survival probabilities calculated using the analytical model with the ionization rate as averaged by time intervals of 1, 3, and 13 Carrington rotation periods before the detection time at 1~AU. 

The relative differences for He and Ne are flat during low solar activity (from $\sim 2005$ to $\sim 2009$) and average slightly above 0. This is because of the electron rate not following the $1/r^2$ dependence. Otherwise, however, the analytic approach is quite satisfactory for He and Ne because variations of the ionization rate are mild. But when the solar activity changes, as before 2005 and after 2009, the analytic model starts to differ from the time dependent and none of the averaging intervals tested seems to be satisfactory.

Another feature visible in Fig.~\ref{fig:surProAveraging} is the systematic difference between the full and analytic models before 2003: the relative differences for He and Ne average above 0. This is due to a stronger contribution to the total ionization rate from electron ionization, caused by the higher solar wind density in that epoch than it is nowadays \citep{mccomas_etal:08a}.

The ionization losses for O are so strong that even the mild variation in the ionization rate during low solar activity make it impossible to reconcile with the analytic model. Each of the lines in the lower panel of Fig.~\ref{fig:surProAveraging} shows a sawtooth behavior with an amplitude of $\sim20\%$ throughout the time interval shown. The amplitude is higher during high solar activity. Additionally, the relatively strong contribution from electron ionization causes the persistent and systematic deviation of the relative difference upward from 0. 

Thus, the simplified formula can be used for the times of prolonged low solar activity for He (as between $\sim2005$ and $\sim2011$) and to a lesser extent to Ne (between $\sim2005$ and $\sim2009$), but it is clearly less accurate when the activity is changing. The simplified approach cannot, however, be used for O for any times at all if the accuracy goal is better than 20$\%$. Therefore we recommend to not use the simplified analytic approach to calculate the survival probabilities of NIS species in the heliosphere.

The results of this study may raise concern if building the time dependent model of ionization losses on the 27.28 day (Carrington rotation) averages of the ionization rate is not too crude since it is known that both solar wind and solar EUV output feature significant variations on time scales from minutes to weeks. This may be important because the survival probability for He at 27.28 days before detection is equal to $\sim 0.73$, and for the moment of detection it goes down to $\sim 0.66$ and for O the change is even stronger, the survival probability changes from $\sim 0.09$ to $\sim 0.03$. 

Ideally, one would be willing to take the variations on a very fine resolution. The problem is that we have available only point measurements at Earth and the NIS atoms close to Earth's orbit, especially those observed by IBEX at their perihelia, travel just before detection (when the ionization is the strongest) with angular velocities almost twice larger than Earth and are running against the direction of Earth motion. This means we do not have credible information on the actual ionization rates experienced by those atoms accurate enough to justify building the ionization rate models on a finer time grid. 

Adoption of Carrington rotation averaging is an attempt to find a balance between the global character of the ionization rate data on one hand and accounting for actual time variations on the other hand. The unavoidable imperfection of this approach is reflected in the relatively large uncertainties of survival probabilities, which we discuss in the following section of this paper. 

\subsection{Uncertainties of survival probabilities and contributions from different ionization reactions}
There are two major sources of uncertainties of survival probabilities: (1) uncertainties of the ionization rates and (2) uncertainty in the velocity vector of the inflow of neutral interstellar gas on the heliosphere. 

Uncertainties resulting from the uncertainty in the NIS gas velocity vector are illustrated in the upper panel of Fig.~\ref{fig:surProHistory}. In addition to survival probabilities calculated for the NIS gas velocity vector obtained by \citet{bzowski_etal:12a} based on observations from IBEX, drawn with solid lines, the probabilities calculated for the velocity vector obtained by \citet{witte:04} based on observations from GAS/Ulysses are shown by broken lines. The two velocity vectors differ by $\sim4~\mathrm{km}~\mathrm{s}^{-1}$ in speed and $\sim4\degr$ in direction. The survival probability differences are systematic in character: for a faster speed, the exposure to ionization is lower and hence the survival probabilities are larger, consistently for all species and all phases of solar activity. The influence of the LIC velocity vector of the gas on survival probabilities and their ratios is further illustrated in Fig.~\ref{fig:surProRatios} and will be discussed in a further part of this section.

Uncertainties of survival probabilities due to the uncertainties in the ionization rates are presented in greater detail below, as, to our knowledge, they have not been discussed in the literature.

The exposures to the three ionization reactions for a species X, introduced in Eq.~(\ref{eq:epsDef}), are defined as follows. For photoionization:
	\begin{equation}
		\epsilon_{\mathrm{ph,X}}=\iint{ F_{\mathrm{ph}}\left(\lambda,t\right)\sigma_{\mathrm{ph,X}}\left(\lambda \right) \mathrm{d}\lambda \mathrm{d}t}
		\label{eq:epsPhoto}
	\end{equation}
where $F_{\mathrm{ph}}\left(\lambda,t\right)$ is the solar spectral flux for a wavelength $\lambda$ at a time $t$ and $\sigma_{\mathrm{ph,X}}$ is the photoionization cross section for species X; for charge exchange:
	\begin{equation}
		\epsilon_{\mathrm{cx,X}}=\int{ F_{\mathrm{sw}}\left(t\right)\sigma_{\mathrm{cx,X}}\left(v_{\mathrm{sw}}\left(t \right) \right) \mathrm{d}t}
		\label{eq:epsCX}
	\end{equation}	
where $F_{\mathrm{sw}}\left(t\right)$ is the solar wind flux for time $t$ and $\sigma_{\mathrm{cx,X}}$ is the cross section for charge exchange for species X for relative speed between the reaction partners approximated by the solar wind speed $v_{\mathrm{sw}}$; for electron impact:
	\begin{equation}
		\epsilon_{\mathrm{el,X}}=\int{n_{\mathrm{sw}}\left(t\right)\int{ f_{\mathrm{el}}\left(E,t\right)\sigma_{\mathrm{el,X}}\left(E \right) E \mathrm{d}E}\mathrm{d}t}
		\label{eq:epsEl}
	\end{equation}	
where $n_\mathrm{sw}\left(t\right)$ is solar wind proton density, $\sigma_{\mathrm{el,X}}$ is the electron-impact ionization cross section for energy $E$, and $f_{\mathrm{el}}\left(E,t\right)$ is the solar wind electron distribution function. This function depends on electron energy and time and is defined so that $n_\mathrm{sw} f_{\mathrm{el}}\left(E,t\right)$ after integration over energy evaluates to the local electron density in the solar wind, which, due to quasi-neutrality, is approximately equal to the local proton density plus twice the local solar wind alpha density.
	
All the parameters $p_i$ affecting the exposures are known from measurements. In the following discussion, we will approximate the uncertainties $\Delta p_i$ of these parameters by their relative errors $\delta p_i$, so that: 
	\begin{equation}
		\Delta p_i=\delta p_i p_i.
		\label{eq:relError}
	\end{equation}	
	The uncertainty of the exposure $\epsilon \left( p_1,\ldots,p_i,\ldots,p_n \right)$ is calculated from the general formula:
	\begin{equation}
		\Delta \epsilon=\left[ \sum_{i=1}^{n}{\left( \frac{\partial \epsilon}{\partial p_i}\Delta p_i \right)^2} \right]^{1/2} > 0
		\label{eq:errorDef}
	\end{equation}
The uncertainty range of survival probability of species X against a given reaction \emph{proc} is defined by the inequality:
	\begin{equation}
		\mathrm{exp}\left[ \epsilon_{\mathrm{proc,X}} - \Delta \epsilon_{\mathrm{proc,X}} \right] < w_{\mathrm{proc,X}} = \mathrm{exp}\left[\epsilon_{\mathrm{proc,X}} \right] < \mathrm{exp}\left[ \epsilon_{\mathrm{proc,X}}+\Delta \epsilon_{\mathrm{proc,X}}\right]
		\label{eq:epsErrorRange}
	\end{equation}
	From the application of formula in Eq.~(\ref{eq:errorDef}) to Eq.~(\ref{eq:epsPhoto}), using also Eq.~(\ref{eq:relError}), we have the following formula for the uncertainty of exposure to photoionization:
	\begin{equation}
		\Delta \epsilon_{\mathrm{ph,X}}=\left( \delta F_\mathrm{ph}^2 + \delta \sigma_{\mathrm{ph,X}}^2 \right)^{1/2}\epsilon_{\mathrm{ph,X}}
		\label{eq:epsPhotoError}
	\end{equation}	
The formula for uncertainty of the charge exchange exposure is:
	\begin{equation}
		\Delta \epsilon_{\mathrm{cx,X}}=\left( \delta F_\mathrm{sw}^2 + \delta \sigma_{\mathrm{cx,X}}^2 \right)^{1/2}\epsilon_{\mathrm{cx,X}}
		\label{eq:epsCXError}
	\end{equation}
and for the uncertainty of exposure to electron impact the following:
	\begin{equation}
		\Delta \epsilon_{\mathrm{el,X}}=\left(\delta \sigma_{\mathrm{el,X}}^2 + \delta f_{\mathrm{el}}^2 + \delta n_{\mathrm{sw}}^2\right)^{1/2}\epsilon_{\mathrm{el,X}}
		\label{eq:epsElError}
	\end{equation}
Note that the uncertainties of exposure to individual ionization processes are expressed as products of the exposures themselves and terms composed of relative errors of the measured quantities. These uncertainties are collected in Table~\ref{tab:auxErrors}.

	\begin{table}
		\caption{Relative uncertainties of parameters adopted for estimates of survival probabilities uncertainties}
		\label{tab:auxErrors}
		\centering
		\begin{tabular}{cccc}
		\hline
		$\delta F_\mathrm{sw}$ & $\delta n_\mathrm{sw}$ & $\delta f_\mathrm{el}$ & $\delta F_\mathrm{ph}$\\ \hline \hline
		5\% & 20\% & 50\% & 10 \% \\ \hline
		& & & \\ \hline
		species & $\delta \sigma_{\mathrm{cx}}$ & $\delta \sigma_{\mathrm{ph}}$ & $\delta \sigma_{\mathrm{el}}$ \\ \hline \hline
		He & 10\% & 5\% & 10\% \\
		Ne & 20\% & 5\% & 15\% \\
		O & 25\% & 5\% & 10\% \\ \hline
		\end{tabular}
	\end{table}

Survival probability against all ionization reactions can be calculated from the sum of exposures to individual reactions:
	\begin{equation}
		w_{\mathrm{X}} = \mathrm{exp}\left[ \epsilon _{\mathrm{X}} \right] = \mathrm{exp}\left[ \epsilon_{\mathrm{ph,X}} + \epsilon_{\mathrm{cx,X}} + \epsilon_{\mathrm{el,X}} \right]
		\label{eq:surProFromProc}
	\end{equation}
and the uncertainty $\Delta \epsilon_{\mathrm{X}}$ of the total exposure $\epsilon_{\mathrm{X}}$ can be obtained from the general Eq.~(\ref{eq:errorDef}). This uncertainty is then inserted to Eq.~(\ref{eq:epsErrorRange}) to assess the uncertainty range of survival probabilities against all ionization processes.

\begin{figure}
		\begin{tabular}{c}
			\resizebox{\hsize}{!}{\includegraphics{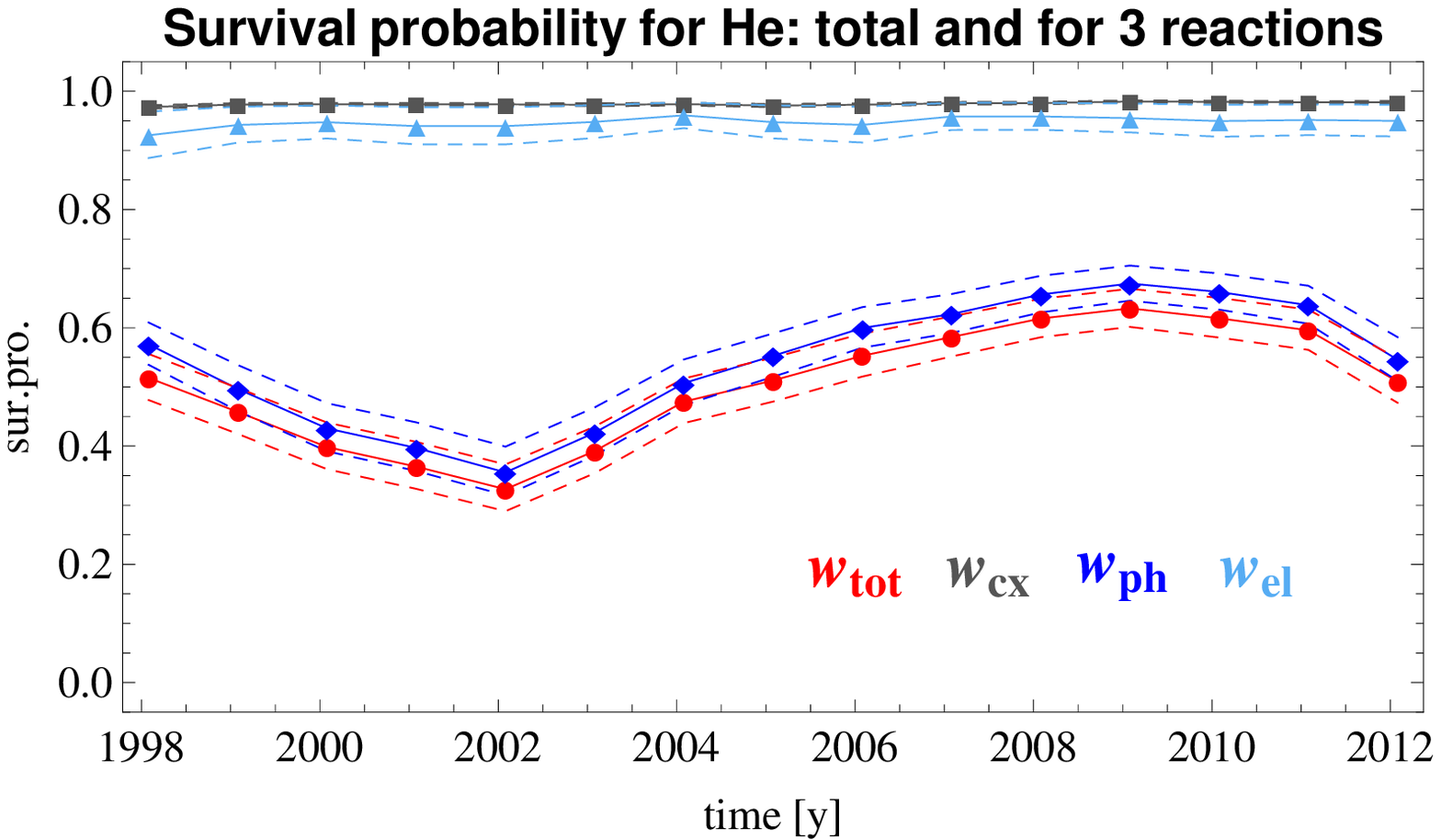}}\\
			\resizebox{\hsize}{!}{\includegraphics{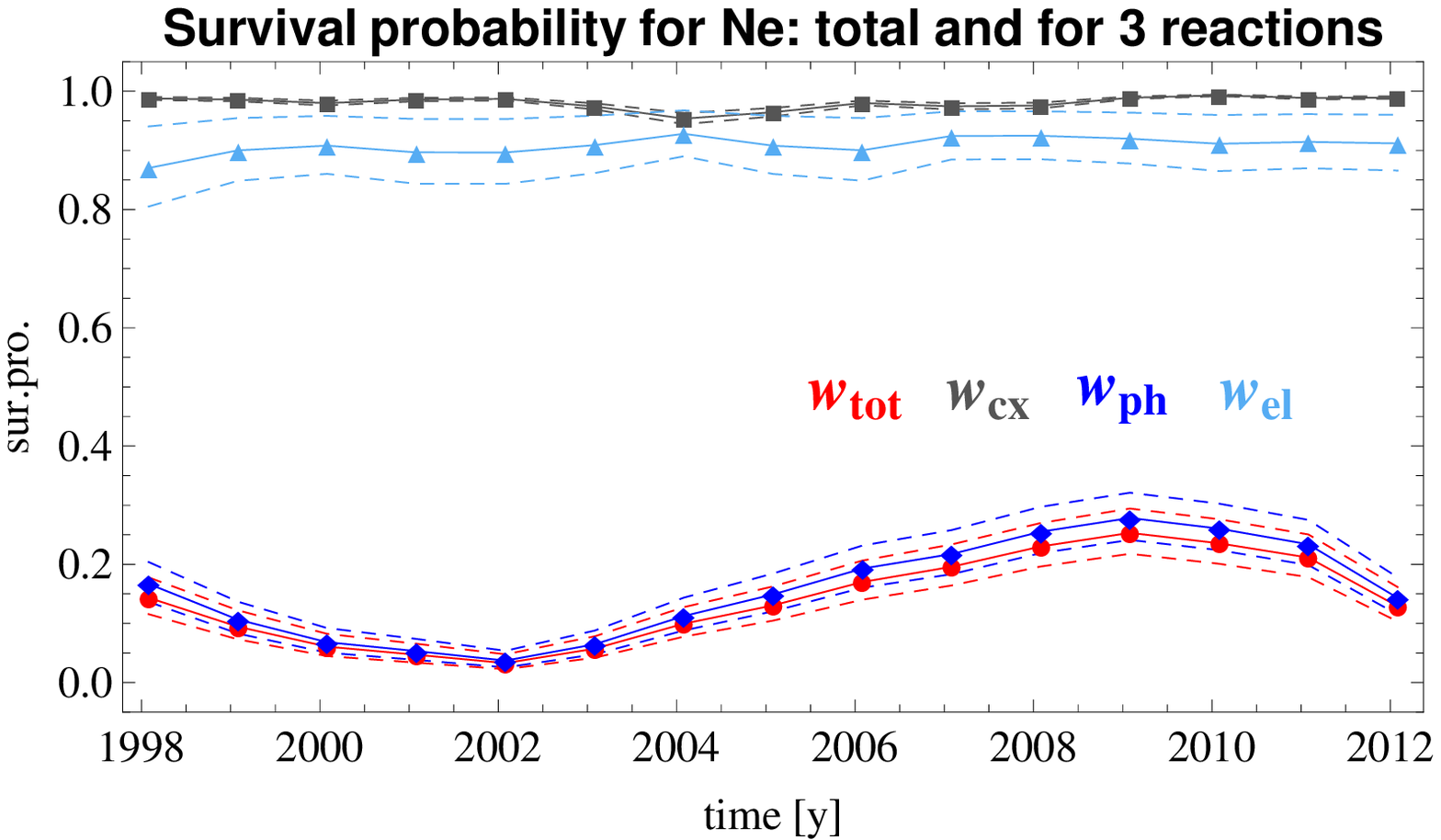}}\\
			\resizebox{\hsize}{!}{\includegraphics{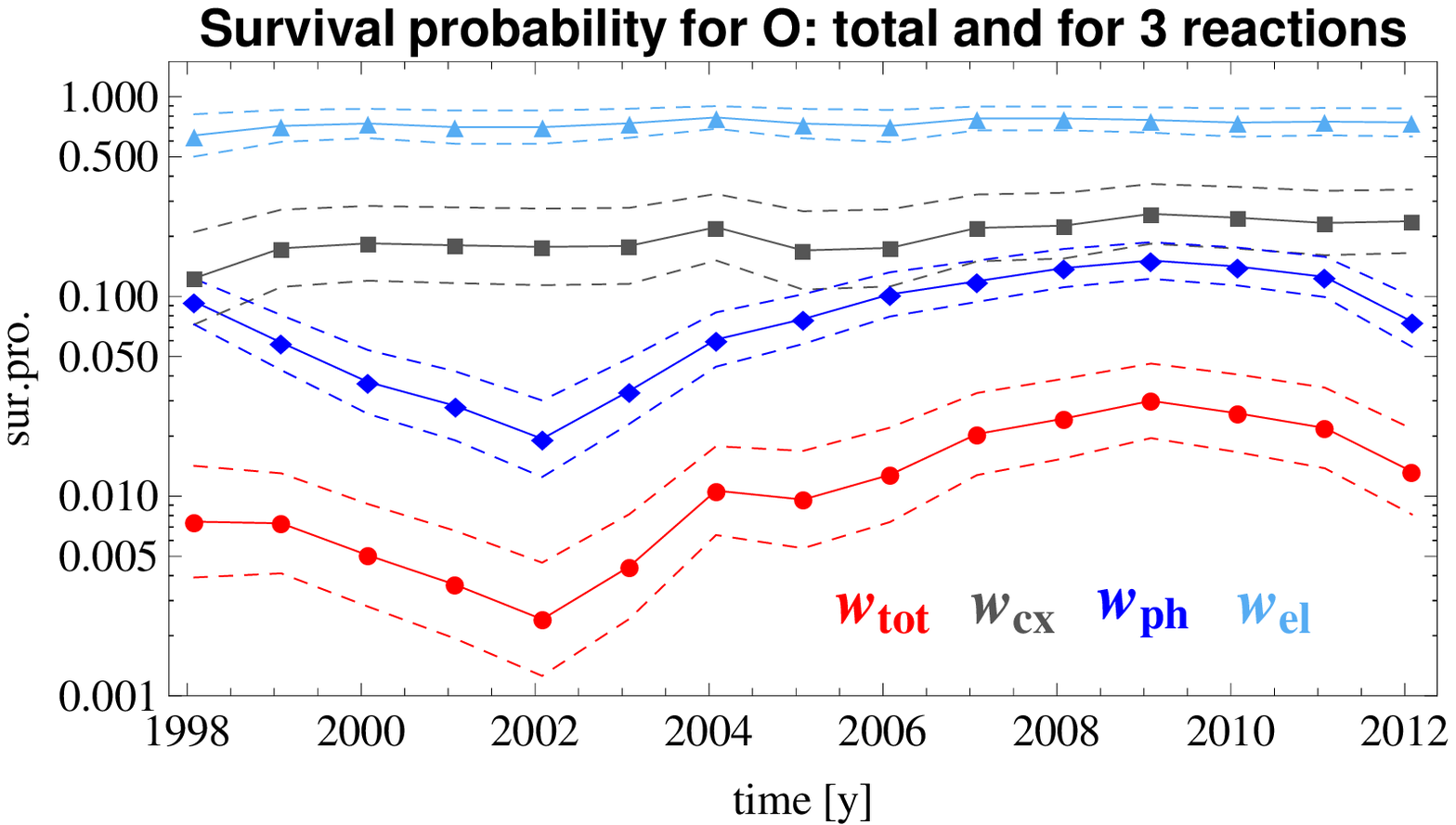}}\\
		\end{tabular}
		\caption{Total survival probabilities (red) and survival probabilities against charge exchange (gray), photoionization (blue) and electron impact ionization (pale blue). The solid lines mark the most probable values, the broken lines mark the uncertainty ranges calculated from Eqs.~(\ref{eq:epsPhotoError}), (\ref{eq:epsCXError}), (\ref{eq:epsElError}) and (\ref{eq:errorDef}). Note that the scale for oxygen (upper right is logarithmic.}
		\label{fig:procSurPro}
	\end{figure}

History of total survival probability of NIS~He, Ne, and O, as well as the breakdown of the probabilities into portions due to individual ionization reactions, is shown in Fig.~\ref{fig:procSurPro}. Note that an additional source of uncertainty, as mentioned above, is due to the uncertainty of the inflow velocity vector. The magnitude of this effect is illustrated by the broken line in Fig.~\ref{fig:surProHistory}. For all species, this uncertainty is less than the uncertainty due to the inaccuracy of the ionization rates.

As evident in Fig.~\ref{fig:procSurPro} (see also Fig.~\ref{fig:ionRates}), practically entire ionization losses of He are due to photoionization. Electron impact eliminates less than 10\% of the atoms from the original population and charge exchange is practically negligible, eliminating just $\sim1\%$ of the atoms, i.e., much less than the uncertainty of the losses. For Ne, photoionization dominates and charge exchange is negligible, but the role of electron impact is larger: it eliminates $\sim10\%$ of the original population. 

For oxygen, the situation is quite different. The ionization losses of O are the strongest, only $\sim 0.3\%$ of the atoms survives to Earth's orbit during high solar activity and up to $\sim 3\%$ during low activity. The main loss source is photoionization, charge exchange with solar wind protons is close second. A little surprising and up to now unrecognized important source of ionization is electron impact. This reaction eliminates as much as $\sim25\%$ of the original NIS~O population. Thus, this reaction cannot be neglected for O and Ne in the abundance studies. Neglecting electron impact for O causes an underestimation of the losses by $\sim25\%$, which may be mistaken for a deficit of NIS~O in the interstellar medium surrounding the heliosphere. 

The uncertainties of survival probabilities are substantial. The main source are uncertainties in the reaction cross sections. An exception is the electron impact ionization. Here, on top of the substantial cross section uncertainty, the main source is the uncertainty of the electron distribution function. The model we use is certainly a broad approximation, based on the limited set of available observations. A discussion of the approximations made will be provided elsewhere (Sok{\'o}{\l} et al., in preparation).
	
\subsection{Ratios of survival probabilities and application to abundance studies}
Using ratios of analytically derived survival probabilities of Ne and O to assess the Ne/O abundance at TS and further in the LIC \citep{bochsler_etal:12a} is equivalent to adopting the so-called cold-gas approximation \citep{blum_fahr:70a, axford:72}. In the cold-gas model the NIS gas far away from the Sun has density $n_{\mathrm{TS}}$ and is composed of individual atoms that all have identical velocity vectors, equal to the NIS gas inflow velocity $\vec{v_B}$. The Sun modifies this homogeneous density distribution by gravitational accretion and ionization. The local density $n_{\mathrm{cold}}\left(\vec{r}, \vec{v_B}\right)$ is due to atoms that follow two different hyperbolic trajectories intersecting at $\vec{r}$ and asymptotically parallel far away from the Sun. The local density is thus an algebraic sum of contributions from the so-called direct and indirect trajectories: $n_{\mathrm{cold}}\left(\vec{r}, \vec{v_B}\right) = n_{\mathrm{dir}}\left(\vec{r}, \vec{v_B}\right) + n_{\mathrm{indir}}\left(\vec{r}, \vec{v_B}\right)$, where the local densities of the direct and indirect components are products of the accretion factors $a\left(\vec{r}, \vec{v_B}\right)$, squares of the respective angular momenta $\vec{p_{\mathrm{dir}}}\left(\vec{r}, \vec{v_B}\right)$, $\vec{p_{\mathrm{indir}}}\left(\vec{r}, \vec{v_B}\right)$, and survival probabilities $w_{\mathrm{dir}}\left(\vec{r}, \vec{v_B}\right)$, $w_{\mathrm{indir}}\left(\vec{r}, \vec{v_B}\right)$:
		\begin{eqnarray}
			n_{\mathrm{cold}}\left(\vec{r},\vec{v_B}\right) & = & n_{\mathrm{TS}}a\left(\vec{r}, \vec{v_B}\right)p_{\mathrm{dir}}^2\left(\vec{r}, \vec{v_B}\right)w_{\mathrm{dir}}\left(\vec{r}, \vec{v_B}\right) \nonumber \\ 
			& + & n_{\mathrm{TS}}a\left(\vec{r}, \vec{v_B}\right) p_{\mathrm{indir}}^2\left(\vec{r}, \vec{v_B}\right)w_{\mathrm{indir}}\left(\vec{r}, \vec{v_B}\right)
		\label{eq:coldDens}
	\end{eqnarray}

For a non-moving observer at $\vec{r}$, the image of the NIS atoms flux will be two point sources (except exactly at the downwind line, where the model has singularity), and the flux will be 
	\begin{equation}
		\vec{F}_{\mathrm{cold}}\left(\vec{r}, \vec{v_B}\right)=\vec{v_{\mathrm{dir}}}n_{\mathrm{dir}}\left(\vec{r}, \vec{v_B}\right)+\vec{v_{\mathrm{indir}}}n_{\mathrm{indir}}\left(\vec{r}, \vec{v_B}\right)
		\label{eq:coldFlux}
	\end{equation}
where $\vec{v}_{\mathrm{dir}}$, $\vec{v}_{\mathrm{indir}}$ are velocity vectors at $\vec{r}$ of the direct and indirect components, respectively. For a given solar distance $r = \mid\vec{r}\mid$, the directions of these vectors depend on $\vec{v_B}$ and on the angle between the bulk velocity $\vec{v_B}$ and \vec{r}, but the magnitudes of velocities are equal and solely depend on $v_B = \mid\vec{v_B}\mid$ and $r$ due to the conservation of energy. The survival probabilities $w_{\mathrm{dir}}$, $w_{\mathrm{indir}}$ are related to each other by the respective angular momenta $\vec{p_{\mathrm{dir}}}$, $\vec{p_{\mathrm{indir}}}$ and by the angles swept by the atoms on the direct and indirect orbits. The survival probabilities for indirect atoms are typically much less than for direct ones; an exception is the cone region, where they become comparable. 

IBEX is able to observe NIS atoms close to their perihelia and thus it can only catch the direct population. The flux observed by IBEX $\vec{F}_{\mathrm{cold,IBEX}}$ will be amplified by the proper velocity of IBEX relative to the inflowing beam of NIS gas:
	\begin{equation}
		\vec{F}_{\mathrm{cold,IBEX}}=n_{\mathrm{TS}}a\left( \vec{r}, \vec{v_B} \right)p_{\mathrm{dir}}^2\left( \vec{r}, \vec{v_B} \right)w_{\mathrm{dir}}\left( \vec{r}, \vec{v_B} \right)\left( \vec{v_{\mathrm{dir}}} - \vec{v_{\mathrm{IBEX}}}\right)
		\label{eq:coldFluxIBEX}
	\end{equation}	
In the approximation that only the atoms at their perihelia can be observed, this equation is valid only for the IBEX orbits for which the Earth's velocity is directed dead against the velocity vector of the direct population:
	\begin{equation}
		\begin{split}
			F_{\mathrm{cold,IBEX,peri}} = & n_{\mathrm{TS}} a\left(  \vec{r_{\mathrm{peri}}},\vec{v_B}\right) p_{\mathrm{dir}}^2\left( \vec{r_{\mathrm{peri}}}, \vec{v_B} \right)w_{\mathrm{dir}}\left( \vec{r_{\mathrm{peri}}}, \vec{v_B} \right) \\
			& \left(\mid \vec{v_{\mathrm{dir,peri}}}\mid + \mid \vec{v_{\mathrm{IBEX}}}\mid \right)
		\end{split}
		\label{eq:coldFluxIBEX1}
	\end{equation}
	
This equation can be evaluated for any NIS species observed by IBEX. Assuming that all species have identical bulk velocity vectors, one immediately notices that the only two terms in Eq.~(\ref{eq:coldFluxIBEX1}) that differ between the species are $n_{\mathrm{TS}}$ and $w_{\mathrm{dir}}$. Hence, when the total counts at IBEX and survival probabilities are known, the abundance $\xi_{\mathrm{dens,TS,X/Y}}=n_{\mathrm{TS,X}}/n_{\mathrm{TS,Y}}$ of two species X,Y at TS is immediately calculated from the formula: 
	\begin{equation}
		\begin{split}
		\xi_{\mathrm{dens,TS,X/Y}} = & \left( F_{\mathrm{cold,IBEX,peri,X}} / F_{\mathrm{cold,IBEX,peri,Y}} \right) / \left( w_{\mathrm{dir,Y}} / w_{\mathrm{dir,X}}\right) = \\
		 = & \left( F_{\mathrm{cold,IBEX,peri,X}} / F_{\mathrm{cold,IBEX,peri,Y}} \right)  \xi_{\mathrm{sur,X/Y}}
		\end{split}
		\label{eq:xiDensCold}
	\end{equation}	
which can be directly applied to observations. The quotient 
	\begin{equation}
		\xi_{\mathrm{sur,X/Y}}=w_{\mathrm{sur,X}}/w_{\mathrm{sur,Y}}
		\label{eq:xiSur}
	\end{equation}
can be used to assess the X/Y species abundance at TS once the ratio of registered counts of respective species is retrieved from the data. In Tables~\ref{tab:abundTabSurProbBz} and \ref{tab:abundTabSurProbWitte} we list these factors for the bulk flow vectors obtained by \citet{bzowski_etal:12a} and \citet{witte:04}, respectively, for the Earth ecliptic longitudes for which the maxima of NIS flux were registered. Derivation of the ratios and their uncertainties, as well as their evolution in time are discussed in the following subsection.

\subsection{Time evolution of survival probabilities ratios and their uncertainties}
While the formula in Eq.~(\ref{eq:xiSur}) for the survival probabilities ratio is simple, we will rewrite it in a more elaborate form to better assess the uncertainties. The reason is that some of the uncertainty contributors are independent between the species, while others are related to each other and this ought to be taken into account in the calculation of the total uncertainty of $\xi_{\mathrm{sur,X/Y}}$:
	\begin{equation}
		\xi_{\mathrm{sur,X/Y}}=\left( w_{\mathrm{ph,X}} / w_{\mathrm{ph,Y}} \right) \left( w_{\mathrm{el,X}} / w_{\mathrm{el,Y}} \right) \left( w_{\mathrm{cx,X}} / w_{\mathrm{cx,Y}} \right)
		\label{eq:xiSurProcDef}
	\end{equation}
which is equivalent to:
	\begin{equation}
		\xi_{\mathrm{sur,X/Y}}=\mathrm{exp}\left[ \left( \epsilon_{\mathrm{ph,X}} - \epsilon_{\mathrm{ph,Y}} \right) + \left( \epsilon_{\mathrm{el,X}} - \epsilon_{\mathrm{el,Y}} \right) + \left( \epsilon_{\mathrm{cx,X}} - \epsilon_{\mathrm{cx,Y}} \right) \right]
		\label{eq:xiSurProc}
	\end{equation}

The term in the exponent can be expressed by a sum of the exposure difference functions $\epsilon_{\mathrm{X/Y}}$ for each of the three ionization processes:
	\begin{equation}
		\epsilon_{\mathrm{X/Y}}=\epsilon_{\mathrm{ph,X/Y}}+\epsilon_{\mathrm{el,X/Y}}+\epsilon_{\mathrm{cx,X/Y}}
		\label{eq:epsXY}
	\end{equation}
where $\epsilon_{\mathrm{proc,X/Y}}$ corresponds to the exposure difference for reaction \emph{proc}. 

We will not repeat the derivation of formulae for uncertainties for all three reactions, we will illustrate the issue of uncertainty correlations on an example of charge exchange.

The exposure difference for charge exchange is equal to:
	\begin{equation}
		\epsilon_{\mathrm{cx,X/Y}}=\int{F_{\mathrm{sw}}\left( t \right) \left( \sigma_{\mathrm{cx,X}}-\sigma_{\mathrm{cx,Y}}\right) \mathrm{d}t}
		\label{eq:epsCxXY}
	\end{equation}
which is a function of three variables: $F_{\mathrm{sw}}$, $\sigma_{\mathrm{cx,X}}$ and $\sigma_{\mathrm{cx,Y}}$. So, based on Eq.~(\ref{eq:errorDef}) and the earlier derivation for the exposure to charge exchange (Eq.~(\ref{eq:epsCXError})), we have the uncertainty of $\epsilon_{\mathrm{cx,X/Y}}$ equal to:
	\begin{equation}\
		\begin{split}
		\Delta \epsilon_{\mathrm{cx,X/Y}}= & \left[ \left( \delta F_{\mathrm{sw}}^2+\delta \sigma_{\mathrm{cx,X}}^2 \right) \epsilon_{\mathrm{cx,X}}^2 + \left( \delta F_{\mathrm{sw}}^2 + \delta \sigma_{\mathrm{cx,Y}}^2\right)\epsilon_{\mathrm{cx,Y}}^2 + \right. \\
		& \left. \left( \delta F_{\mathrm{sw}} \epsilon_{\mathrm{cx,X/Y}} \right)^2 \right]^{1/2}
		\end{split}
		\label{eq:epsCxXYError}
	\end{equation}

Similarly, the uncertainties of the exposure differences for photoionization is given by:
	\begin{equation}\
		\Delta \epsilon_{\mathrm{ph,X/Y}}= \left[ \left( \delta \sigma_{\mathrm{ph,X}} \epsilon_{\mathrm{ph,X}} \right)^2+ \left( \delta \sigma_{\mathrm{ph,Y}} \epsilon_{\mathrm{ph,Y}}\right)^2 +  \left( \delta F_{\mathrm{ph}} \epsilon_{\mathrm{ph,X/Y}} \right)^2 \right]^{1/2}
		\label{eq:epsPhXYError}
	\end{equation}
and the exposure difference for electron impact by:
	\begin{equation}\
		\Delta \epsilon_{\mathrm{el,X/Y}}= \left[ \left( \delta \sigma_{\mathrm{el,X}} \epsilon_{\mathrm{el,X}} \right)^2+ \left( \delta \sigma_{\mathrm{el,Y}} \epsilon_{\mathrm{el,Y}}\right)^2 +  \left( \delta f_{\mathrm{el}}^2 +\delta n_{\mathrm{sw}}^2 \right) \epsilon_{\mathrm{el,X/Y}}^2 \right]^{1/2}
		\label{eq:epsElXYError}
	\end{equation}
	
The total uncertainty of the exposure difference is given by 
	\begin{equation}\
		\Delta \epsilon_{\mathrm{el,X/Y}}= \left[ \Delta \epsilon_{\mathrm{ph,X/Y}}^2 + \Delta \epsilon_{\mathrm{cx,X/Y}}^2 + \Delta \epsilon_{\mathrm{el,X/Y}}^2  \right]^{1/2}
		\label{eq:epsXYError}
	\end{equation}
and the lower $\xi^{-}_{\mathrm{sur,X/Y}}$ and upper $\xi^{+}_{\mathrm{sur,X/Y}}$ uncertainty ranges for the survival probability ratio $\xi_{\mathrm{sur,X/Y}}$ are given by:
	\begin{equation}
		\begin{split}
		\xi^{-}_{\mathrm{sur,X/Y}} = & \mathrm{exp}\left[ \epsilon_{\mathrm{X/Y}} - \Delta \epsilon_{\mathrm{X/Y}}  \right] < \xi_{\mathrm{sur,X/Y}} < \\
		& < \mathrm{exp}\left[ \epsilon_{\mathrm{X/Y}} + \Delta \epsilon_{\mathrm{X/Y}}  \right] = \xi^{+}_{\mathrm{sur,X/Y}}
		\end{split}
		\label{eq:xiSurRange}
	\end{equation}
	
\begin{figure}
		\begin{tabular}{c}
			\resizebox{\hsize}{!}{\includegraphics{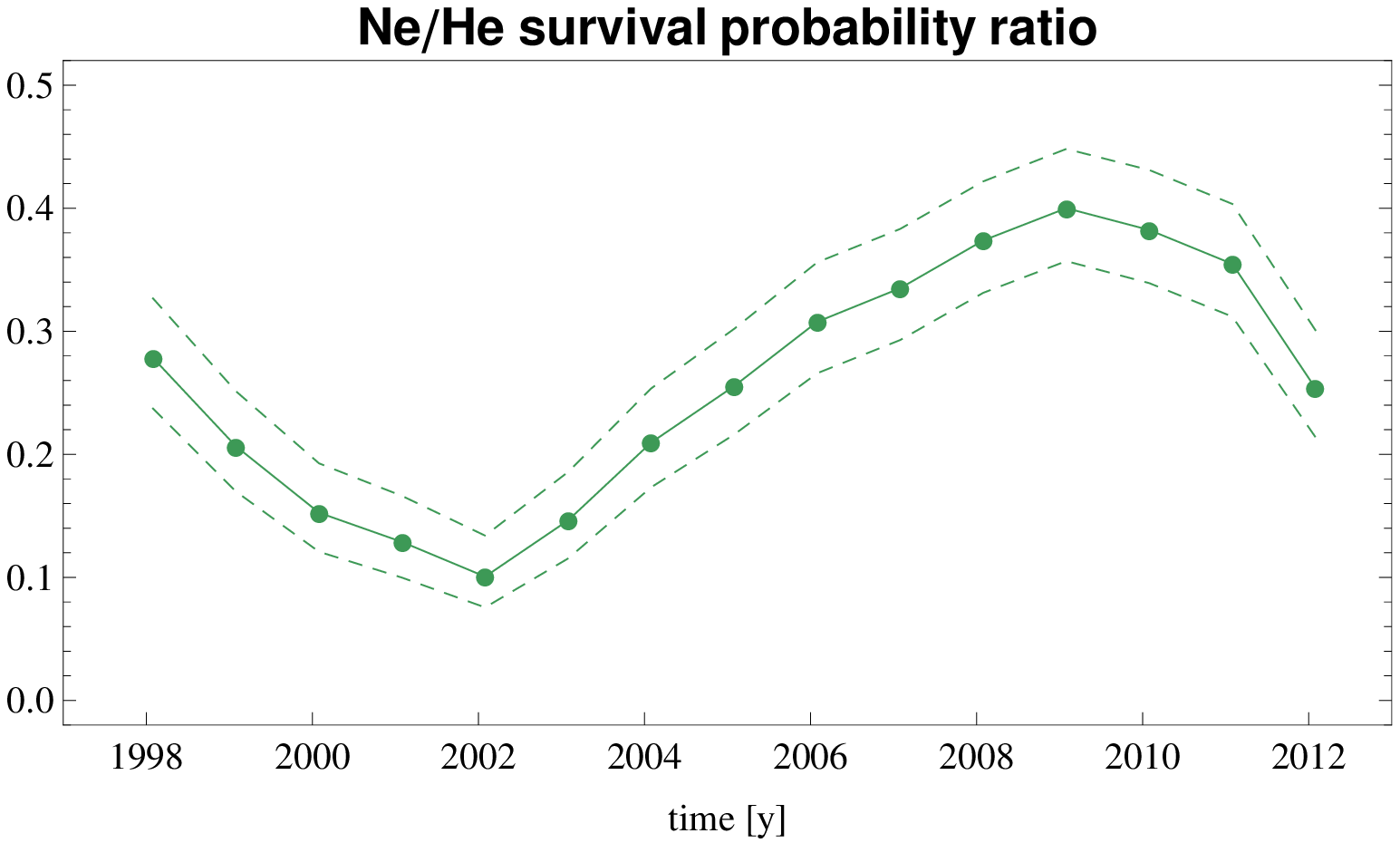}}\\
			\resizebox{\hsize}{!}{\includegraphics{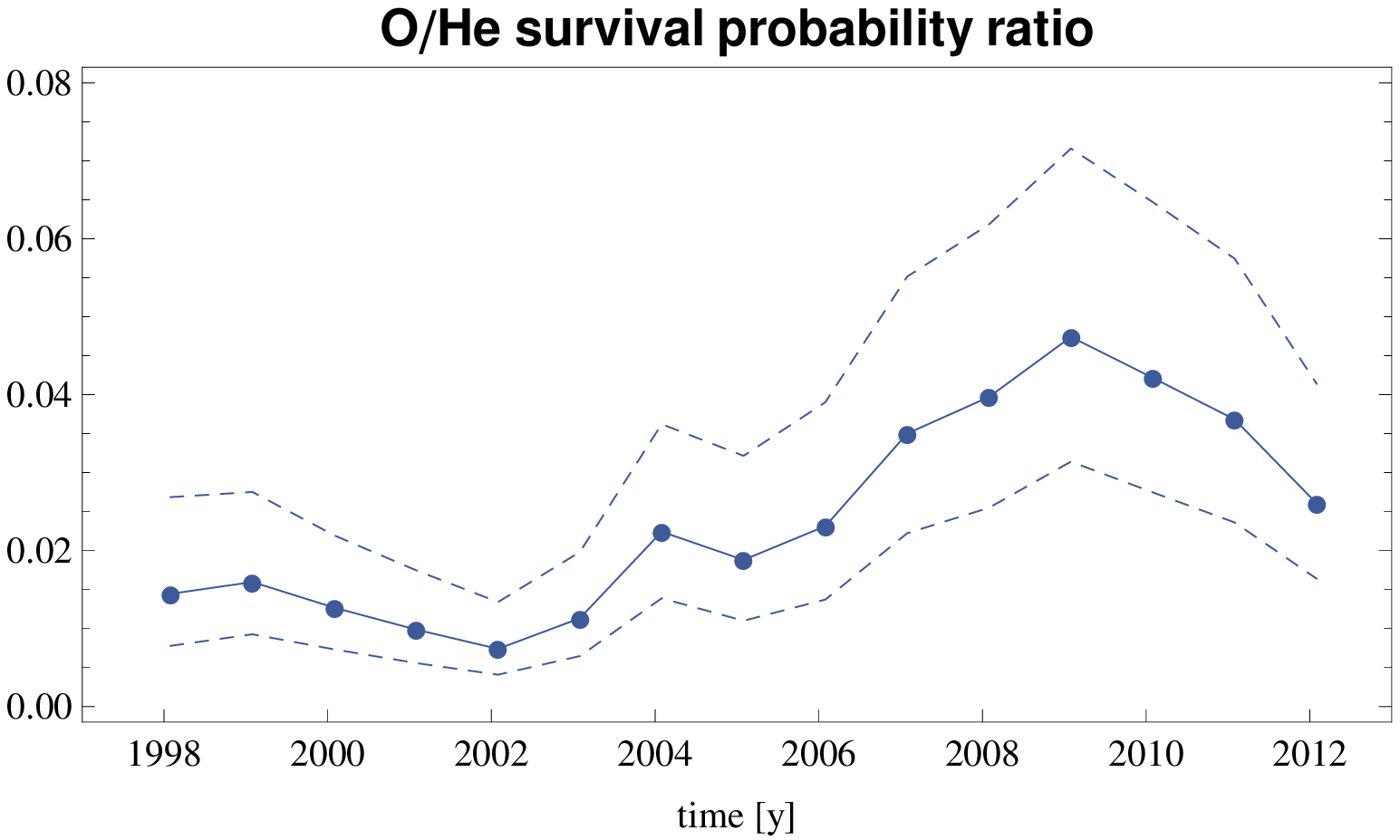}}\\
			\resizebox{\hsize}{!}{\includegraphics{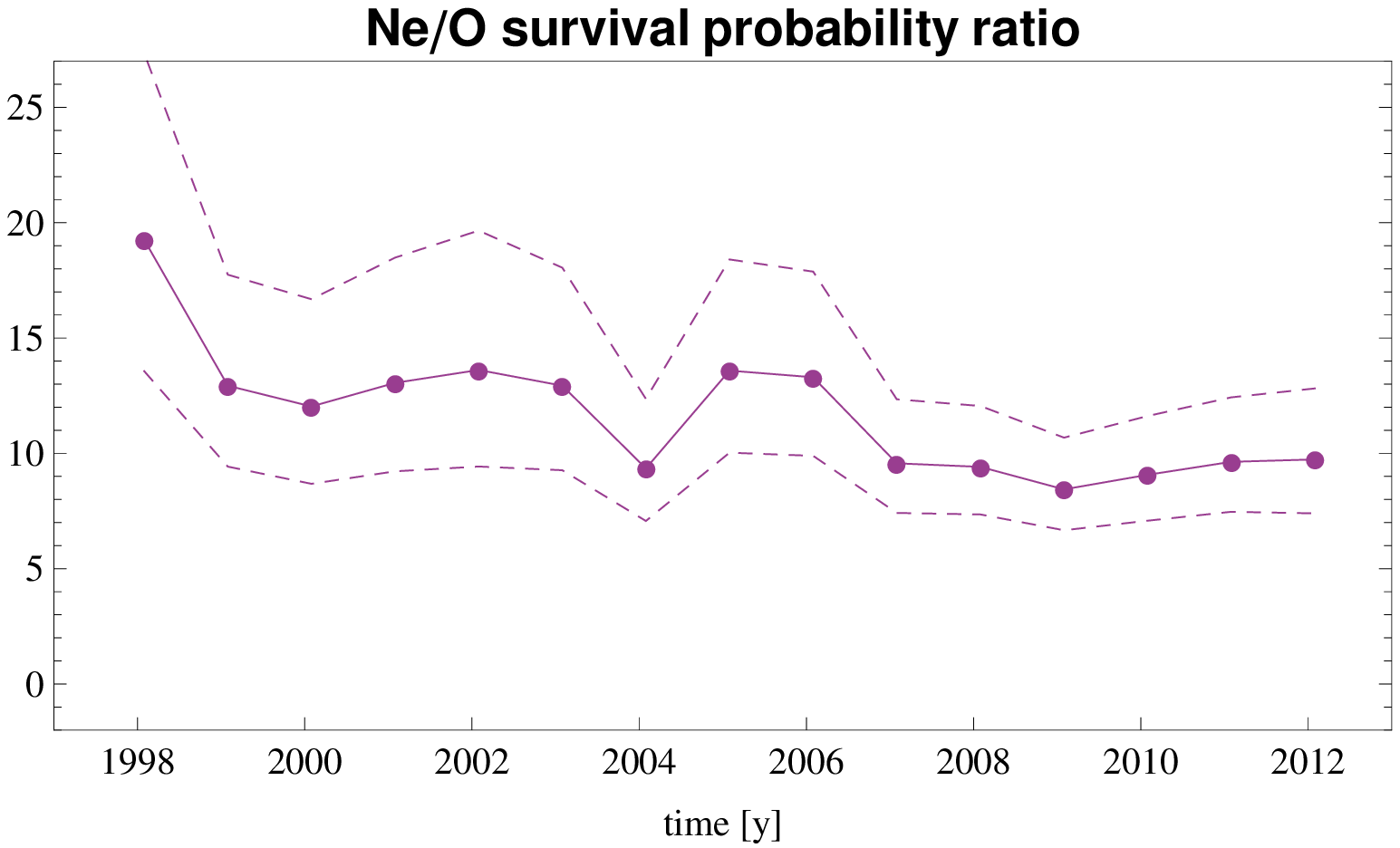}}\\
		\end{tabular}
		\caption{Modulation of the survival probability ratios for the Ne/He (upper left), O/He (lower left) and Ne/O (upper right) species pairs, drawn with solid lines. The broken lines mark the uncertainties calculated using Eqs.~(\ref{eq:epsCxXYError})--(\ref{eq:xiSurRange}).}
		\label{fig:xiSurHistory}
	\end{figure}

History of survival probability ratios for the Ne/He, O/He, and Ne/O pairs is shown in Fig.~\ref{fig:xiSurHistory}. The mean values of the ratios starkly differ among the three pairs of species, which is understandable given very different typical ionization rates for the three species. All three survival probability ratios feature significant modulation during the solar cycle, but the character of the modulation differs among the pairs. This is because the proportions between the ionization rates due to the three relevant ionization reactions differ among the species. 

The dominant ionization reactions for He and Ne is photoionization, with some addition from electron impact, which is larger in the case of Ne. Since the photoionization rate features a clear modulation as a function of the solar cycle phase and since the magnitude of photoionization is larger for Ne than for He, the modulation of the $\xi_{\mathrm{sur,Ne/He}}$ ratio follows the solar cycle modulation of photoionization in antiphase. The amplitude of the modulation is relatively mild: $0.1 < \xi_\mathrm{sur,Ne/He} < 0.4$.

Oxygen is ionized much stronger than helium and among the contributing reactions only photoionization features a quasi-periodic solar cycle behavior (cf Fig.~\ref{fig:ionRates}). The contribution to the losses from electron ionization and charge exchange are (jointly) comparable to the contribution from photoionization and they are not periodical: they show a drop (and thus an increase in survival probability) since the previous maximum, which seems to have leveled off about 2008. Thus, $\xi_\mathrm{sur,O/He}$ has increased since the previous maximum of solar activity and shows a superimposed modulation in antiphase with the solar cycle. Since the total ionization rates for O are much larger than for He, the $\xi_\mathrm{sur,O/He}$ ratio is an order of magnitude less than $\xi_\mathrm{sur,Ne/He}$ and varies during the solar cycle almost by an order of magnitude, between $\sim0.006$ and $\sim0.04$. 

The $\xi_\mathrm{sur,Ne/O}$ ratio (where, by the way, $\xi_\mathrm{sur,Ne/O}=\xi_\mathrm{sur,Ne/He}/\xi_\mathrm{sur,O/He}$) varies during the solar cycle, but the quasi-periodic modulation is hidden by the solar wind-related effects discussed earlier. Between 1999 and 2012, it varied between 8.3 and 13 and during the time after IBEX launch remained relatively stable. 

The uncertainties of the survival probabilities ratios are considerable. They vary in time, but typically are approximately $\pm15\%$ for $\xi_\mathrm{sur,Ne/He}$, $\left(-35\%, +50\% \right)$ for $\xi_\mathrm{sur,O/He}$, and $\pm30\%$ for $\xi_\mathrm{sur,Ne/O}$. This estimate is obtained taking into account solely the uncertainties of the ionization rates. In reality, as indicated earlier, they are also affected by uncertainties of the gas inflow vector parameters. Differences between the survival probabilities in time for two sets of the (speed, direction) pairs are illustrated by broken lines in Fig.~\ref{fig:surProHistory}, where $w$ for ecliptic longitudes and speeds of NIS~He inferred by \citet{bzowski_etal:12a} and by \citet{witte:04} are shown (respectively, $\left(v = 22.8~\mathrm{km}~\mathrm{s}^{-1}, \lambda = 259.2\degr \right)$ and $\left(v = 26.3~\mathrm{km}~\mathrm{s}^{-1}, \lambda = 255.4\degr \right)$). In the following, we will assess the magnitude of this additional uncertainty and the total uncertainty due to the uncertainties of the ionization rate and inflow parameters together. 

\begin{figure*}
		\begin{tabular}{ccc}
		\includegraphics[width=.3\textwidth]{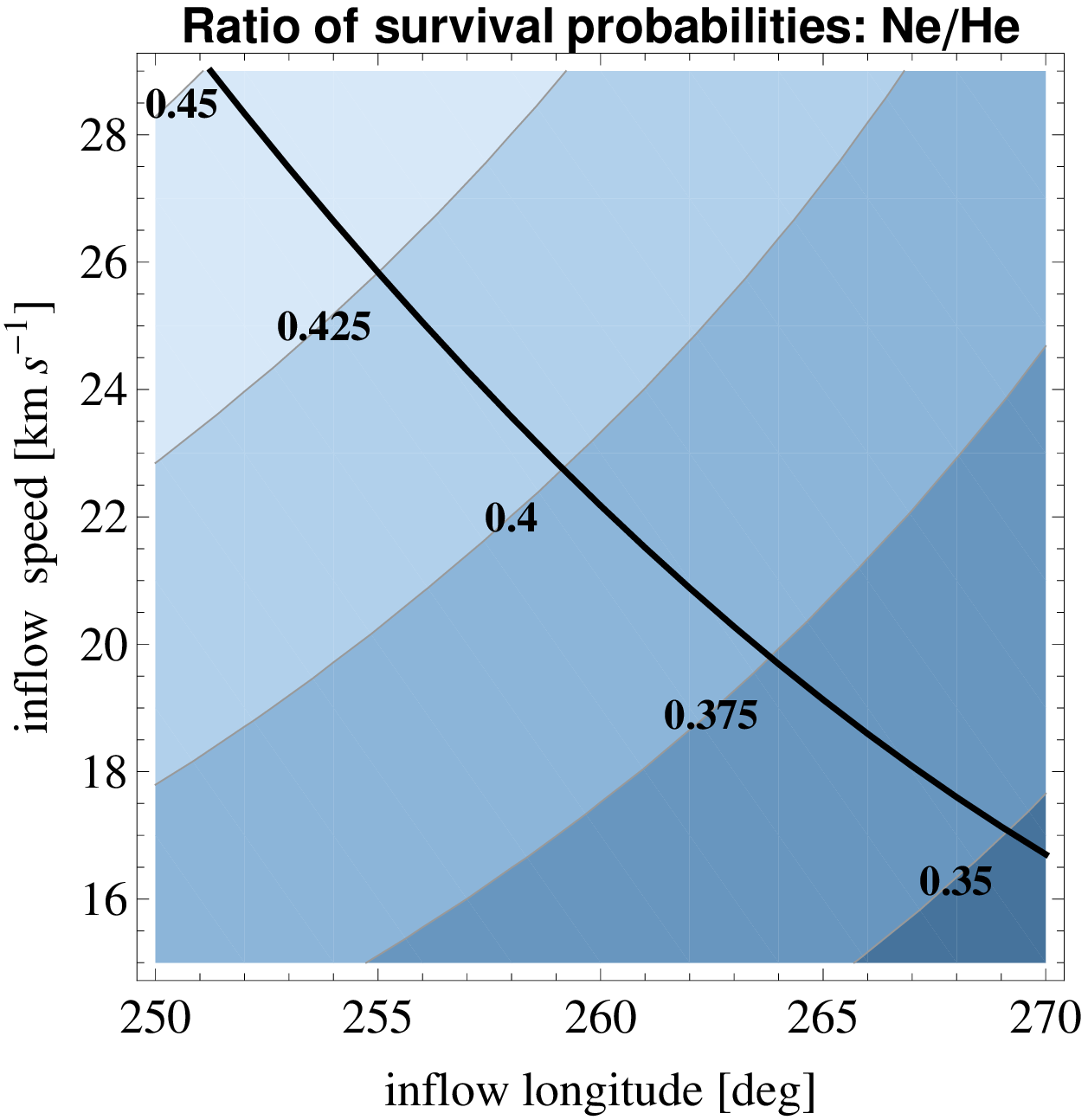} & 	\includegraphics[width=.3\textwidth]{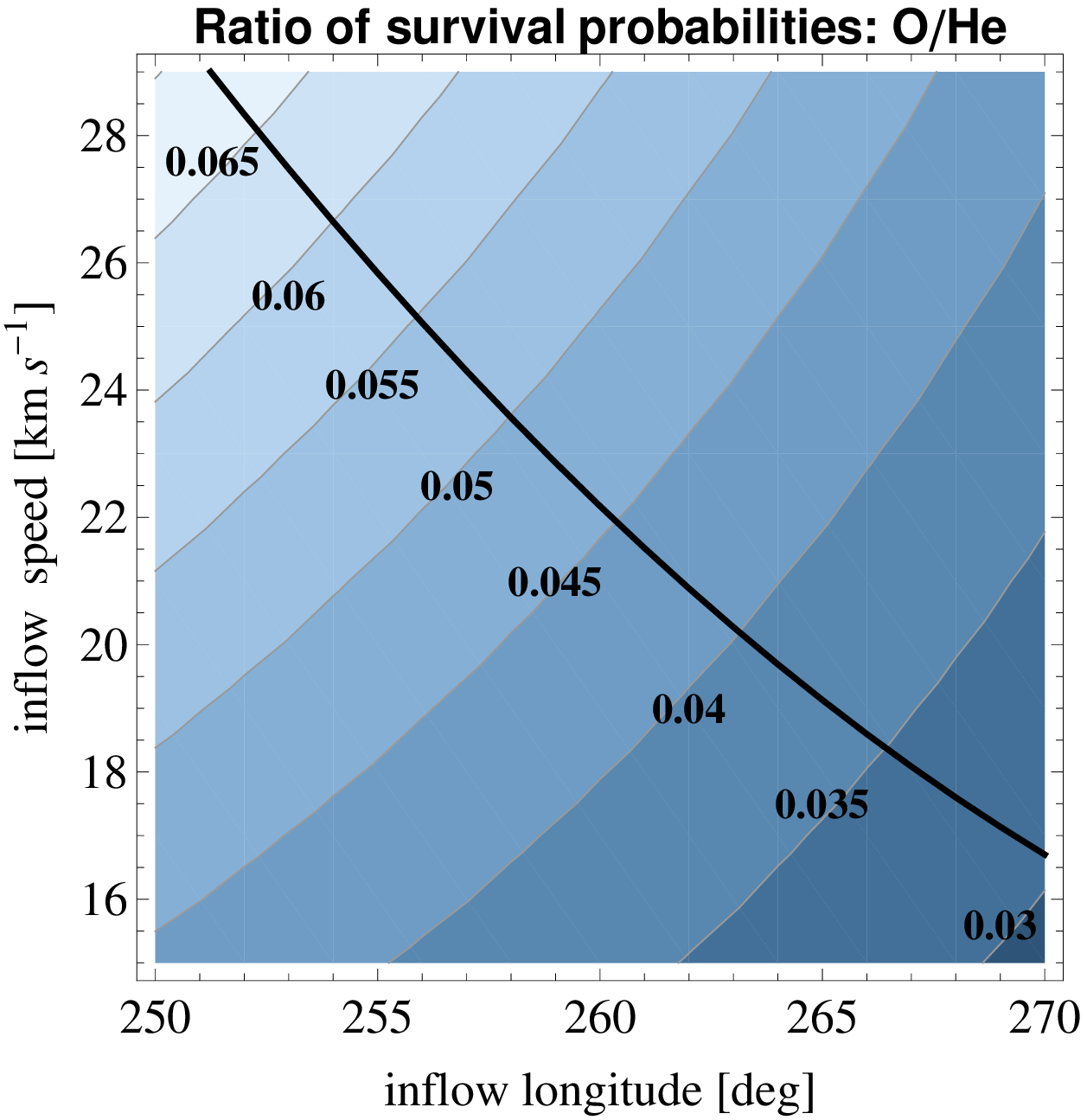} & 	\includegraphics[width=.3\textwidth]{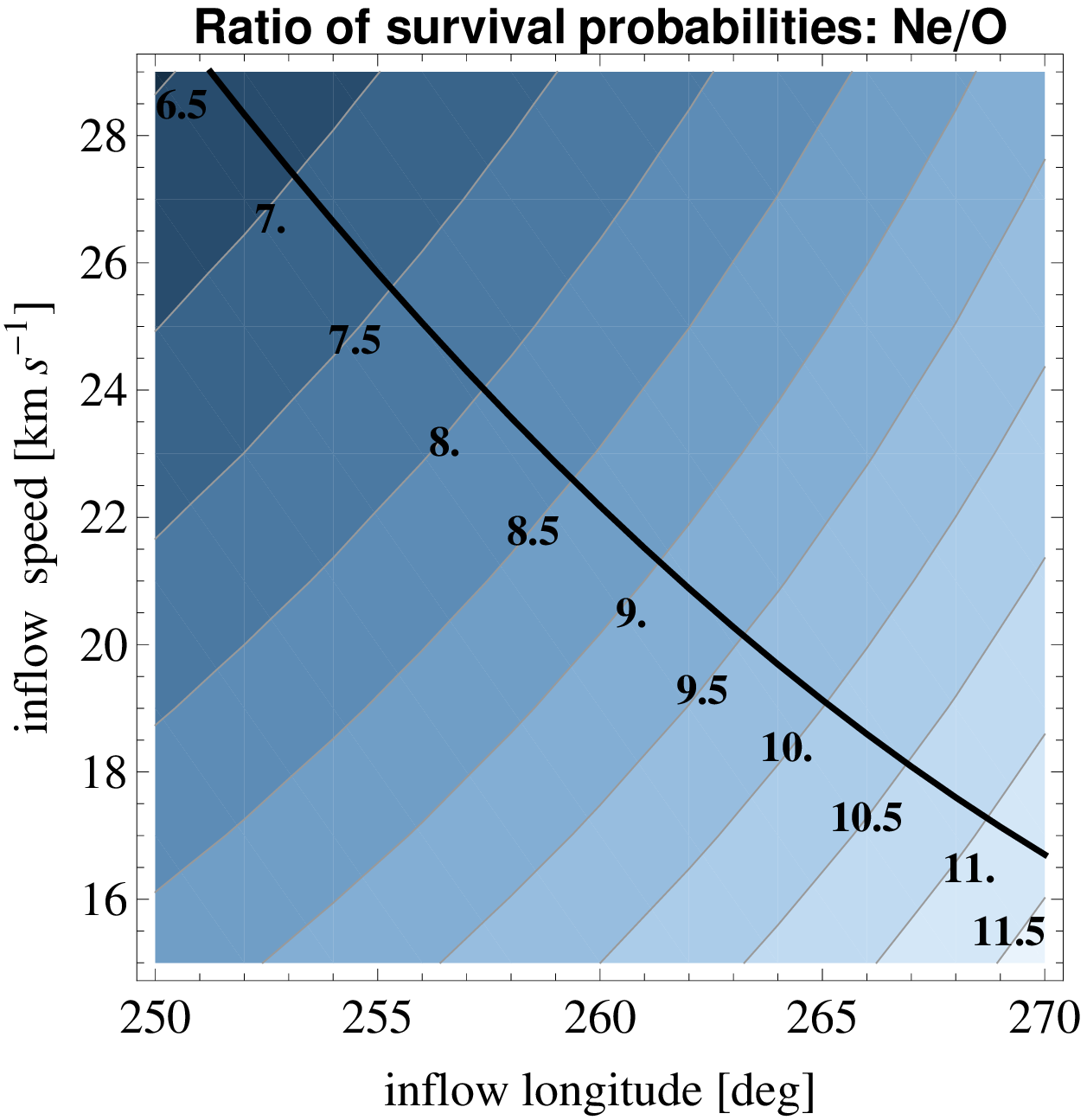} \\
		\end{tabular}
		\caption{Ratios of survival probabilities for Ne/He, O/He, and Ne/O, equivalent to the quotients $\xi_{\mathrm{sur,X/Y}}$ defined in Eq.~(\ref{eq:xiSur}), for various combinations of the flow vector of the NIS gas in the LIC. The solid lines indicate the acceptable parameter region established by \citet{bzowski_etal:12a} for the NIS gas flow parameters based on IBEX observations.}
		\label{fig:surProRatios}
	\end{figure*}

IBEX showed that acceptable solutions for the inflow parameters are correlated with each other \citep{bzowski_etal:12a, mobius_etal:12a}. These correlation lines are drawn with the black lines in Fig.~\ref{fig:surProRatios}. For our assessment of the uncertainty of $\xi_\mathrm{sur,X/Y}$, we thus take two $\xi_\mathrm{sur,X/Y}$ values for the parameters at the black lines in Fig.~\ref{fig:surProRatios}: one for the inflow parameters obtained from IBEX (i.e., the nominal case discussed in the paper up to now: $\lambda = 259.2\degr$), and the other one corresponding to the longitude of inflow direction as obtained by \citet{witte:04} from GAS/Ulysses (i.e., $\lambda = 255.4\degr$). The additional uncertainty will be assessed from the $\xi_\mathrm{sur,X/Y}$ values obtained from Fig.~\ref{fig:surProRatios} for these two parameter sets. We denote the two $\xi$ values as $\xi_0$ and $\xi_1$, respectively.

To assess the uncertainty range of $\xi_\mathrm{sur,X/Y}$, we calculate the exposure differences
	\begin{equation}\
		\epsilon_{\mathrm{X/Y,0}}=\ln{\xi_0}, \quad \epsilon_{\mathrm{X/Y,1}}=\ln{\xi_1}
		\label{eq:epsDiffParam}
	\end{equation}
The total uncertainty of the exposure difference, resulting from both the uncertainty of the ionization rate and uncertainty of the inflow parameters, will be given by 
	\begin{equation}\
		\Delta \epsilon_{\mathrm{X/Y,tot}}=\left[ \Delta \epsilon_{\mathrm{X/Y}}^2 + \left( \epsilon_{\mathrm{X/Y,0}} - \epsilon_{\mathrm{X/Y,1}} \right)^2 \right]^{1/2}
		\label{eq:epsDiffTot}
	\end{equation}
and the uncertainty is calculated by inserting this value to Eq.~(\ref{eq:xiSurRange}). The increase of uncertainties is relatively small: the full uncertainties are equal to $\pm20\%$ for $\delta \xi_{\mathrm{sur,Ne/He}}$,  $\left( -40\%, +60\% \right)$ for $\delta \xi_{\mathrm{sur,O/He}}$, $\pm35\%$ for $\delta \xi_{\mathrm{sur,Ne/O}}$.

In reality, however, the cold gas approximation is not fully adequate for the abundance studies because the thermal velocities of the NIS species $u_T = \sqrt{2kT/m}$ (where $T$ is temperature, $k$ the Boltzmann constant, and $m$ atomic mass) are not 0 and thus the velocities of individual atoms in the LIC are neither identical, nor parallel. Indeed, the temperature of the gas is finite and the thermal spread of velocities of individual atoms in the LIC is substantial when compared with the bulk flow speed. For $T = \sim6000$~K, the thermal speed is equal to $4.9~\mathrm{km}~\mathrm{s}^{-1}$ for He, $2.5~\mathrm{km}~\mathrm{s}^{-1}$ for O and $2.2~\mathrm{km}~\mathrm{s}^{-1}$ for Ne. Our calculations showed that the directions of velocity vectors of the atoms actually observed by IBEX vary in the LIC by more than $15\degr$. Combinations of speeds and velocity directions of atoms in the LIC cover practically the whole range of parameters shown in Fig.~\ref{fig:surProRatios}. 

The survival probabilities depend not only on the details of the ionizing factors but also on details of the trajectories, which are governed by their velocity vectors in the LIC. The ratio of survival probabilities of species X and Y is a function of the velocity vectors in the LIC, as illustrated in Fig.~\ref{fig:surProRatios}. In this figure, we illustrate the values of  $\xi_{\mathrm{sur,X/Y}}$ pairwise for He, Ne, and O at 1~AU from the Sun at the ecliptic longitude $130\degr$, corresponding to the peak flux registered by IBEX in 2009, for various combinations of speed and ecliptic longitude of the inflow direction. 

The range of $\xi_{\mathrm{sur,X/Y}}$ values for various combinations of speed and inflow direction is quite substantial. This indicates that studies of NIS abundances by direct sampling at 1~AU from the Sun require an accurate knowledge of the gas inflow speed and direction, but also raises concern whether the ratios of survival probabilities calculated as presented above credibly represent the true change in the NIS abundances between TS and 1~AU from the Sun. This will be verified in the next two sections.

	\begin{table*}
		\caption{Survival probabilities $w$ of He, Ne, and O atoms reaching their perihelia at $\sim130\degr$  ecliptic longitude, and respective heliospheric abundance quotients $\xi_{\mathrm{X/Y}}$ calculated from respective survival probabilities, for the velocity vector of the atoms in the LIC as established by \citet{bzowski_etal:12a} from IBEX observations.}
		\label{tab:abundTabSurProbBz}
		\centering
		\begin{tabular}{llllllll}
		\hline
			year & $\lambda_E$ & $w_{\mathrm{He}}$ & $w_{\mathrm{Ne}}$ & $w_{\mathrm{O}}$ & $\xi_{\mathrm{sur,Ne/He}}$ & $\xi_{\mathrm{sur,O/He}}$ & $\xi_{\mathrm{sur,Ne/O}}$ \\
   \hline \hline
 			2009.08 & 130.61 & 0.633 & 0.253 & 0.0300 & 0.400 & 0.0474 & 8.44 \\ 
 			2010.08 & 130.35 & 0.616 & 0.236 & 0.0260 & 0.382 & 0.0422 & 9.07 \\
 			2011.08 & 130.09 & 0.596 & 0.211 & 0.0219 & 0.355 & 0.0368 & 9.63 \\
 			2012.08 & 129.83 & 0.509 & 0.129 & 0.0133 & 0.254 & 0.0261 & 9.74 \\ \hline
		\end{tabular}
	\end{table*}
	
	\begin{table*}
		\caption{Same as Table~\ref{tab:abundTabSurProbBz}, but for the velocity vectors in the LIC as established for He by \citet{witte_etal:04a}.}
		\label{tab:abundTabSurProbWitte}
		\centering
		\begin{tabular}{llllllll}
		\hline
			year & $\lambda_E$ & $w_{\mathrm{He}}$ & $w_{\mathrm{Ne}}$ & $w_{\mathrm{O}}$ & $\xi_{\mathrm{sur,Ne/He}}$ & $\xi_{\mathrm{sur,O/He}}$ & $\xi_{\mathrm{sur,Ne/O}}$ \\
   \hline \hline
 			2009.08 & 130.61 & 0.652 & 0.278 & 0.0375 & 0.426 & 0.0575 & 7.41 \\ 
 			2010.08 & 130.35 & 0.635 & 0.258 & 0.0325 & 0.407 & 0.0512 & 7.94 \\
 			2011.08 & 130.09 & 0.562 & 0.233 & 0.0277 & 0.379 & 0.0450 & 8.42 \\
 			2012.08 & 129.83 & 0.529 & 0.146 & 0.0171 & 0.275 & 0.0323 & 8.53 \\
 			\hline
		\end{tabular}
	\end{table*}
	
		\begin{table*}
		\caption{Times and ecliptic longitudes for the calculation local relative densities $n_{\mathrm{X,rel}}$ (Eq.~(\ref{eq:relDens})) for NIS~He, Ne, and O and the heliospheric abundance quotients $\xi_{\mathrm{dens,X/Y}}$ calculated for respective pairs of relative densities (Eq.~(\ref{eq:xiDens})) for IBEX NIS flow parameters.}
		\label{tab:abundTableDensIbex}
		\centering
		\begin{tabular}{llllllll}
		\hline
			year & $\lambda_E$ & $n_{\mathrm{He,rel}}$ & $n_{\mathrm{Ne,rel}}$ & $n_{\mathrm{O,rel}}$ & $\xi_{\mathrm{dens,Ne/He}}$ & $\xi_{\mathrm{dens,O/He}}$ & $\xi_{\mathrm{dens,Ne/O}}$ \\
   \hline \hline
 			2009.09 & 132.64 & 1.30 & 0.391 & 0.0400 & 0.308 & 0.0309 & 9.97\\ 
 			2010.09 & 133.40 & 1.24 & 0.368 & 0.0356 & 0.296 & 0.0286 & 10.3 \\
 			2011.09 & 134.15 & 1.18 & 0.333 & 0.0309 & 0.281 & 0.0261 & 10.8 \\
 			2012.08 & 129.83 & 1.01 & 0.205 & 0.0177 & 0.203 & 0.0175 & 11.6 \\
 			\hline
		\end{tabular}
	\end{table*}

		\begin{table*}
		\caption{Times and ecliptic longitudes for the calculation the flux abundance quotients pairwise for He, Ne, and O averaged over the IBEX NIS sampling seasons and ($\xi_{\mathrm{flux,X/Y}}$; Eq.~(\ref{eq:xiFlux})) and the quotients integrated over the orbits when Earth was crossing the NIS beam $\xi_{\mathrm{flux,X/Y}}^{\mathrm{peak}}$.}
		\label{tab:abundTableFluxIBEX}
		\centering
		\begin{tabular}{llllllll}
		\hline
			year & $\lambda_E$ & $\xi_{\mathrm{flux,Ne/He}}$ & $\xi_{\mathrm{flux,O/He}}$ & $\xi_{\mathrm{flux,Ne/O}}$ & $\xi_{\mathrm{flux,Ne/He}}^{\mathrm{peak}}$ & $\xi_{\mathrm{flux,O/He}}^{\mathrm{peak}}$ & $\xi_{\mathrm{flux,Ne/O}}^{\mathrm{peak}}$ \\
   \hline \hline
 			2009.09 & 132.64 & 0.492 & 0.0517 & 9.52 & 0.578 & 0.0602 & 9.61 \\ 
 			2010.09 & 133.40 & 0.483 & 0.0490 & 9.85 & 0.551 & 0.0552 & 9.98 \\
 			2011.09 & 134.15 & 0.468 & 0.0459 & 10.2 & 0.522 & 0.0495 & 10.6 \\
 			2012.08 & 129.83 & 0.289 & 0.0302 & 9.59 & 0.323 & 0.0271 & 11.9 \\
 			\hline
		\end{tabular}
	\end{table*}

		\begin{table*}
		\caption{Same as Table~\ref{tab:abundTableFluxIBEX}, but for the inflow parameters of the gas as from \citet{witte:04}.}
		\label{tab:abundTableFluxWitte}
		\centering
		\begin{tabular}{llllllll}
		\hline
			year & $\lambda_E$ & $\xi_{\mathrm{flux,Ne/He}}$ & $\xi_{\mathrm{flux,O/He}}$ & $\xi_{\mathrm{flux,Ne/O}}$ & $\xi_{\mathrm{flux,Ne/He}}^{\mathrm{peak}}$ & $\xi_{\mathrm{flux,O/He}}^{\mathrm{peak}}$ & $\xi_{\mathrm{flux,Ne/O}}^{\mathrm{peak}}$ \\
   \hline \hline
   		2009.09 & 132.64 & 0.518 & 0.0632 & 8.21 & 0.592 & 0.0701 & 8.44 \\ 
 			2010.09 & 133.40 & 0.505 & 0.0598 & 8.44 & 0.565 & 0.0646 & 8.74 \\
 			2011.09 & 134.15 & 0.508 & 0.0571 & 8.88 & 0.542 & 0.0586 & 9.23 \\
 			2012.08 & 129.83 & 0.366 & 0.0444 & 8.23 & 0.413 & 0.0391 & 10.6 \\
 			\hline
		\end{tabular}
	\end{table*}

\section{Evolution of densities of NIS~He, Ne, and O at Earth}

To assess the quality of approximating the local abundance of NIS species by ratios of survival probabilities for the atoms on the special orbits, we calculate the local densities of NIS species along the Earth positions in space and their change from year to year during the IBEX observations of NIS gas. To calculate the local NIS densities, we use a time-dependent calculation scheme extensively discussed by \citet{rucinski_etal:03}, with the ionization rates from the present paper linearly interpolated between the nodes of the time and heliolatitude grid. The calculations were performed using the most recent version of the Warsaw Test Particle Model, successfully employed by \citet{bzowski_etal:12a} to analyze IBEX observations of NIS He gas. This model is an extension of the model used by \citet{rucinski_etal:03} to model NIS He density evolution during the solar cycle, by \citet{tarnopolski_bzowski:09} for modeling of NIS H, and by \citet{tarnopolski_bzowski:08a, kubiak_etal:13a} for NIS D. 
	
We believe that our calculation of NIS He, Ne, and O in the inner heliosphere, carried out using realistic, measurement-based ionization rates and sophisticated test-particle models of NIS gas distribution in the heliosphere, is the first in the literature. \citet{rucinski_etal:03} discussed time variations in NIS He gas in the inner heliosphere neglecting the electron ionization and assuming a simplified, sinusoidal evolution of the photoionization rate, even though the absolute values and extremes of their ionization model featured values are close to ours.

For the abundance study, the most relevant is not the absolute density $n_{\mathrm{X}}\left( \vec{r},t \right)$ of a given species X, but the magnitude of change in density due to the interaction with inner heliosphere, defined as the ratio of local density to the density at TS: 
	\begin{equation}\
		n_{\mathrm{X,rel}} \left( \vec{r},t \right) = n_{\mathrm{X}} \left( \vec{r},t \right) / n_{\mathrm{X,TS}} 
		\label{eq:relDens}
	\end{equation}
	
We will refer to these quantities as relative densities. The ratio of relative densities of two species X,Y at a given time moment and location in space 
	\begin{equation}\
		\xi_{\mathrm{dens,X/Y}}=n_{\mathrm{X,rel}}/n_{\mathrm{Y,rel}}
		\label{eq:xiDens}
	\end{equation}
is the change of the absolute abundance of the two species due to the interaction with the heliosphere. We will refer to the $\xi_{\mathrm{dens}}$ factors as the abundance change quotients.

\begin{figure}
		\begin{tabular}{c}
			\resizebox{\hsize}{!}{\includegraphics{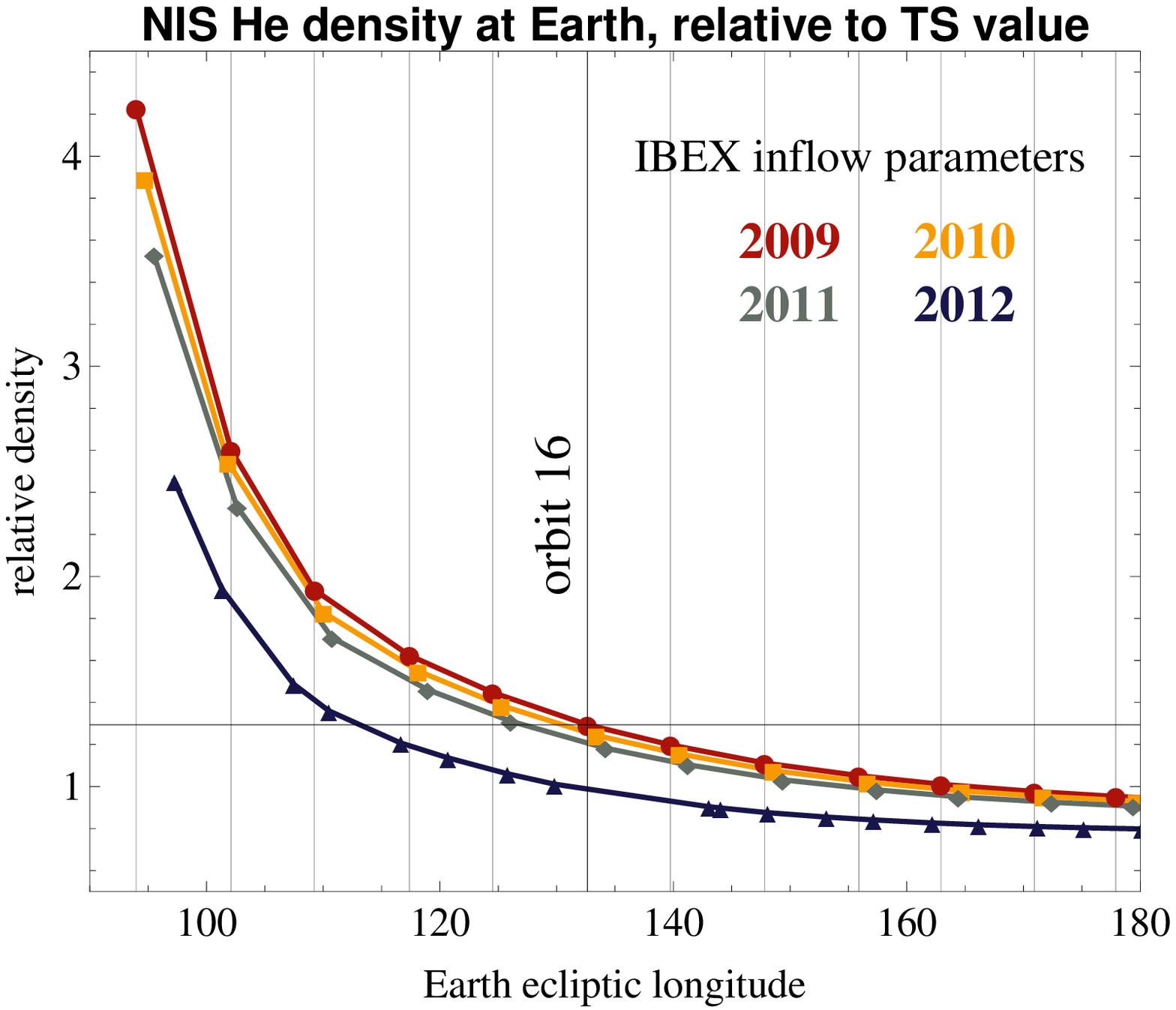}}\\
			\resizebox{\hsize}{!}{\includegraphics{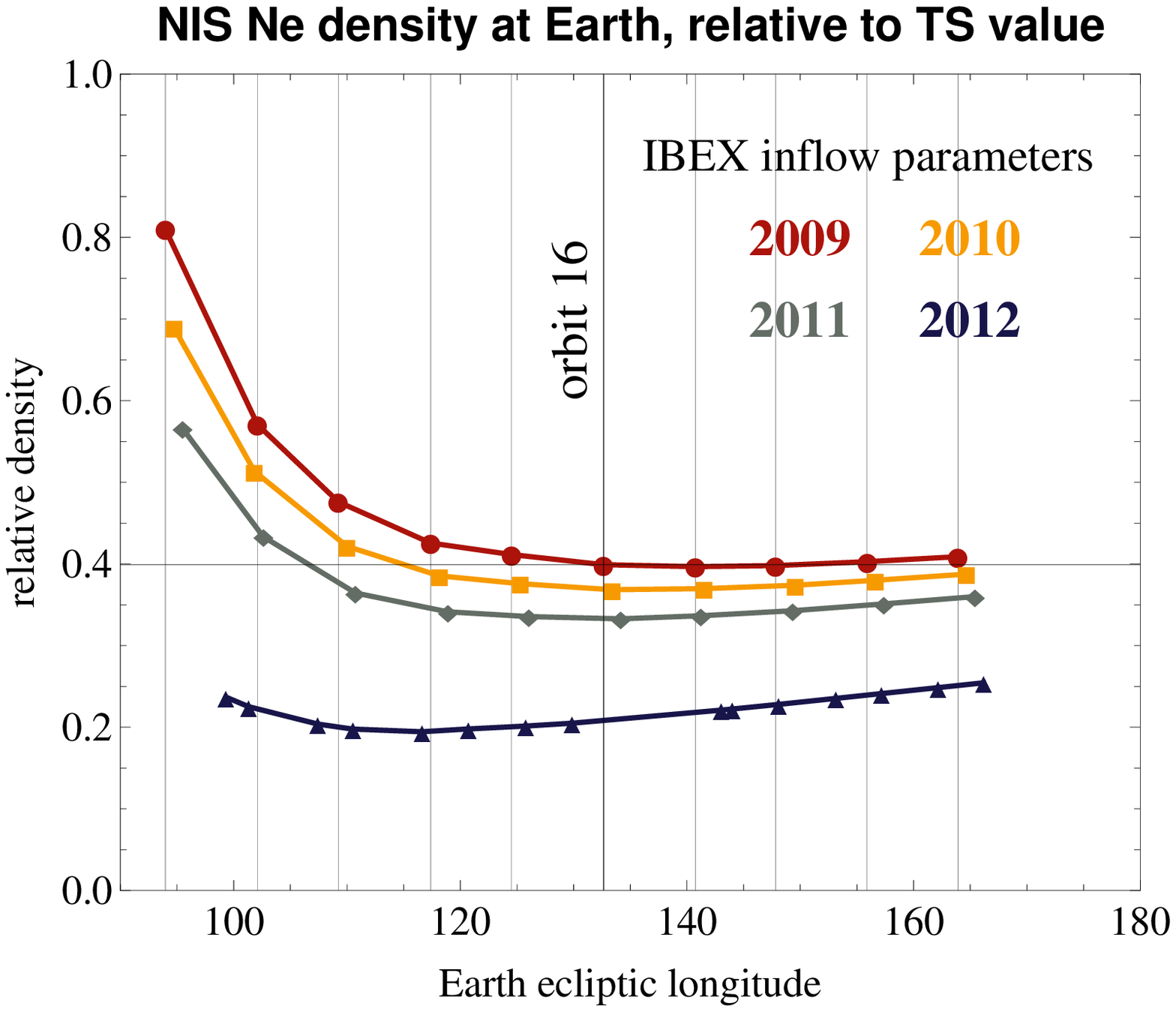}}\\
			\resizebox{\hsize}{!}{\includegraphics{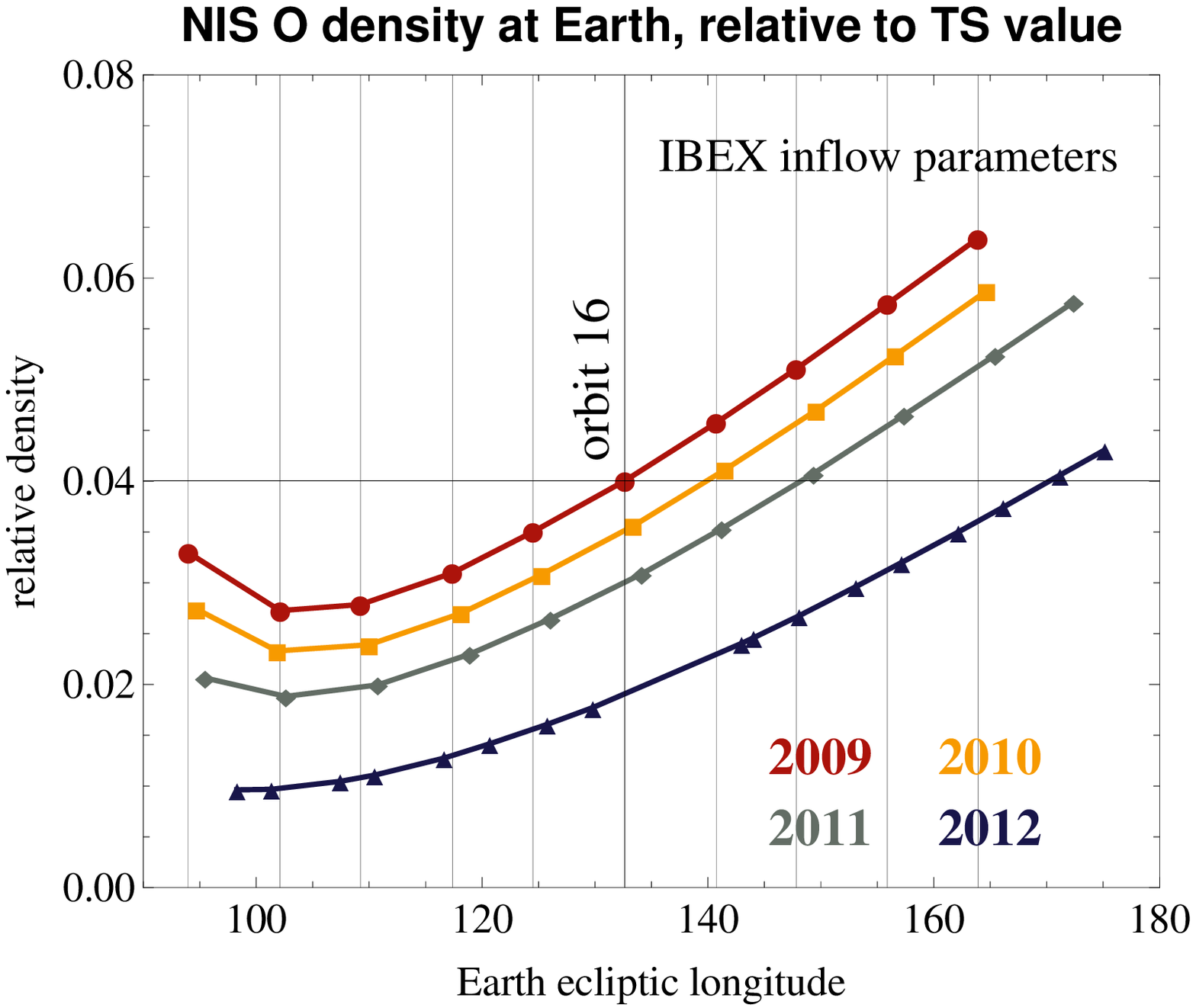}}\\
		\end{tabular}
		\caption{Relative densities (scaled to the values at TS, Eq.~(\ref{eq:relDens})) for NIS~He, Ne, and O for the four IBEX sampling seasons, calculated for the apogee times of IBEX orbits for the inflow parameters obtained by \citet{bzowski_etal:12a} from IBEX NIS He observations. The vertical bars mark the longitudes of IBEX apogees in 2009, the horizontal bars marks the relative density levels for this time and location in space.}
		\label{fig:relDensities}
	\end{figure}

\begin{figure}
		\begin{tabular}{c}
			\resizebox{\hsize}{!}{\includegraphics{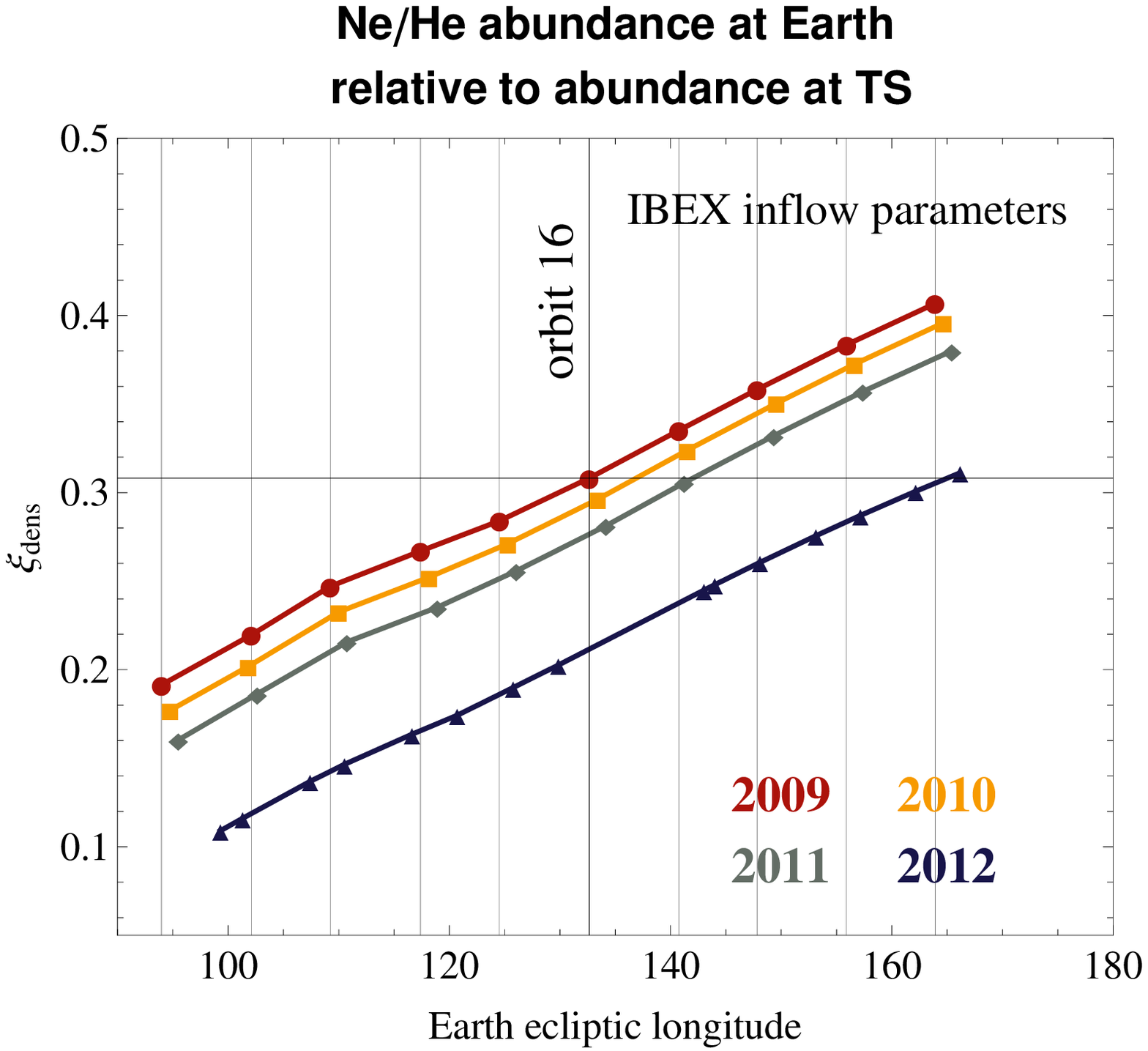}}\\
			\resizebox{\hsize}{!}{\includegraphics{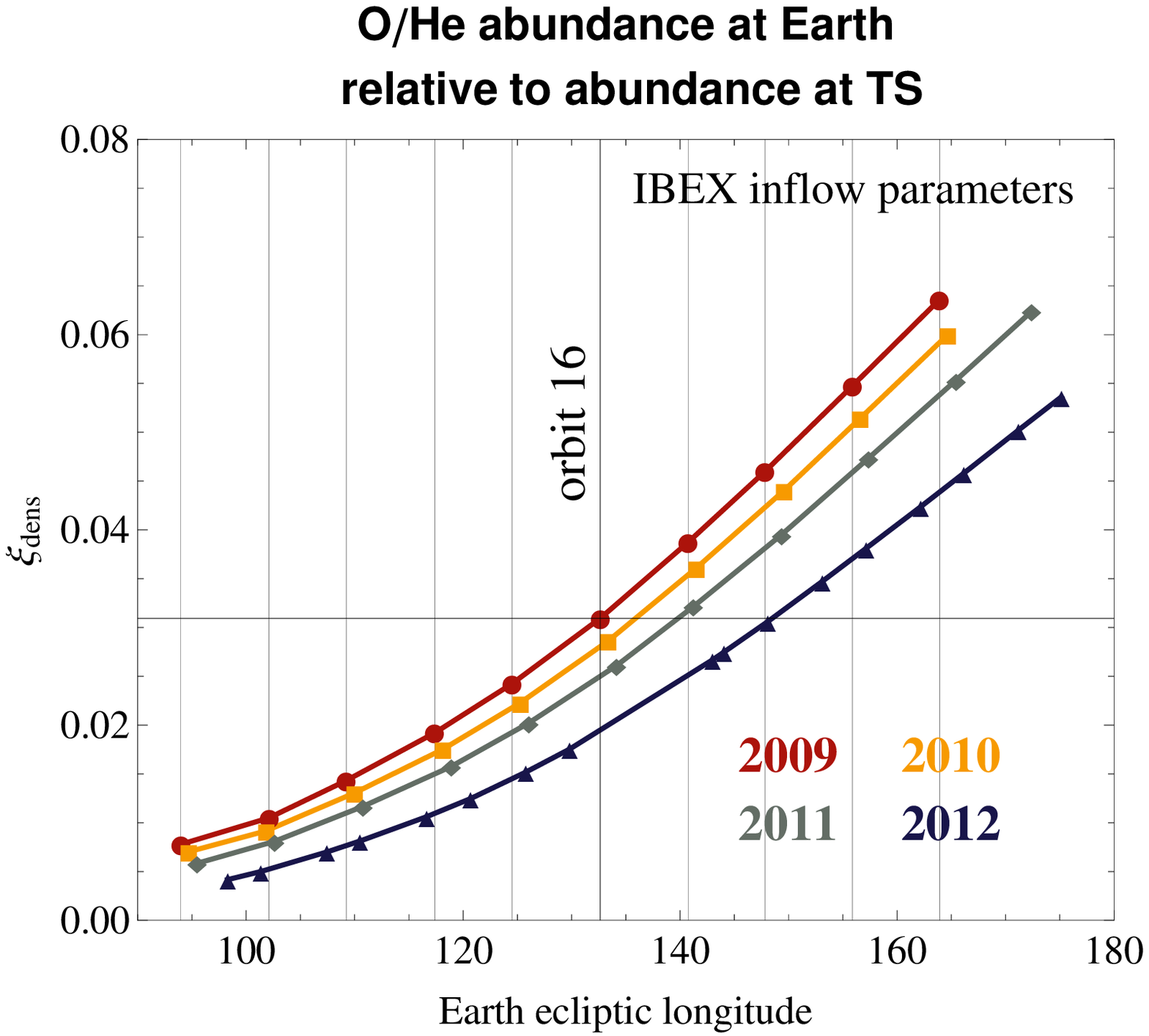}}\\
			\resizebox{\hsize}{!}{\includegraphics{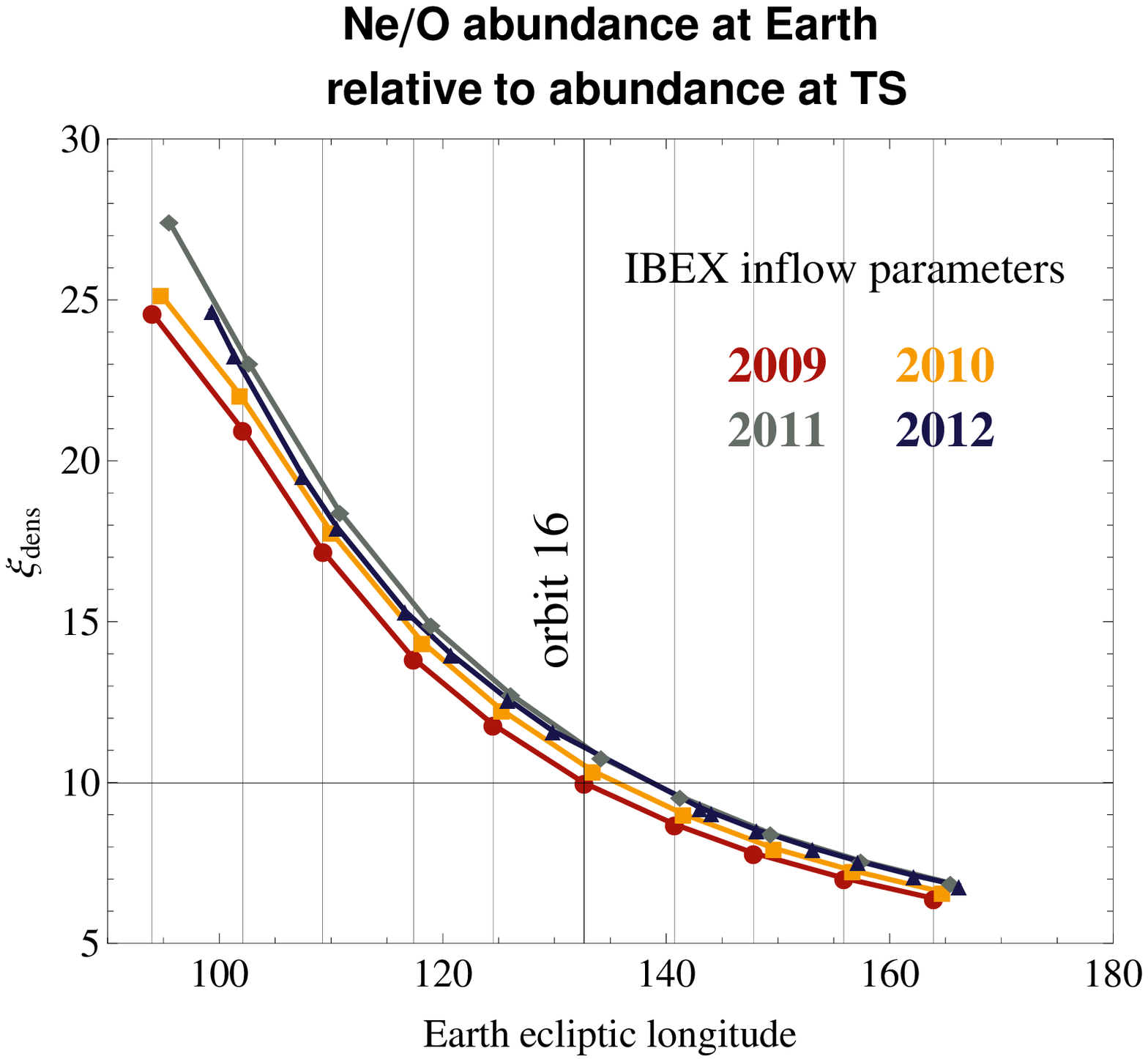}}\\
		\end{tabular}
		\caption{Heliospheric abundance quotients calculated from the relative densities shown in Fig.~\ref{fig:relDensities} using Eq.~(\ref{eq:relDens}) for the times and locations as for Fig.~\ref{fig:relDensities}.}
		\label{fig:densAbund}
	\end{figure}

To facilitate comparison with IBEX, we have calculated the relative densities of NIS~He, Ne, and O along the Earth orbit during the IBEX observation seasons from 2009 to 2012, as shown in Fig.~\ref{fig:relDensities}, and additionally pairwise the abundance change quotients $\xi_{\mathrm{dens,X/Y}}$ (i.e., ratios of relative densities Ne/O, Ne/He, and O/He, see Fig.~\ref{fig:densAbund}). 

The densities of NIS~He, Ne, and O along the Earth orbit where the IBEX observations are carried out show different patterns between the species. While NIS~He, which is weakly ionized in the heliosphere and has the largest thermal speed of $\sim5~\mathrm{km}~\mathrm{s}^{-1}$, features a steady decrease of relative densities with ecliptic longitude when going away from the cone region, NIS~Ne, which has an intermediate ionization rate and more than twice lower thermal speed, shows an almost constant density after the Earth leaves the cone region. NIS~O, due to its strongest ionization of the three species considered, features a moderate drop immediately past the cone and afterwards a relatively sharp increase with increasing longitude. 

The densities of all three species vary relatively weakly from year to year during 2009, 2010, and 2011, but in 2012 they are appreciably reduced because of the increased ionization rate due to the increase in solar activity. This behavior of densities explains well the pattern of abundance changes along the Earth orbit in Fig.~\ref{fig:densAbund}. 
	
\begin{figure}
			\resizebox{\hsize}{!}{\includegraphics{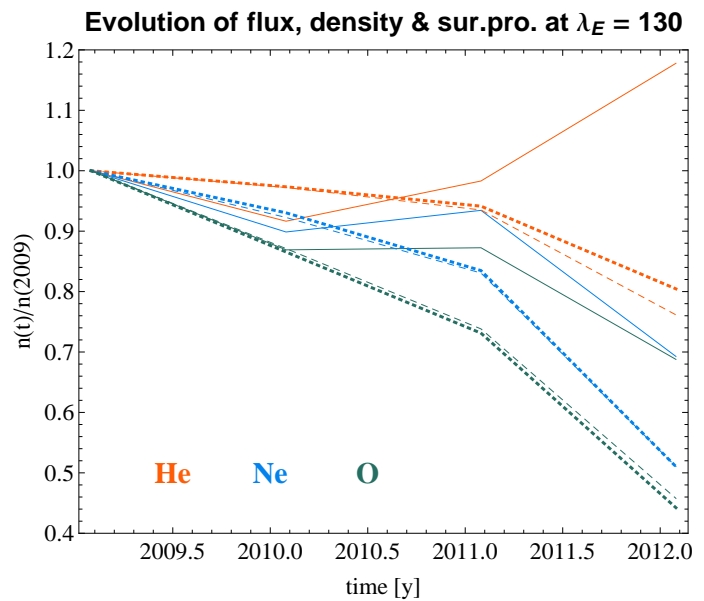}}\\
		\caption{Relative change of the season-integrated flux (solid lines), density at the ecliptic longitude where the NIS gas flow intersects the Earth's orbit (broken lines) and time-dependent survival probabilities of the NIS atoms reaching IBEX at this location (dotted lines). All quantities are shown as ratios to the respective values in 2009. The relative change of the flux is equivalent to the change of total counts per orbit. The color code for species is indicated in the panel.}
		\label{fig:TimeEvol}
	\end{figure}

The relative year-to-year change of densities at the ecliptic longitude of Earth during the IBEX peak orbits $\left( \lambda_{E} \simeq 130\degr \right)$ is shown in Fig.~\ref{fig:TimeEvol}. It is evident that the percentage change in density relative to the density in 2009 (shown for Ne, He, and O by broken lines) very closely follows the relative change in the survival probabilities, calculated using the time-dependent approach (shown with the dotted lines). Discrepancies are typically on a level of a few percent and are smallest for Ne, which has the highest atomic mass from the three species discussed in this paper and thus has the lowest thermal spread. The discrepancies increase a little for O and reach a $\sim5\%$ level for He in 2012, after a time interval of increasing ionization. It is this condition: a relatively high thermal spread (the highest from the three species in question) and strongly time-varying ionization conditions that are the cause of this discrepancy. 

A conclusion from this part of study is that during the first four years of IBEX operation, the percentage change of densities at the location on Earth orbit where the peak of NIS gas flux occurs closely follows the percentage change in survival probabilities of the NIS species, but the survival probabilities must be calculated taking into account the actual history of the ionization rates. 

The Ne/He and O/He abundances increase with increasing ecliptic longitude, but from year to year they gradually go down with the increasing solar activity. In 2012, this overall decrease is expected to be much stronger than before, especially for the Ne/He abundance, because of the sharp reduction in the local O and Ne density. But the Ne/O abundance shows just an opposite behavior: it decreases with increasing ecliptic longitude and slightly increases from year to year (see the broken lines in Fig.~\ref{fig:AbundEvol}). The Ne/He abundance at 1~AU is reduced by a factor of $\sim3$ relative to the abundance at TS. The drop in the O/He abundance is stronger by an order of magnitude. By contrast, the Ne/O abundance at 1~AU is increased by a factor of $\sim10$ where IBEX samples the atoms, and even stronger in the cone region. This is partly due to the higher ionization of O and partly to stronger focusing of Ne than O because of the lower thermal speed of Ne.

As can be seen from this comparison, the abundance change is similar, but not identical to the survival probabilities ratio for the atoms that IBEX potentially can see, discussed in the previous section. The reason is that analysis of survival probabilities of individual atoms cannot take into account the fact that because of the finite temperature of the NIS gas in the LIC the atoms reaching a given space location at a given time moment (e.g., IBEX) have various velocity vectors in the LIC and thus have various survival probabilities. Typically, they do not reach this given region exactly at the perihelia of their orbits. Therefore studies of survival probabilities of individual atoms can provide only a limited insight into the change of NIS species abundances in the inner heliosphere.

Because of the differences in the thermal velocities $u_T$ in the LIC, the widths of the focusing cones of NIS He, Ne, and O are quite different, and the big differences in the ionization rates are responsible for different density profiles at the remaining portions of Earth's orbit. Because of the differences in relative densities of the NIS species, also the abundances strongly vary along the Earth orbit. Overall, however, the increase in the Ne/O abundance due to the interaction with the inner heliosphere is lower by $\sim10\%$ than suggested by analysis of survival probabilities of the atoms that IBEX can see, presented in the former section. 

\begin{figure*}
		\centering
		\begin{tabular}{ccc}
			\includegraphics[width=.3\textwidth]{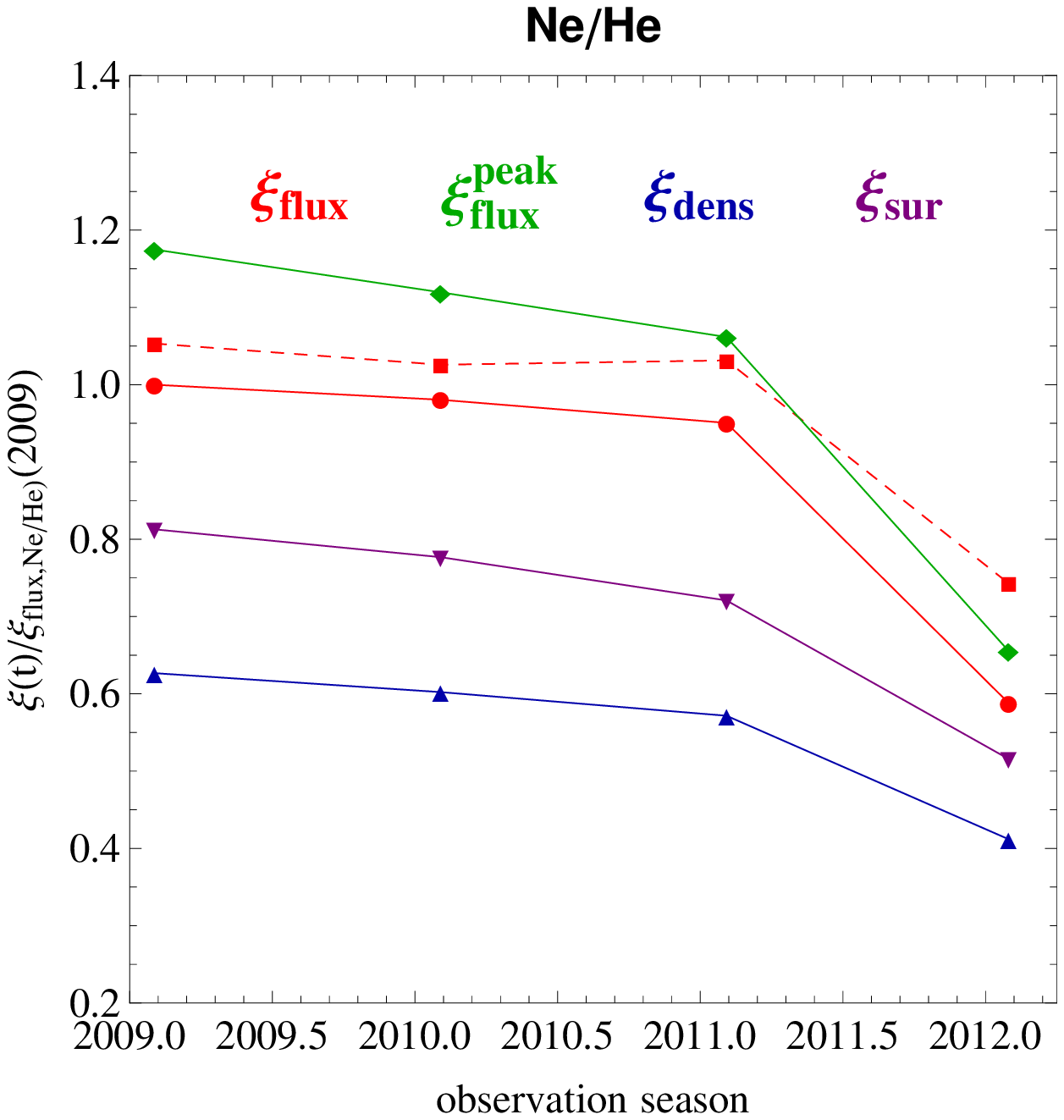}&
			\includegraphics[width=.3\textwidth]{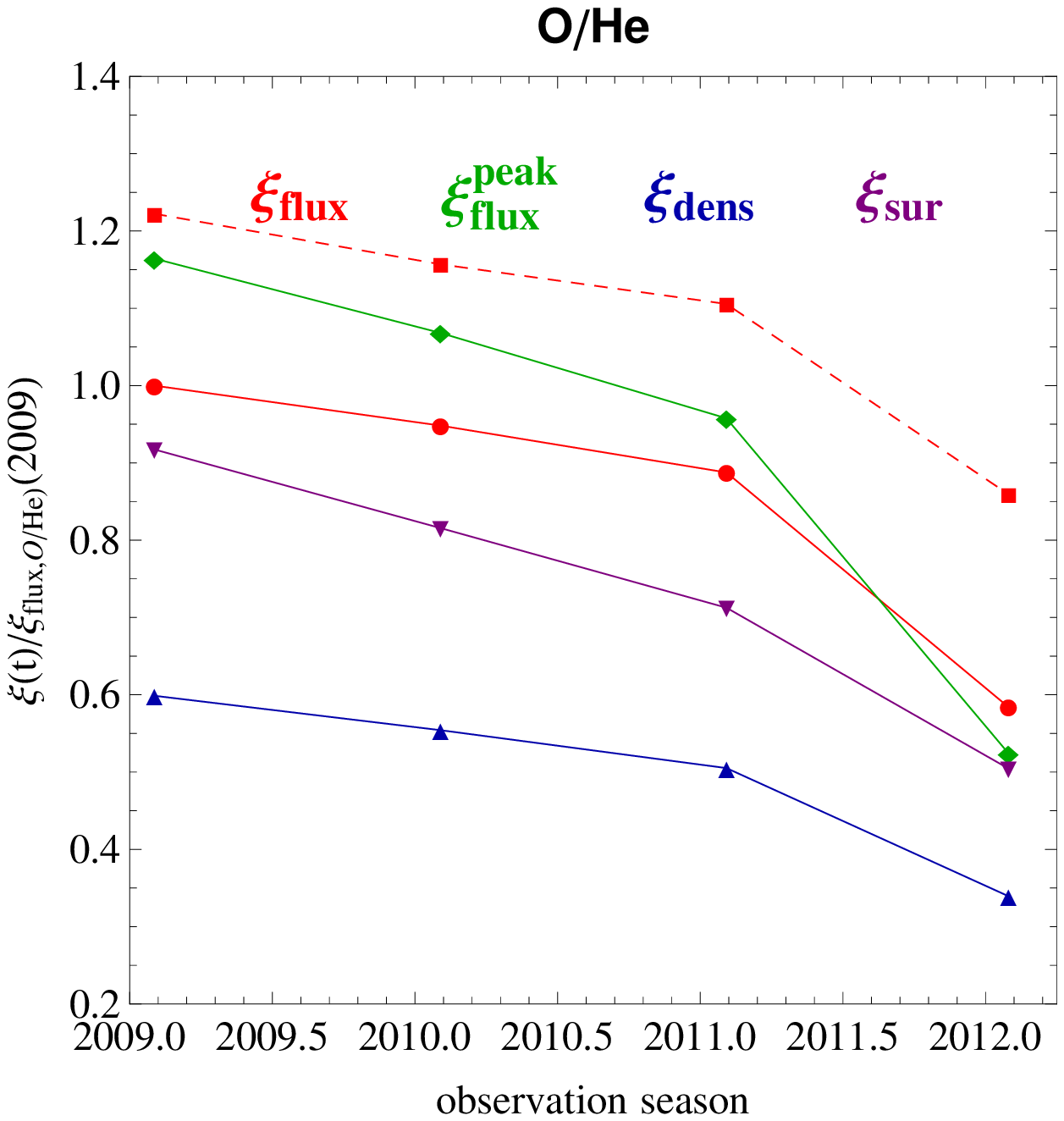}&
			\includegraphics[width=.3\textwidth]{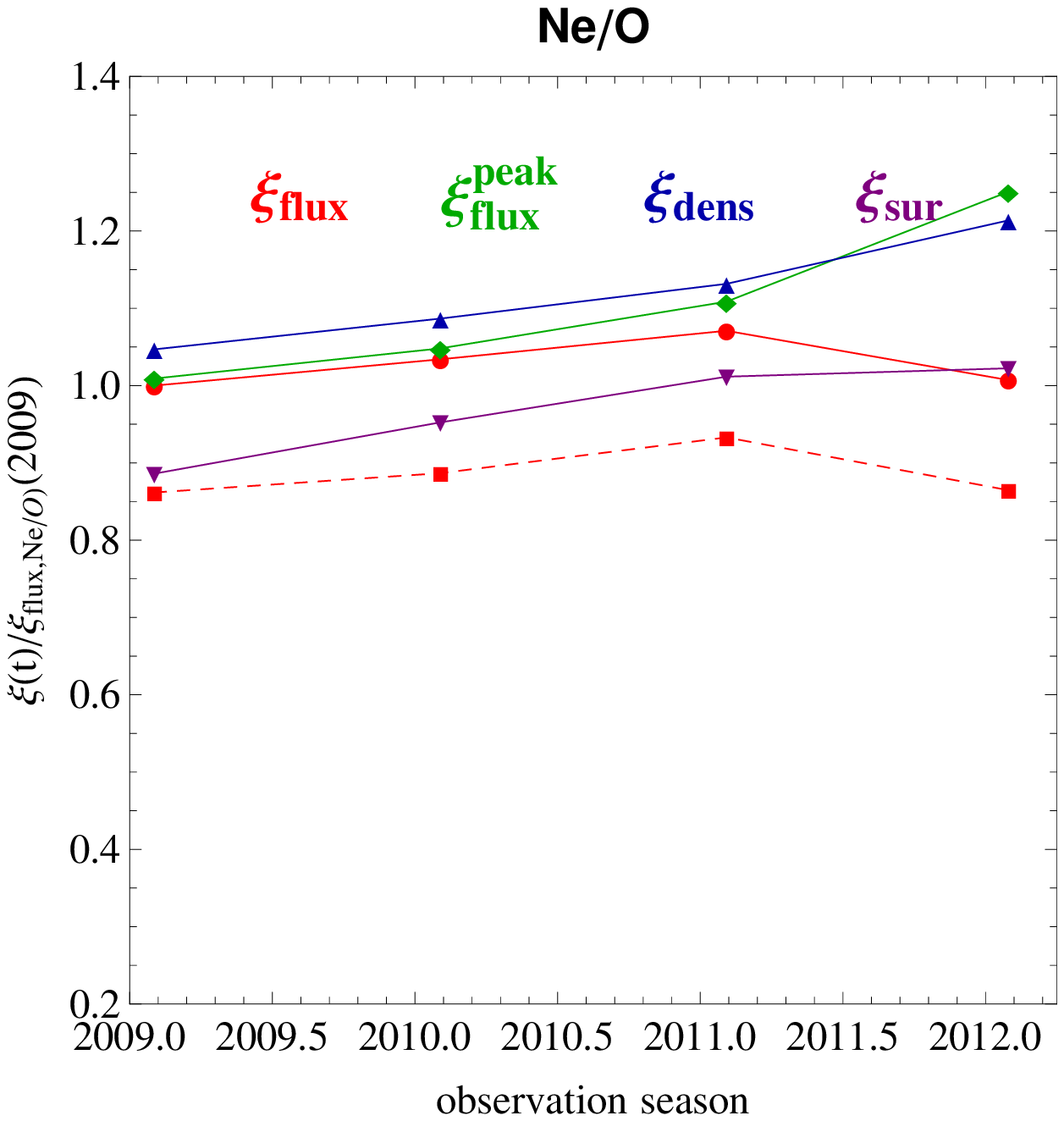}\\
		\end{tabular}
		\caption{Abundance quotients calculated from various quantities discussed in the paper, divided by the flux quotient for the 2009 season, for all three pairs of NIS gas species. The figure illustrates the temporal evolution of the abundance quotients and differences between the quotients calculated using different definitions. Solid lines mark the NIS gas flow parameters obtained from IBEX \citep{bzowski_etal:12a}, the broken line marks the flux quotient calculated for the gas inflow parameters obtained by \citet{witte:04} from GAS/Ulysses observations.}
		\label{fig:abundEffects}
	\end{figure*}

The survival probability ratios $\xi_{\mathrm{sur,X/Y}}$ and the relative density ratios $\xi_{\mathrm{dens,X/Y}}$ are different for respective pairs of species (cf. Tables~\ref{tab:abundTabSurProbBz} and \ref{tab:abundTableDensIbex} and Fig.~\ref{fig:abundEffects}). In addition to the mixing of atoms with different velocity vectors in the LIC, the $\xi_{\mathrm{sur,X/Y}}$ factor does not take into account the local change in density due to gravitational focusing of the gas and thus is not recommended as a measure of the local abundance of NIS species at 1 AU from the Sun. An exception is the $\xi_{\mathrm{dens,Ne/O}}$ ratio, which can be approximated by the easier-to-calculate $\xi_{\mathrm{sur,Ne/O}}$ within a few percent. 

\section{Evolution of NIS He, Ne, and O fluxes and their ratios at IBEX}

IBEX samples individual atoms and with the tallies of atoms of different species, the local abundances can be inferred. Since the tallies of directly sampled atoms are proportional to the fluxes of the direct population atoms relative the IBEX detectors, integrated over relevant intervals of time, the measured ratios of NIS atoms represent the local ratios of NIS atom fluxes averaged over the times of observations and the positions of Earth on its orbit around the Sun. 

IBEX is a spin-stabilized spacecraft whose rotation axis was originally reoriented once per orbit and maintained within $7\degr$ from the Sun. In the middle of 2011, the IBEX orbit was raised \citep{mccomas_etal:11a} and thus the orbital period increased. To maintain the field of view sufficiently far from the Sun, a scheme of reorienting of the spin axis twice per orbit was implemented. The IBEX operations during the yearly NIS sampling campaigns were discussed by \citet{mobius_etal:12a} and details of the spin axis orientation and discussion of its uncertainty was presented in detail by \citet{hlond_etal:12a}.

	\begin{figure*}
		\centering
		\begin{tabular}{cc}
			\includegraphics[width=.45\textwidth]{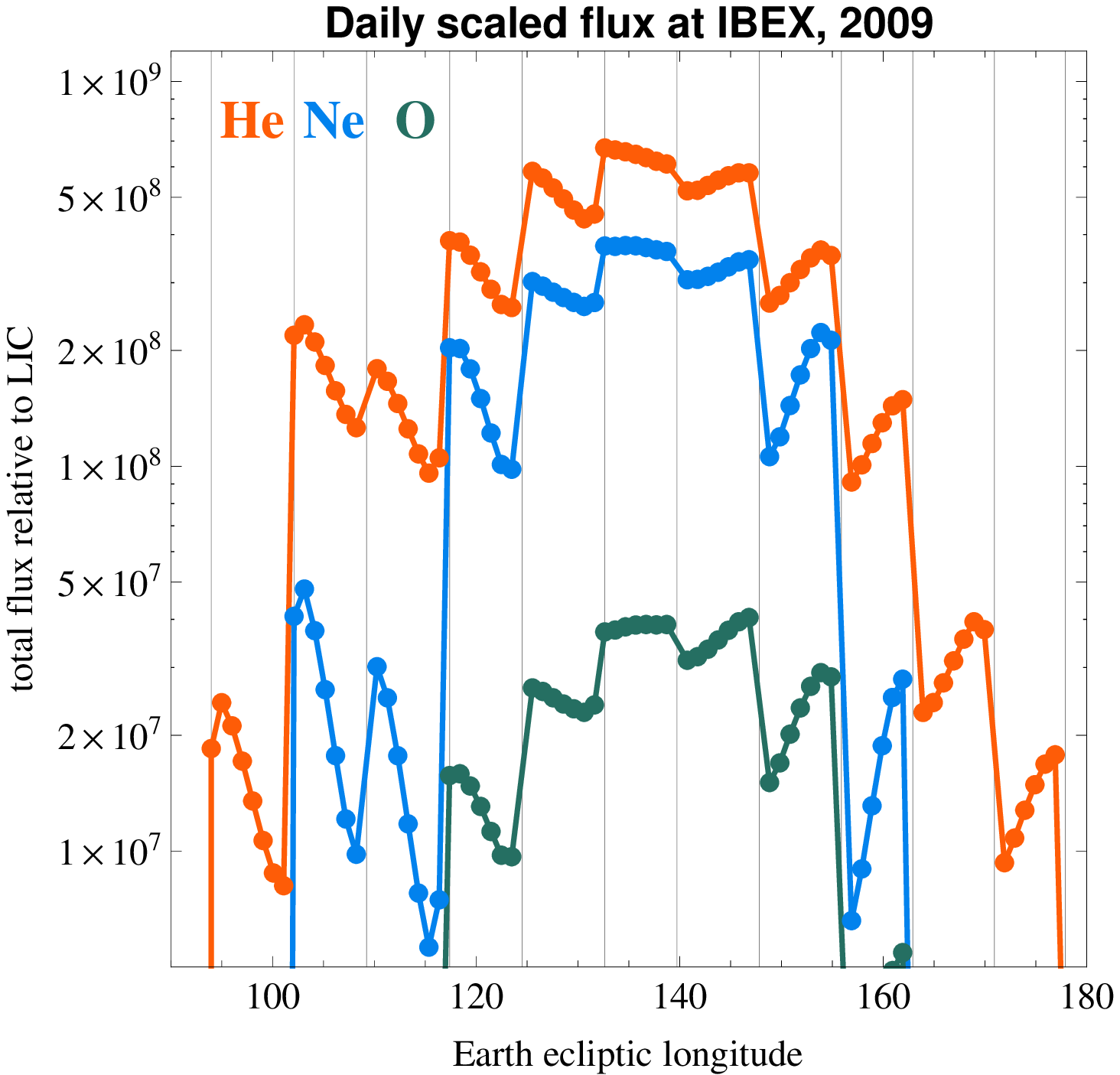}&
			\includegraphics[width=.45\textwidth]{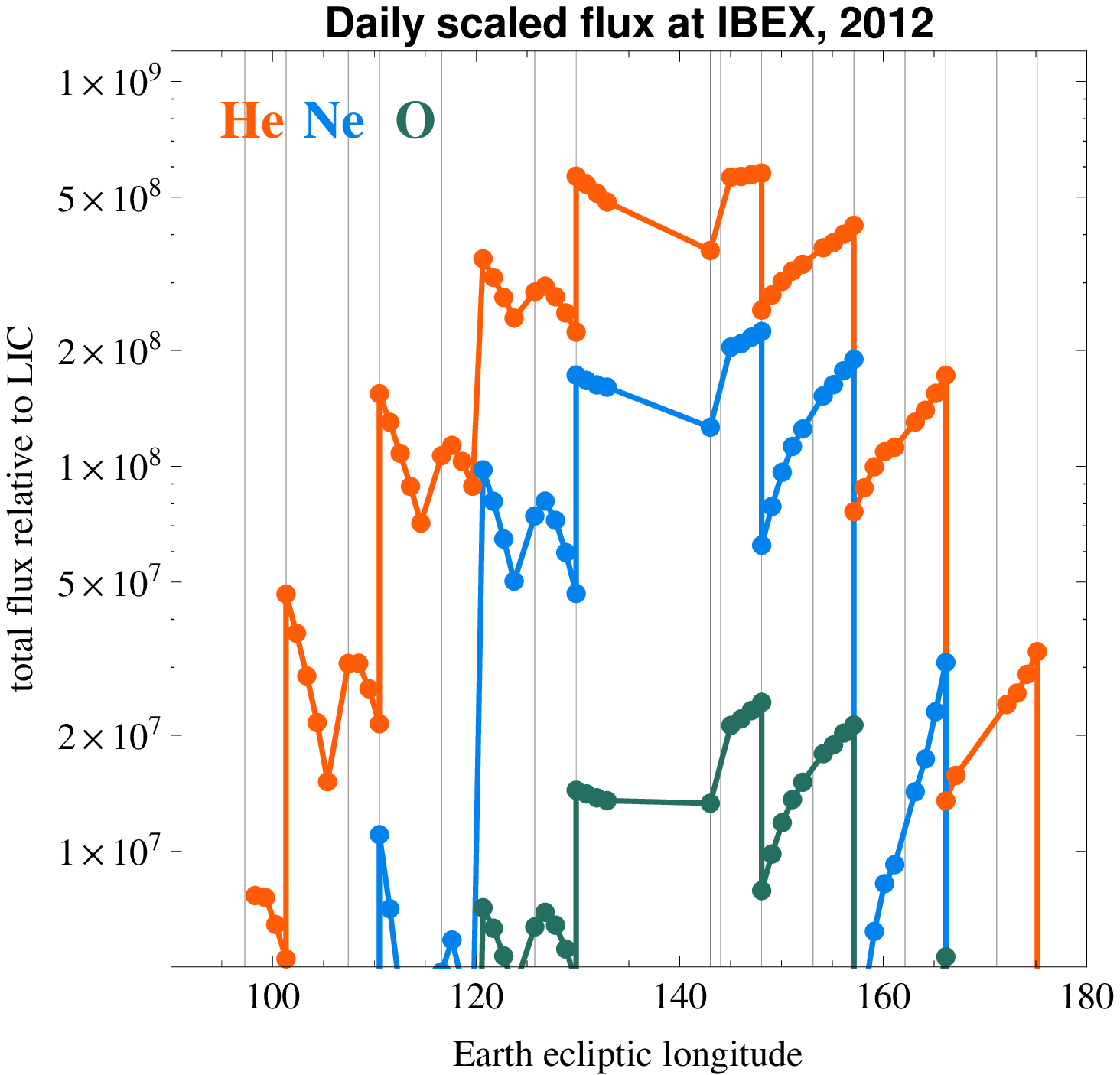}\\
		\end{tabular}
		\caption{Daily fluxes of NIS He, Ne, and O scaled by respective density values at TS for the 2009 and 2012 observation seasons (Eq.~(\ref{eq:scaledFlux})). The vertical bars mark the boundaries of orbital arcs, when the spin axis of the IBEX spacecraft was reoriented. The fluxes calculated assuming the NIS gas flow parameters obtained from IBEX analysis \citep{bzowski_etal:12a}.}
		\label{fig:dailyfluxes}
	\end{figure*}

The aperture of the IBEX-Lo detector, dedicated for the observation of NIS species, scans a great-circle $\sim15\degr$ strip on the sky (the equivalent width due to the  collimator transmission function is about $7\degr$). The strip does not change between the spin axis reorientations. Since the spacecraft is traveling with Earth around the Sun approximately $1\degr$ per day, the beam of NIS atoms is moving across the field of view (FOV) and can be observed at its peak intensity during only one IBEX orbit each year. Thus the flux actually registered is expected to considerably vary both during individual orbits and from orbit to orbit. 

	\begin{figure*}
		\centering
		\begin{tabular}{ccc}
			\includegraphics[width=.3\textwidth]{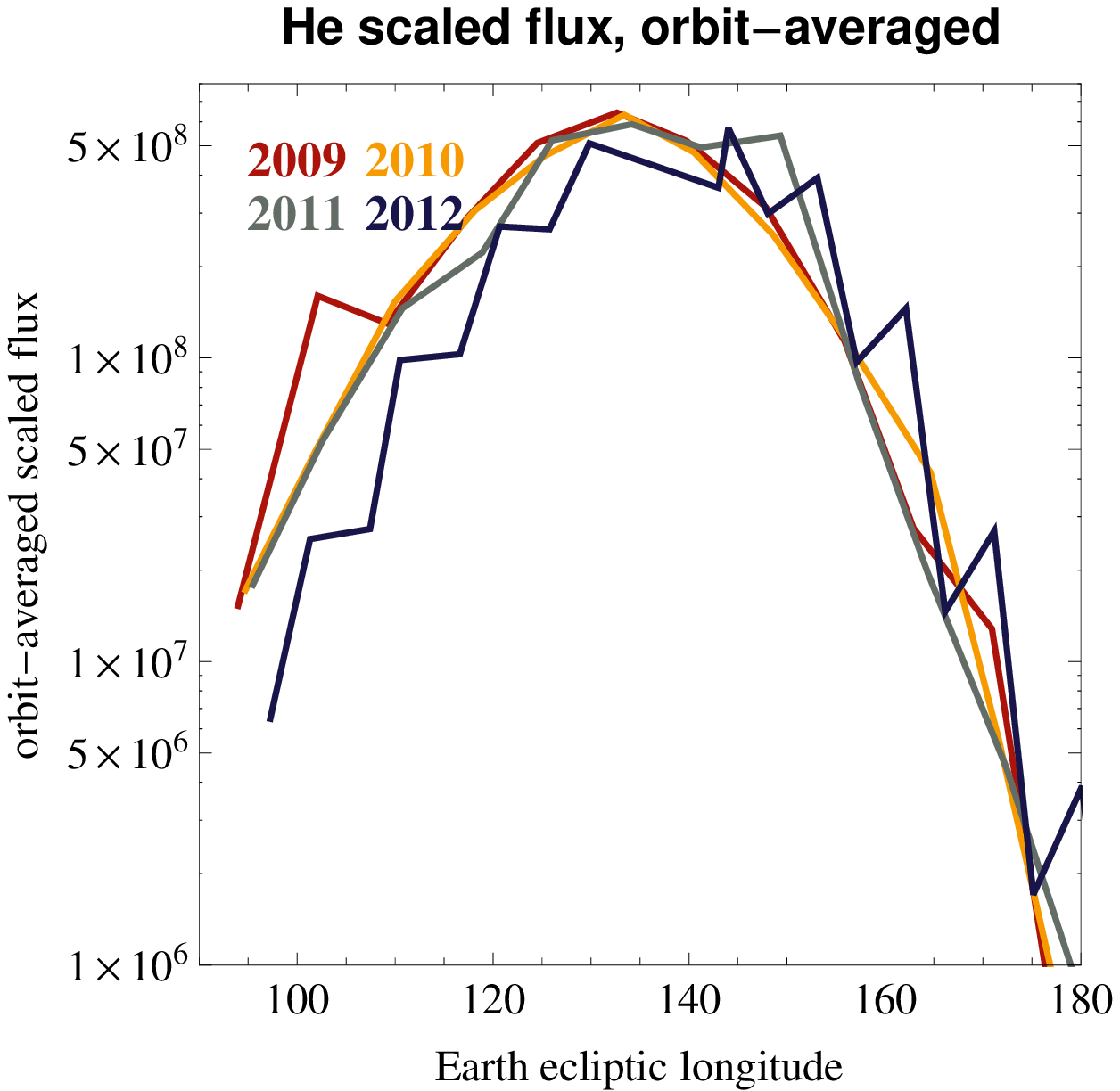}&
			\includegraphics[width=.3\textwidth]{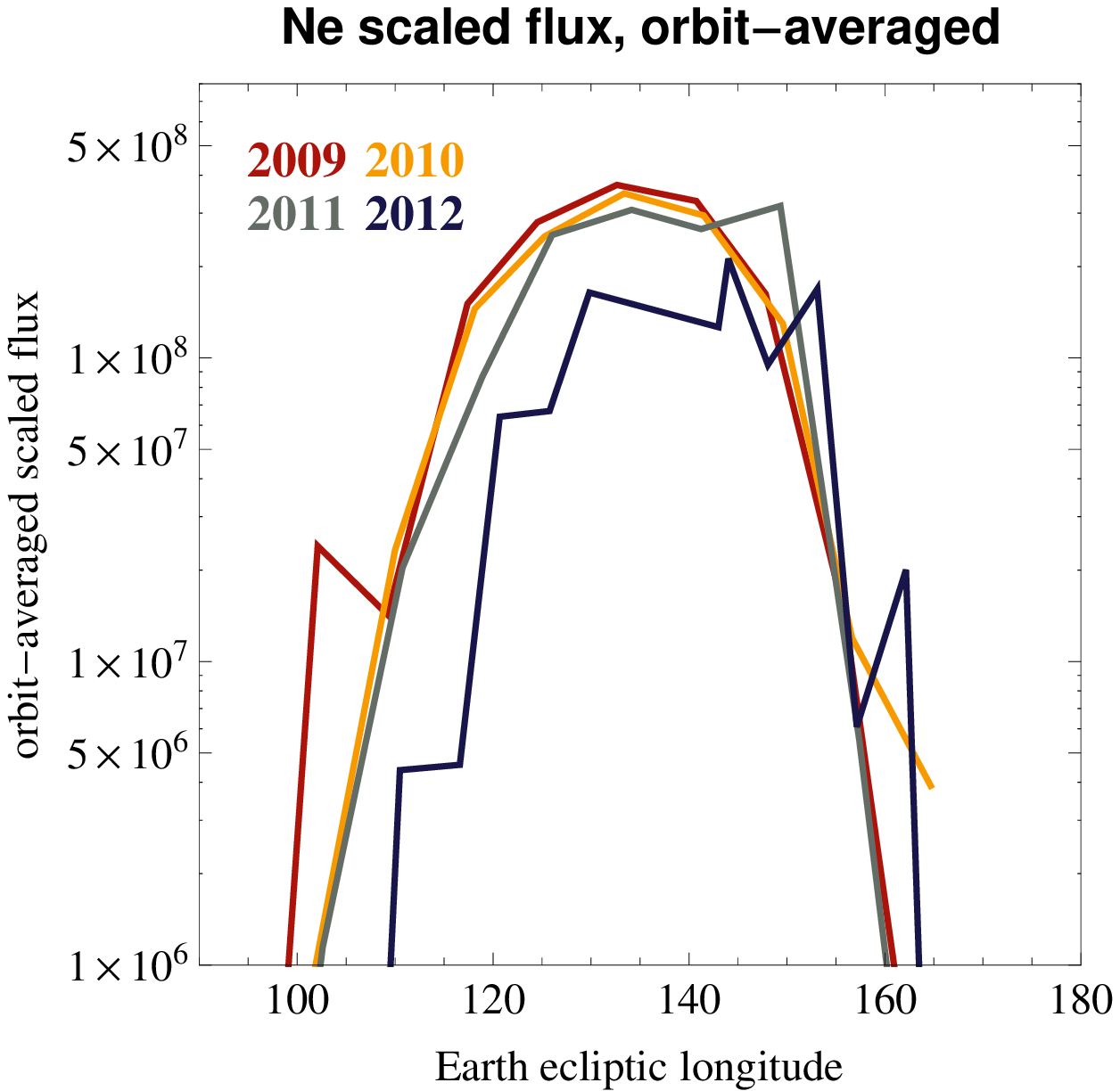}&
			\includegraphics[width=.3\textwidth]{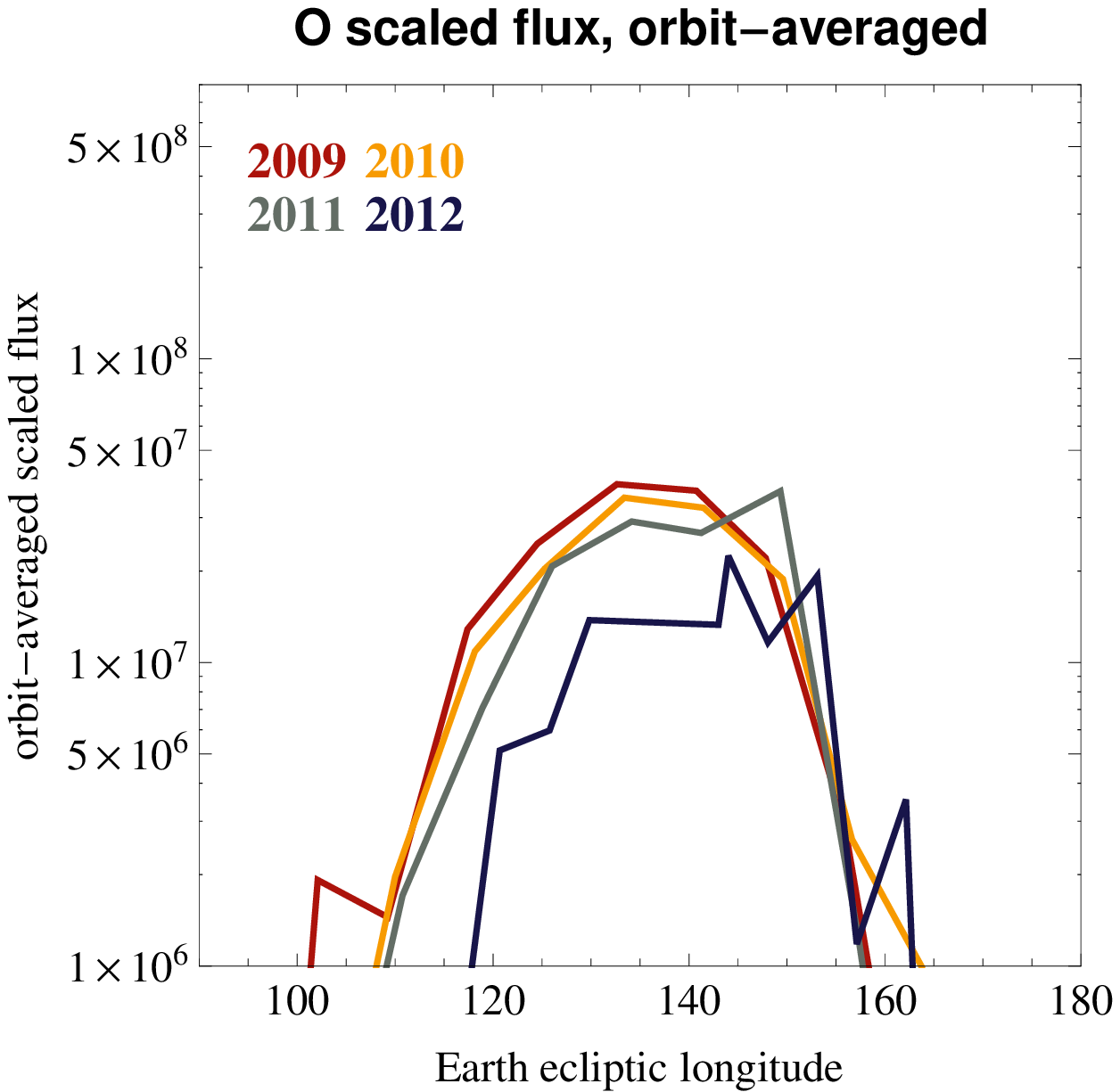}\\
		\end{tabular}
		\caption{Orbit-averaged scaled fluxes for the conditions as for Fig.~\ref{fig:dailyfluxes} for He, Ne, and O, simulated for the first four NIS gas observation seasons by IBEX. The sawtooth features are due to IBEX spin axis reorientation.}
		\label{fig:scaledFluxes}
	\end{figure*}
	
To investigate the flux abundance of NIS species at IBEX we simulated the fluxes at IBEX (i.e., in the IBEX-inertial frame) for each day of the 4 seasons of NIS gas observations following Eqs.~(1--3) in \citet{bzowski_etal:12a}. The fluxes are scaled to the density at TS; we will refer to these quantities as the scaled flux $F_{\mathrm{scaled,X}}$:
	\begin{equation}\
		F_{\mathrm{scaled,X}}=F_{\mathrm{X}}/n_{\mathrm{TS,X}}
		\label{eq:scaledFlux}
	\end{equation}
where $F_{\mathrm{X}}$ is the absolute flux in the IBEX-inertial frame for species X. From these, we calculated the fluxes integrated over spin phase and averaged over the High Altitude Science Operations (HASO) times for each orbit.

The time series of daily scaled fluxes for NIS He, Ne, and O at IBEX, integrated over the FOV, spin axis and each day of the IBEX NIS gas campaigns in 2009 and 2012 are shown in Fig.~\ref{fig:dailyfluxes}. They illustrate the characteristic features of the observed flux. Because of the motion of IBEX with Earth around the Sun the beam of NIS gas wanders through the fixed FOV of IBEX-Lo. Thus, the daily counts registered by the instrument during the HASO times at a specific orbit systematically change. First, they decrease from one day to another, until the spin axis of the spacecraft is changed and the NIS beam is re-aligned with the FOV. Such a sequence repeats until Earth passes ecliptic longitude of about $130\degr$, when the FOV is aligned with the peak of the beam. During this one specific orbit, the daily counts change very little; those are the orbits when the total counts registered are maximum.

	\begin{figure*}
		\centering
		\begin{tabular}{ccc}
			\includegraphics[width=.3\textwidth]{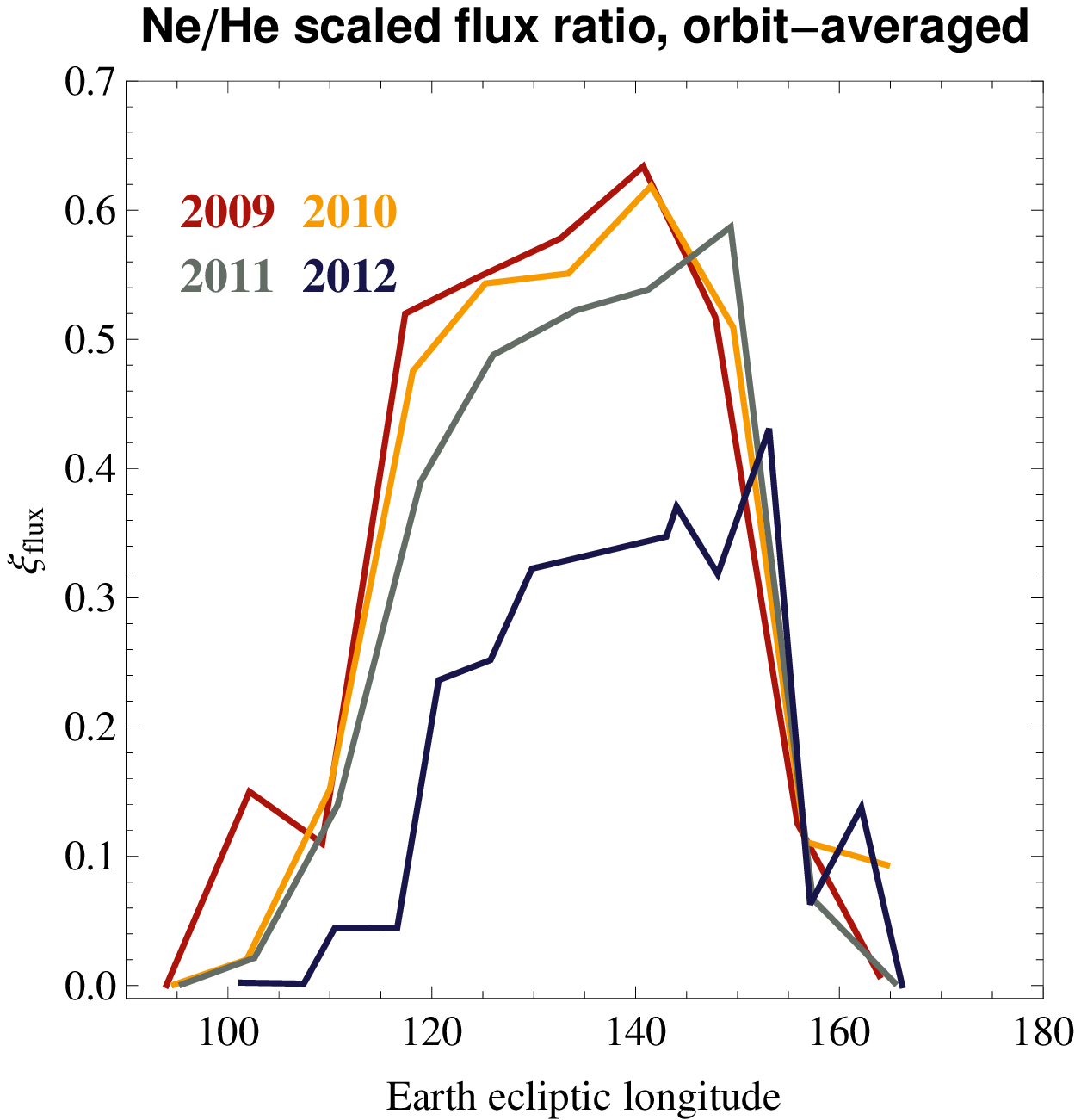}&
			\includegraphics[width=.3\textwidth]{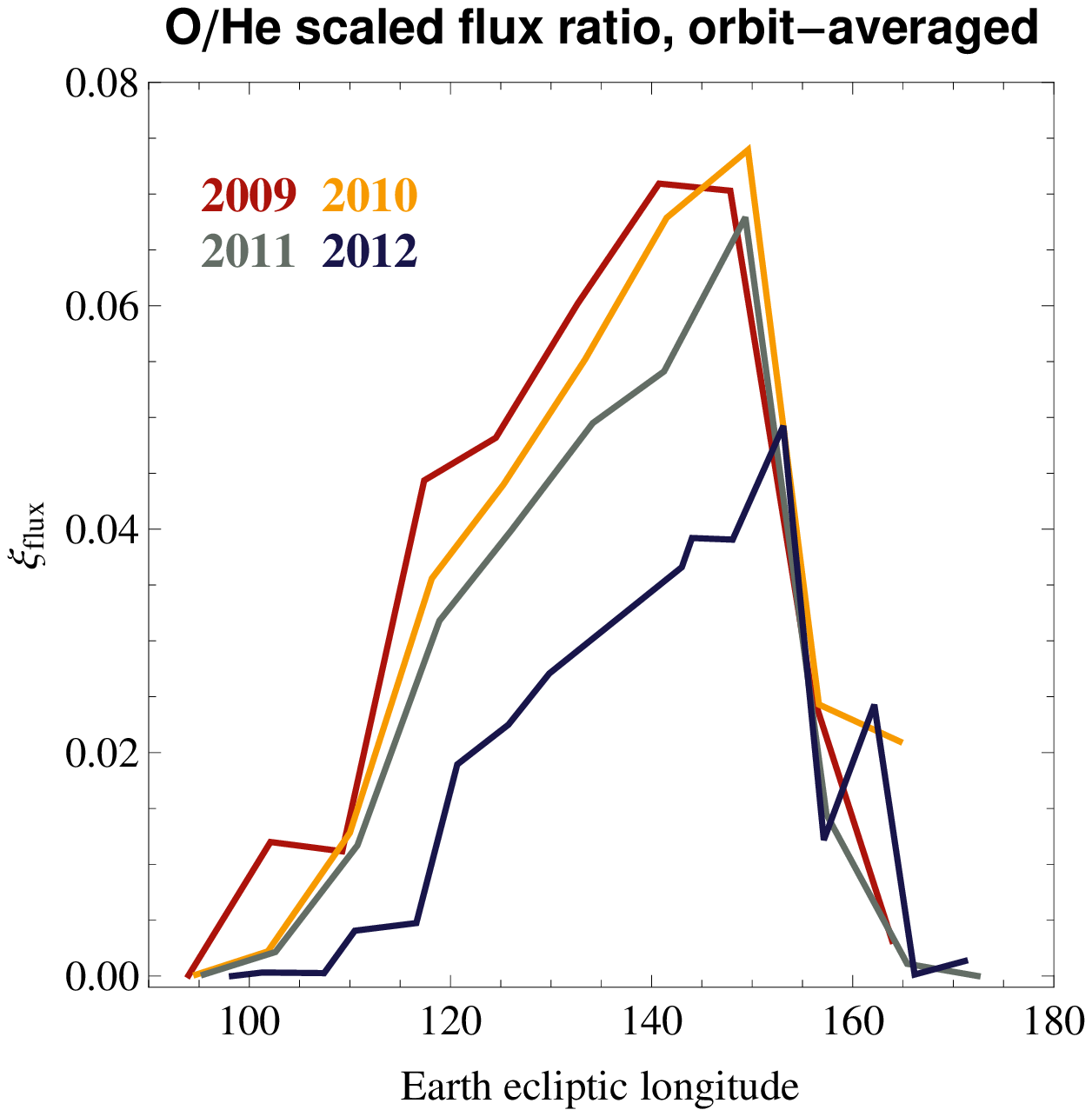}&
			\includegraphics[width=.3\textwidth]{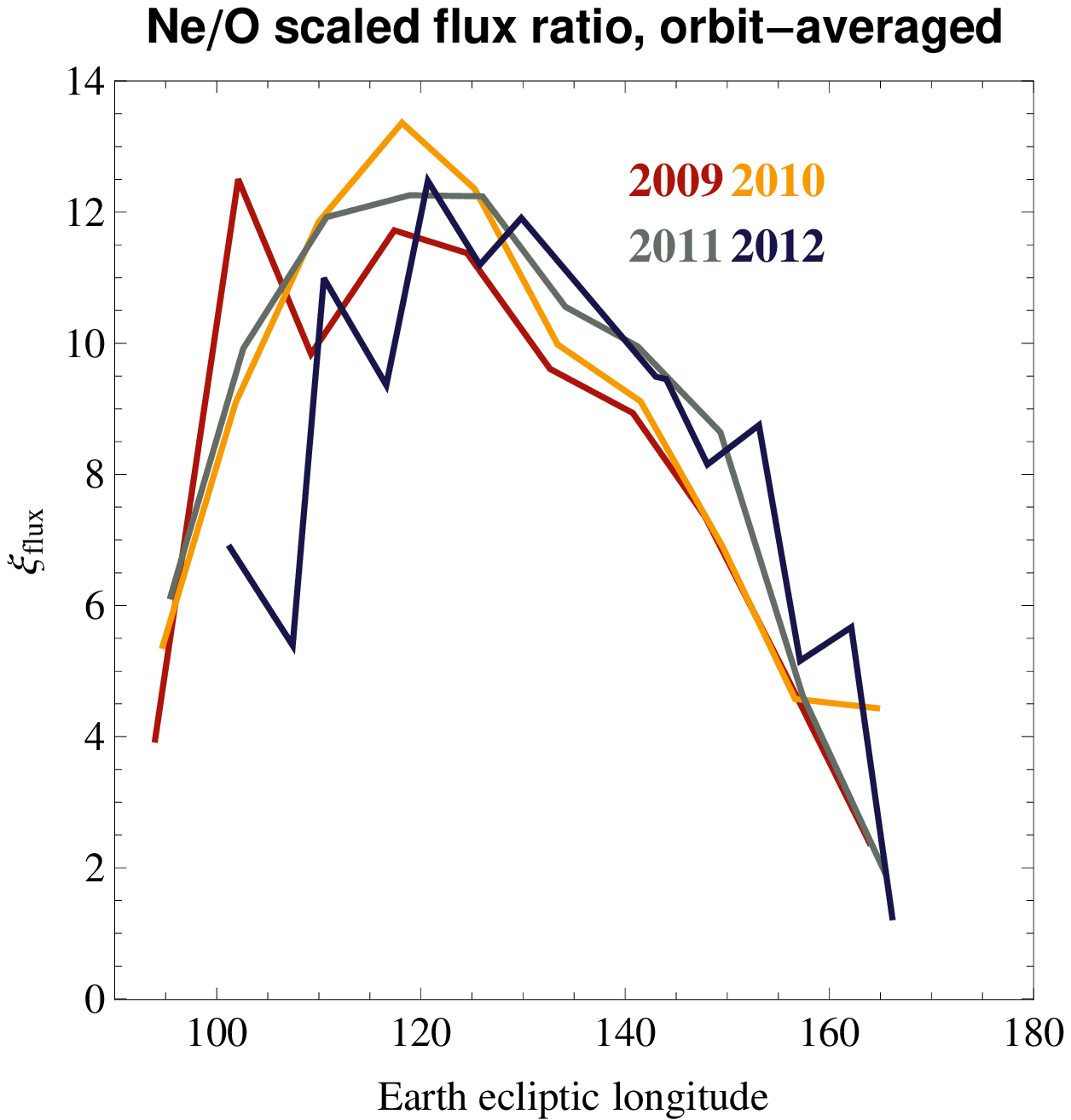}\\
		\end{tabular}
		\caption{Flux abundance quotients $\xi_{\mathrm{flux,X/Y}}$ for the Ne/He, O/He, and Ne/O pairs calculated from the scaled fluxes integrated over IBEX spin phase and averaged over the orbits for the first four years of IBEX observations.}
		\label{fig:fluxAbund}
	\end{figure*}

After the peak, the orbital sequence of daily counts reverses: it is low at the beginning of the orbit and gradually increases, until the spin axis is realigned to avoid sunlight in the aperture. This behavior of the flux, presented in simulation, has actually been observed by \citet{saul_etal:12a} for H and He and by \citet{mobius_etal:12a} and \citet{bzowski_etal:12a} for He. 

The reader will observe a few characteristic features in Fig.~\ref{fig:dailyfluxes}. The daily changes in the total flux appear to be approximately linear in log scale as a function of ecliptic longitude of Earth. The slopes of the daily sequences vary from orbit to orbit and between the species even though the flow parameters in the LIC were assumed identical. This is because of different masses of the atoms, which result in different thermal speeds for the same temperature. A combination of different natural widths of the beams with different ionization rates and some selection effects in the ionization losses (slower atoms attenuated stronger than the faster ones) result in different slopes between the species and asymmetry of the orbit-averaged flux relative to the peak. 

Differences between the spin axis pointing readjustments between seasons also affect the daily fluxes, and consequently the orbit-averaged fluxes. The change in the spin axis reorientation scheme in 2011 affected the data collection. The daily fluxes for two seasons with different spin axis readjustment policy can be compared in the two panels of Fig.~\ref{fig:dailyfluxes}. 

In Fig.~\ref{fig:scaledFluxes} we show orbit-averaged fluxes of NIS He, Ne, O at IBEX during the four observation seasons, scaled to respective densities at TS according to Eq.~(\ref{eq:scaledFlux}). The quantities shown in this figure are directly proportional to the expected numbers of counts on specific orbits divided by densities of respective species at TS.

The change of the total count numbers for given species between 2009 and 2010 are small, especially for Ne and He (see Table~\ref{tab:abundTableFluxIBEX} and Fig.~\ref{fig:TimeEvol}). This is because of the small year to year differences in the total ionization rates and relatively weak overall ionization losses. In 2011, one observes that a drop in the total counts starts, and it becomes clearly evident for all species in 2012. Another interesting feature is the shift towards larger longitudes of the increasing branch of the flux (the portion to the left from the peak in Fig.~\ref{fig:scaledFluxes}). This effect, as we have verified, is related to the change of IBEX orbit and resulting change in the spin axis reorientation strategy after the 2011 season. 

The year-to-year changes in the count numbers do not track the changes in survival probability and local density (Fig.~\ref{fig:TimeEvol}). This is because in addition to the varying ionization rate, also details of the observation process (i.e., the history of field of view changes, details of the observation times, which in general do not cover the whole time interval on a given orbit) strongly affect the averaged count numbers. Therefore, assessing the expected year-to-year changes in the counts requires carrying out detailed simulations of the observations. The predictions or conclusions drawn from analysis of survival probabilities, which are relatively easy to calculate, are not credible and the errors become particularly large in the conditions of rapidly changing solar activity. 

Taking pairwise the quantities shown in Fig.~\ref{fig:scaledFluxes} and calculating their quotients between species one obtains ratios $\xi_{\mathrm{flux,X/Y}}$ of fluxes scaled to the values at TS:
	\begin{equation}\
		\xi_{\mathrm{flux,X/Y}} = F_{\mathrm{scaled,X}}/F_{\mathrm{scaled,Y}} = \left( F_{\mathrm{X}}/n_{\mathrm{TS,X}} \right) / \left( F_{\mathrm{Y}}/n_{\mathrm{TS,Y}} \right)
		\label{eq:xiFlux}
	\end{equation}
They are similar to the quantities shown in Fig.~\ref{fig:densAbund}, i.e., to the modification of the abundances of NIS species relative to their respective values at TS. We show them in Fig.~\ref{fig:fluxAbund} and in Table~\ref{tab:abundTableFluxIBEX} for Ne/He, O/He, and Ne/O ratios. These quantities, multiplied by respective count ratios actually retrieved from IBEX data, would return the abundances of NIS species at TS. In addition to the $\xi_{\mathrm{flux,X/Y}}$ quotients for the fluxes integrated over the whole seasons, in Table~\ref{tab:abundTableFluxIBEX} we present the quotients $\xi_{\mathrm{flux,X/Y}}^{\mathrm{peak}}$ for the ``peak orbits'' (i.e. for the orbits on which the observed NIS flux was maximum: 16, 64, 112, 154).

	\begin{figure}
		\resizebox{\hsize}{!}{\includegraphics{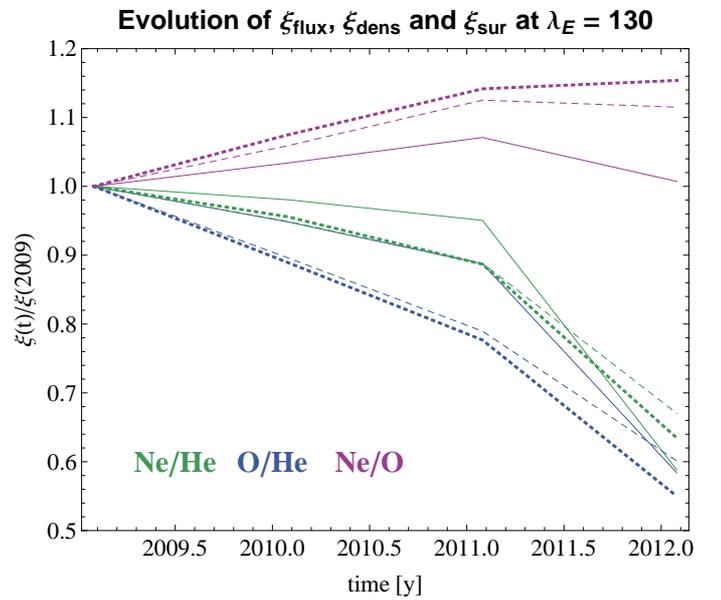}}
		\caption{Relative change of the $\xi$ quotients for the season-integrated flux (solid lines), densities at the ecliptic longitude where the NIS gas flow intersects the Earth's orbit (broken lines) and time-dependent survival probabilities of the NIS atoms reaching IBEX at this location (dotted lines). All quantities are shown as ratios to the respective values in 2009. The color code for species pairs is indicated in the panel.}
		\label{fig:AbundEvol}
	\end{figure}
	
The time series of abundance quotients $\xi_{\mathrm{flux}}$ for the Ne/He and O/He have peaks at ecliptic longitudes greater than the longitude of maximum of the He flux. This behavior is consistent among all four observation seasons. The quotient for the Ne/O abundance, on the other hand, peaks before the longitude of the peak He flux. 
	
The abundances obtained from the fluxes averaged over the full seasons are listed in Table~\ref{tab:abundTableFluxIBEX} and the evolution of the quotient relative to the value from 2009 is illustrated in Fig.~\ref{fig:AbundEvol}. They feature a monotonic decrease between 2009 and 2012 for Ne/He and O/He, which can be explained by the increase of ionization losses because of the increase in solar activity. An exception is the Ne/O quotient, which increases from 2009 to 2011 and starts to decrease in 2012. 

The evolution of the $\xi_{\mathrm{flux,X/Y}}$ quotients differs from the evolution of the survival probabilities quotients $\xi_{\mathrm{sur,X/Y}}$, which is not surprising when one remembers that already the flux evolution did not agree too well with the evolution of $w$. 

The total count numbers from given orbits are sums of daily counts, but in reality it frequently happens that some time intervals are not suitable for analysis and consequently the actually observed sequence of orbit-integrated counts deviates from a model with full time coverage. Therefore, the following analysis must be regarded as qualitative. For comparison with actual IBEX measurements, the actual observation times must be taken into account, as well as calibration and instrument setting details that we omitted from this analysis.

\section{Summary and conclusions}

We were mostly interested in the modification of the abundances of NIS species He, Ne, and O in the heliosphere due to solar accretion and ionization losses. We have carefully investigated the change in abundance of NIS gas species at 1 AU relative to the abundance at TS of the solar wind due to the interaction of NIS gas with the Sun. All three NIS gas species investigated have identical bulk speed vectors and temperatures in the LIC. Equal temperatures imply different thermal velocities. Together with different ionization losses, this results  in differences in the transmission of NIS species through the inner heliosphere and changes in the abundances relative to the respective abundances at TS. 

We have analyzed the history of ionization rates of He, Ne, and O during the years of NIS species observation seasons by IBEX. The weakest ionization is for He, the strongest for O. The main ionization reaction for all three species is photoionization, whose rate varies during the solar cycle by a factor of $\sim2$. For O also important is charge exchange with solar wind protons. For Ne and He, charge exchange is practically negligible. Electron impact ionization is of a quite significant importance, as it is responsible for ionization of $\sim25\%$ of NIS~O atoms and of about $10\%$ for Ne and He. However, electron impact ionization is the poorest known ionization process because of limited information on the evolution of the distribution function of solar wind electrons with the distance from the Sun.

The simplest way to assess the abundance change between TS and IBEX is to calculate the ratio of survival probabilities of individual atoms on orbits that reach perihelia at the detector. With the ionization rates on hand, we have investigated the quality of the approximation of the analytic model of the survival probabilities, where the ionization rate is assumed invariable with time and falling off with the square of solar distance with electron impact ionization excluded. 

We have shown that this approach is acceptable only for He and Ne during low solar activity (see Fig.~\ref{fig:surProAveraging}), when discrepancies with the calculation performed using the full time-dependent ionization rate are within a few percent. For O the discrepancies are greater. Furthermore, we were unable to identify an averaging interval of the actual ionization rate that would guarantee reliable results from the analytical model at all times. The conclusion is that with the exception of NIS~He during the low solar activity, calculating the survival probabilities of NIS species using the time dependent approach gives more reliable results.

\begin{figure}
		\resizebox{\hsize}{!}{\includegraphics{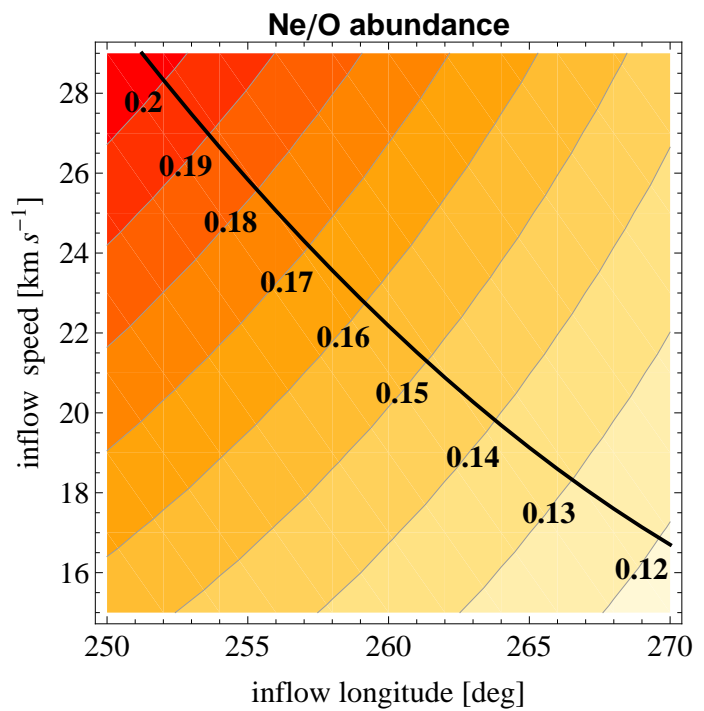}}
		\caption{The Ne/O abundance in the LIC calculated from Eq.~(\ref{eq:xiSur}) for various combinations of NIS gas inflow longitude and speed, based on filtration factors and the Ne/O flux ratio measured at IBEX by \citet{bochsler_etal:12a}. The LIC abundances are listed at the line which is the correlation between the inflow longitude and speed obtained by \citet{bzowski_etal:12a} from analysis of NIS He observations by IBEX-Lo. For the best-fit parameters obtained by \citet{bzowski_etal:12a} the LIC abundance is 0.16, for the flow parameters obtained by \citet{witte_etal:04a} from analysis of NIS He observations by GAS/Ulysses the abundance is 0.18.}
		\label{fig:abundMatrix}
	\end{figure}

The abundance change quotients $\xi_{\mathrm{sur}}$ calculated from survival probabilities for all four IBEX NIS gas sampling seasons are listed in Table~\ref{tab:abundTabSurProbBz}. We have also investigated the sensitivity of the $\xi_{\mathrm{sur}}$ to the inflow parameters of NIS gas (Fig.~\ref{fig:surProRatios}, Tables~\ref{tab:abundTabSurProbBz} and \ref{tab:abundTabSurProbWitte}). The spread in the ratios for various inflow parameters is $\sim20\%$ for O/He, $\sim12\%$ for Ne/O, and $\sim7\%$ for Ne/He. This suggests that the studies of NIS abundances in the LIC and of the NIS gas flow vector must go in parallel (see Figs.~\ref{fig:surProRatios} and \ref{fig:abundMatrix}) because the results can potentially influence each other. 

We have performed a careful analysis of uncertainties of survival probabilities and their ratios. We took into account the uncertainties of the measurements of the solar quantities (solar wind and EUV fluxes) and the reaction cross sections, and the NIS gas inflow velocity. The resulting uncertainties of the survival probabilities and their ratios vary between the pairs of species in question, from $\sim20\%$ for Ne/He through $\sim35\%$ for Ne/O to $\sim50\%$ for O/He. We believe these uncertainties are valid also for the density and flux ratios. 

We also investigated the evolution of densities of NIS species at the location on Earth's orbit where IBEX sees the maximum of NIS flux. The relative variation of densities and survival probabilities of NIS species generally agree with each other, even though systematic differences do exist. This suggests that the changes in survival probabilities (calculated using the full time-dependent model) are sufficient to model the expected relative variations of NIS density, but not the absolute values of the densities, which should be evaluated using a time-dependent thermal (``hot'') model. 

The most realistic, but also the most calculations-intensive approach to the calculation of the abundance quotients is the modeling based on a fully time-dependent model of the ionization losses, where the local fluxes at IBEX are integrated over the field of view, spin phase, and time. To infer the NIS gas abundance at TS, one can use either the flux quotients $\xi_{\mathrm{flux}}$ calculated from the total counts of the NIS species registered at IBEX during the whole observation season, or the quotient $\xi_{\mathrm{flux}}^{\mathrm{peak}}$ obtained from counts registered on the peak orbit of each season. They are not identical (see Tables~\ref{tab:abundTableFluxIBEX} and \ref{tab:abundTableFluxWitte}). Using the latter method requires less calculations, but the former one potentially offers a better statistics. 

We studied differences between the local flux change quotients $\xi_{\mathrm{flux}}$ and density change quotients $\xi_{\mathrm{dens}}$, which potentially might be useful for studies of NIS abundances. The patterns of the relative densities of NIS species at Earth are strong functions of ecliptic longitude. They are quite different between He, Ne, and O because of their different thermal speeds and different ionization rates (see Fig.~\ref{fig:relDensities}). This results in different patterns of the density change quotients for Ne/He, O/He, and Ne/O (Eq.~(\ref{eq:xiDens}), Fig.~\ref{fig:densAbund}). 

Generally, $\xi_{\mathrm{flux}}$ is greater by $\sim40\%$ than $\xi_{\mathrm{dens}}$ for Ne/He and O/He; in the case of Ne/O, $\xi_{\mathrm{dens}}$ is greater than $\xi_{\mathrm{flux}}$ by about $5\%$ (Fig.~\ref{fig:abundEffects}). This is because the accretion factors (Eqs.~(\ref{eq:coldDens}) and (\ref{eq:coldFlux})) for Ne and O are very similar to each other. See Table~\ref{tab:abundTableDensIbex} and Fig.~\ref{fig:abundEffects} for details. The differences between the abundance quotients $\xi_{\mathrm{flux}}$ (Eq.~(\ref{eq:xiFlux})) and $\xi_{\mathrm{flux}}^{\mathrm{peak}}$ are on the order of $12\%$ for Ne/He and O/He and a few percent for Ne/O (see Fig.~\ref{fig:abundEffects} and Table~\ref{tab:abundTableFluxIBEX}). The qoutients $\xi_{\mathrm{flux}}$ and $\xi_{\mathrm{dens}}$ also different from $\xi_{\mathrm{sur}}$ (Eqs.~(\ref{eq:xiDensCold}) and (\ref{eq:xiSur}), the comparison is shown in Fig.~\ref{fig:abundEffects}. 

For the 2009 season, $\xi_{\mathrm{flux,Ne/O}}$ quotient is $\sim9.52$. Adopting the measured at IBEX Ne/O abundance and the filtration factors beyond TS from \citet{bochsler_etal:12a}, one immediately obtains the neutral Ne/O abundance at TS equal to 0.042 and 0.14 in the LIC. Using our estimate of the $\xi_{\mathrm{sur,Ne/O}}$, the LIC Ne/O abundance equals 0.16 for the NIS gas inflow parameters from IBEX \citep{bzowski_etal:12a} and 0.18 for the inflow parameters from GAS/Ulysses \citep{witte:04}. Fig.~\ref{fig:abundMatrix} presents the dependence of the NIS Ne/O abundance as a function of various inflow longitude and speed. The uncertainty of this result is approximately $40\%$. One has to realize, however, that this calculation does not take into account the actual time intervals that \citet{bochsler_etal:12a} used in their analysis, so this estimate is approximate. A more thorough analysis of NIS Ne/O abundance based on IBEX observations is ongoing and will be a subject of a future paper.

\begin{acknowledgements}

M.B. and J.M.S. would like to thank Peter Bochsler for exciting discussion and valuable remarks to the paper draft. The authors from SRC PAS were supported by grant N-N203-513-038 from the Polish Ministry for Science and Higher Education, managed by the Polish National Science Centre. M.A.K was stipendist of the START program of the Polish Science Foundation for 2012. The authors acknowledge the use of NASA/GSFC's Space Physics Data Facility's ftp service for Ulysses/SWOOPS and TIMED/SEE data, SOHO/CELIAS/SEM (\url{http://www.usc.edu/dept/space_science/semdatafolder/long/daily_avg/}), and OMNI2 data collection (\url{ftp://nssdcftp.gsfc.nasa.gov/spacecraft_data/omni/}). 
The F10.7 solar radio flux was provided by the NOAA and Pentincton Solar Radio Monitoring Programme operated jointly by the National Research Council and the Canadian Space Agency (\url{ftp://ftp.ngdc.noaa.gov/STP/SOLAR_DATA/SOLAR_RADIO/FLUX/Penticton_Adjusted/}
and \url{ftp://ftp.geolab.nrcan.gc.ca/data/solar_flux/daily_flux_values/}). 
The composite Lyman-$\alpha$ flux and MgII$_{\mathrm{c/w}}$ were obtained from LASP, accessed through the LISIRD Web page at  (\url{http://lasp.colorado.edu/lisird/lya/})

\end{acknowledgements}

\appendix
\label{sec:appendix}
\section{Modulation of the ionization rates of neutral He, Ne, O in the heliosphere}
\subsection{Photoionization rates of Ne, He, and O}

The model of photoionization rates of NIS species used in this paper calculates Carrington rotation averages of H, He, O, Ne photoionization rates based on directly measured solar spectra from TIMED/SEE and CELIAS/SEM onboard SOHO and a hierarchy of solar EUV radiation proxies. The model of photoionization rates for hydrogen in the heliosphere was described by \citet{bzowski_etal:12b}, here we focus on He, Ne, and O. Details of the derivation and discussion of instrumental and statistical uncertainties will be provided elsewhere (Sok{\'o}{\l} et al., in preparation, Bochsler et al., in preparation).

\begin{figure}
		\resizebox{\hsize}{!}{\includegraphics{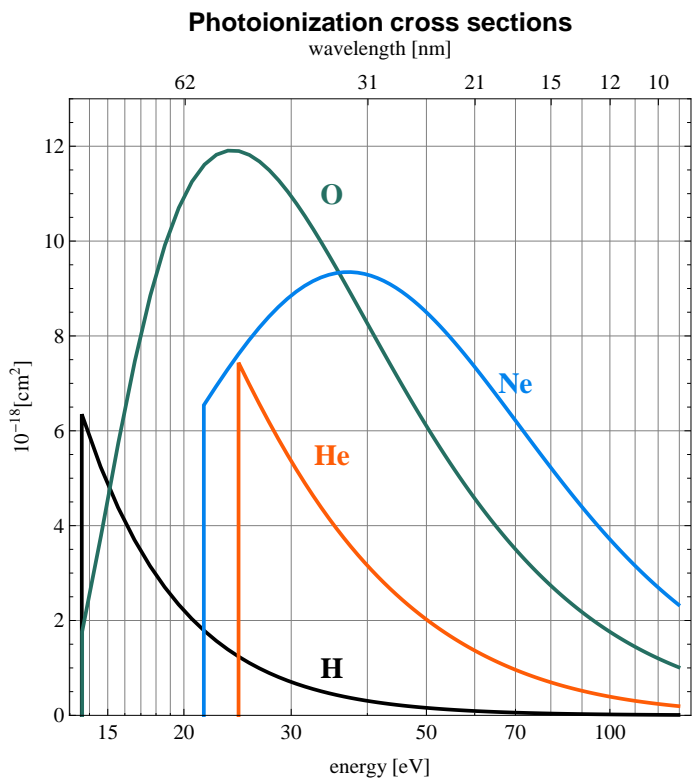}}
		\caption{Cross sections for photoionization of neutral H, He, Ne, and O used in the calculations \citep{verner_etal:96}. The accuracy is at a $\sim5\%$ level. Note that even though the ionization threshold energy for O and H are very similar, the cross section values for the low energy portion, where the solar spectral flux is greatest, are substantially different.}
		\label{fig:photoRatesCrossSection}
	\end{figure}

Basically, calculating photoionization rate for a given species is simple: one integrates the spectral flux of the Sun $F\left( \lambda,t \right)$ for a wavelength $\lambda$ and time $t$, multiplied by the cross section $\sigma\left( \lambda \right)$ for a given species over an integral starting from $0$ to the ionization threshold wavelength $\lambda_0$:
	\begin{equation}
		\beta_{\mathrm{ph}}\left( t \right)=\int\limits_0^{\lambda_0}{F\left( \lambda, t \right)\sigma\left( \lambda \right) \mathrm{d}\lambda}
		\label{eq:photoIonRate}
	\end{equation}
For the cross sections we adopted results from \citet{verner_etal:96} (see Fig.~\ref{fig:photoRatesCrossSection}).
	
Regular measurements of the solar spectral flux are available only starting from 2002 owing to the TIMED experiment \citep{woods_etal:05a}. Earlier, measurements in several spectral bands in the EUV above the ionization threshold of H, He, Ne, and O region (SEM/CELIAS/SOHO, \citet{hovestadt_etal:95a}) were available since 1996. Fortunately, it has been demonstrated that the photoionization rates of many species are correlated with a number of other routinely measured quantities (proxies), including the MgII~core-to-wing index \citep[MgII$_{\mathrm{c/w}}$,][]{heath_schlesinger:86, viereck_puga:99}, solar Lyman-$\alpha$ flux \citep[Ly$\alpha$,][]{woods_etal:96a, woods_etal:00} and the solar radio 10.7~cm flux \citep[F10.7,][]{covington:47, tapping:87}. Thus, to infer a sufficiently long time series of photoionization rates of NIS species one can use proxies obtained from statistical analysis of the correlations. 

\begin{figure*}
		\resizebox{\hsize}{!}{\includegraphics{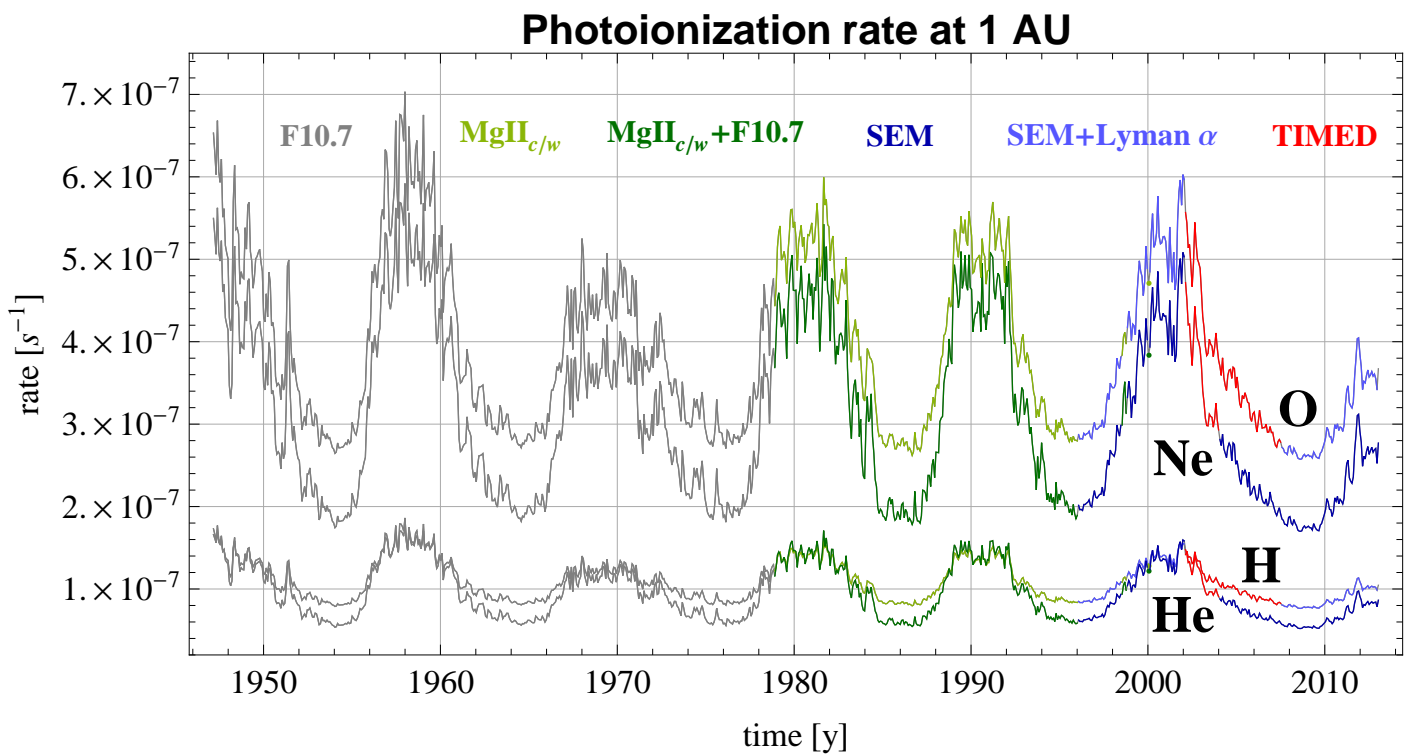}}
		\caption{Time series of Carrington period-averaged photoionization rates of H, He, Ne, and O at 1~AU from the Sun. The color code denotes the various sources of data and proxies used, as discussed in the text.}
		\label{fig:photoRates}
	\end{figure*}

The baseline of the model is the solar spectrum measured by TIMED/SEE \citep{woods_etal:05a} during a few years after launch, when the solar activity was going down from maximum to minimum value and the absolute calibration of the instrument was the most reliable. These portions of the time series (calculated separately for each species) are shown in red in Fig.~\ref{fig:photoRates}. 
We used measurements of 2-hours resolution in time and calculated the photoionization rates using Eq.~(\ref{eq:photoIonRate}). 
We filtered the results against flares and particle background and calculated Carrington rotation averages. Due to suspected degradation of the TIMED/SEE detector we decided to use only a few first years of the TIMED measurements, namely the first 27 Carrington rotations from 22~January~2002 for He and Ne, and 74 Carrington rotations for H and O. These data are shown in red in Fig.~\ref{fig:photoRates}.

For other times we developed a set of proxies by taking statistical correlation between the directly calculated values and the solar EUV radiation proxy. The hierarchical set of proxies is based on the chromospheric MgII$_{\mathrm{c/w}}$ index available from 1978 as released by LASP \citep{heath_schlesinger:86, viereck_puga:99, viereck_etal:04a}, the F10.7 flux released by NOAA and DRAO \citep{tapping:87} available from 1947, and the composite Lyman-$\alpha$ flux published by LASP (available from 1947, but based on actual UV observations from mid-1970-ties, with some gaps filled by proxies developed by the LASP team, see \citet{woods_etal:00}). As the most reliable proxy we have chosen the CELIAS/SEM data \citep{hovestadt_etal:95a}, which are measurements of a fragment of the solar spectrum responsible for most of helium ionization.

The hierarchy of photoionization rates retrieval we used was similar for He and Ne, as they show similar properties, and a different one for H and O, as they have the same ionization threshold energy (see photoionization cross sections in question in Fig.~\ref{fig:photoRatesCrossSection}). 

The procedure of building the time series for the noble gases was as follows. In the first step we used the EUV flux in the two spectral bands measured by CELIAS/SEM onboard SOHO, which is available since 1996. The first order flux (26--34~nm, here referred to as Ch1) and the central order flux (0.1--50~nm, here referred to as Ch2) were considered. We found, for each species separately, a linear correlation between the daily time series of photoionization rates from the TIMED and SEM time series for the chosen time intervals from the TIMED data. 

The general formula to calculate photoionization rate from SEM data (darker blue line in Fig.~\ref{fig:photoRates}) is the following:
	\begin{equation}\
		\beta_{\mathrm{ph}}^{\mathrm{SEM}} = a + b \, \mathrm{Ch1} + c \, \mathrm{Ch2}
		\label{eq:betaSEM}
	\end{equation}
with the parameter values for He and Ne shown in Table~\ref{tab:phModelSEMHeNe}. From this formula we calculate daily photoionization rates, which are further used to compute the Carrington rotation averaged time series that we use as a basis for further studies.

	\begin{table*}
		\caption{Parameter values for He and Ne, Eq.~(\ref{eq:betaSEM})}
		\label{tab:phModelSEMHeNe}
		\centering
		\begin{tabular}{llll}
		\hline
		Species & $a$ & $b$ & $c$ \\ \hline \hline
		He & $1.42000 \times 10^{-8}$ & $3.41831 \times 10^{-19}$ & $2.01850 \times 10^{-18}$ \\ 
		Ne & $5.53033 \times 10^{-8}$ & $-3.33550 \times 10^{-18}$ & $8.43235 \times 10^{-18}$ \\ \hline
		\end{tabular}
	\end{table*}

The time coverage of the SEM data is not fully sufficient for modeling heliospheric ionization processes. The travel of NIS atoms from 150 AU takes about 30 years or more, even though most of the ionization losses occur inside $\sim10$~AU from the Sun. Thus it is desirable to have a homogeneously prepared time history of ionization rates in the heliosphere dating back as far as possible, so that one model can be used to analyze both present and past measurements, as, e.g., obtained from Ulysses. 

\begin{figure*}
		\centering
		\begin{tabular}{|c|c|c|}
		\hline
			\includegraphics[width=.3\textwidth]{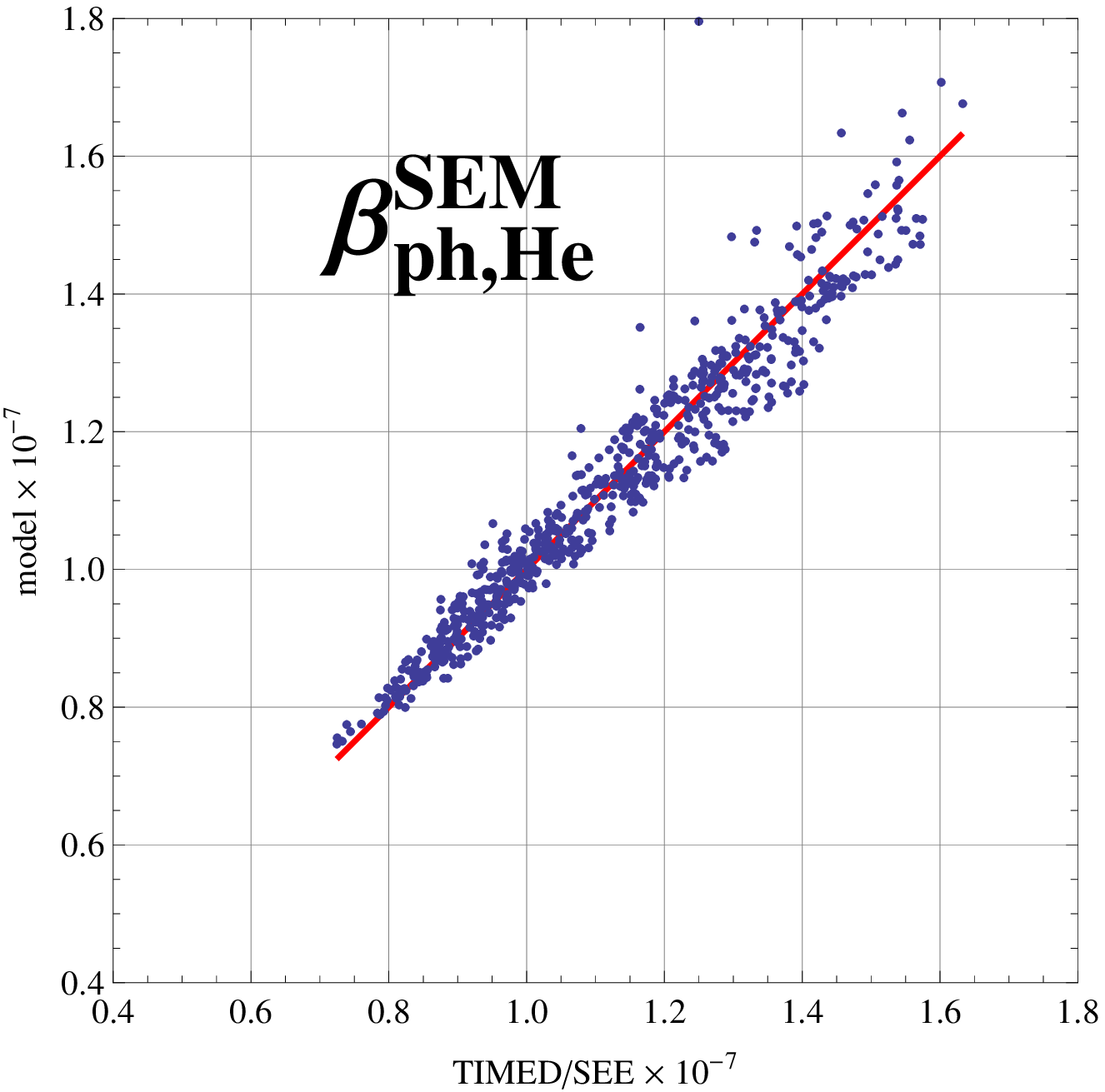}&
			\includegraphics[width=.3\textwidth]{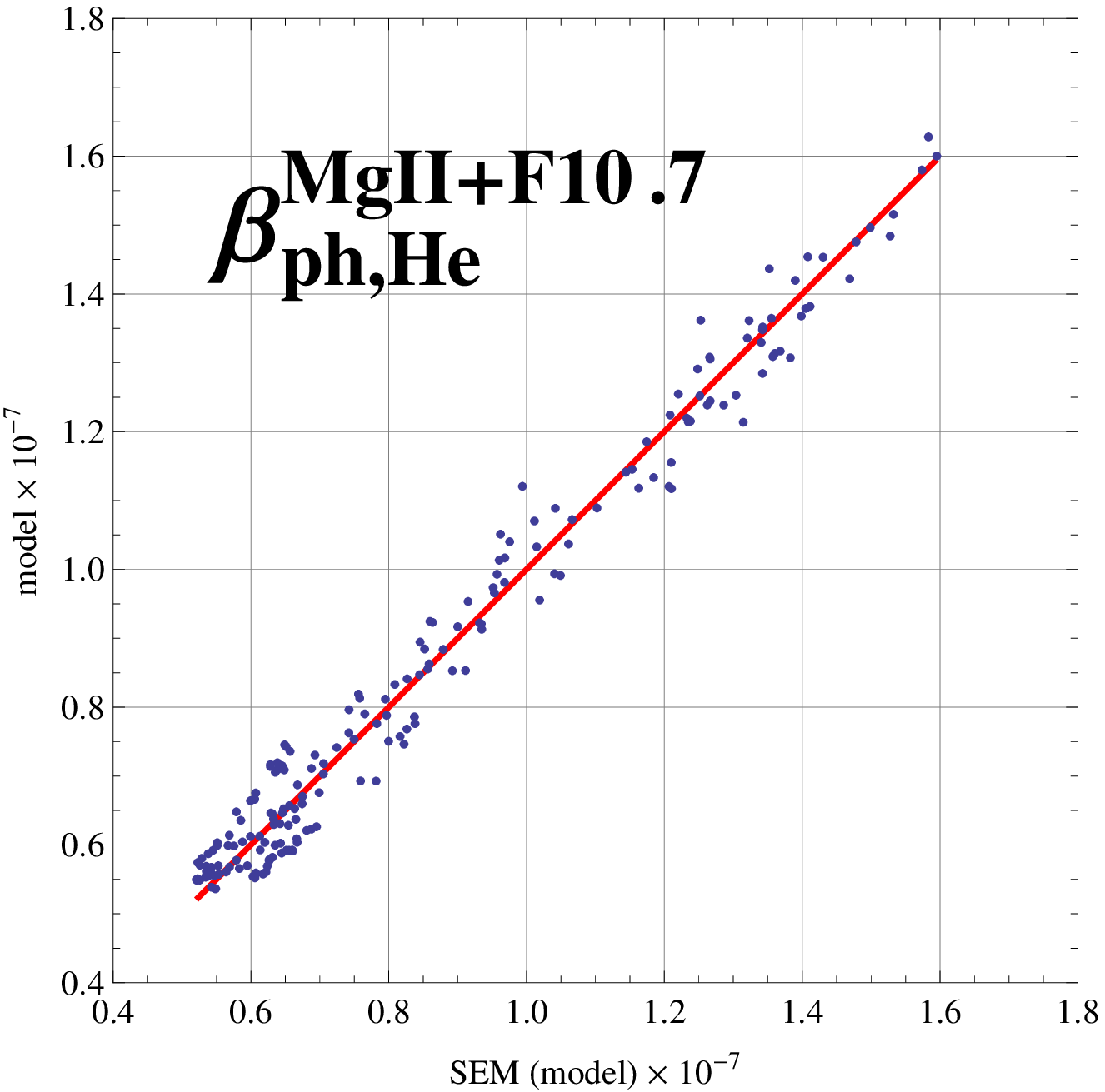}&
			\includegraphics[width=.3\textwidth]{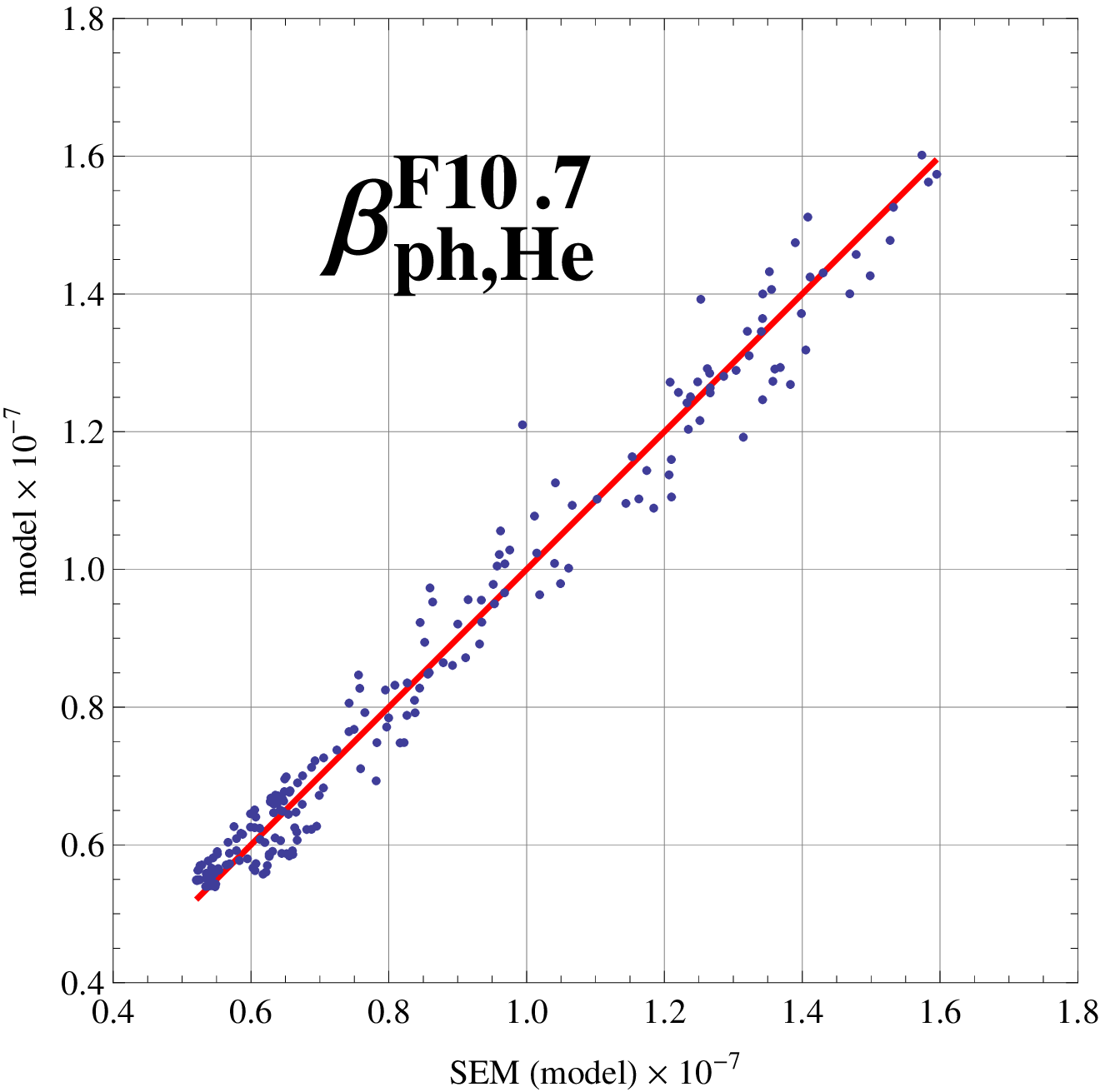}\\ \hline
			\includegraphics[width=.3\textwidth]{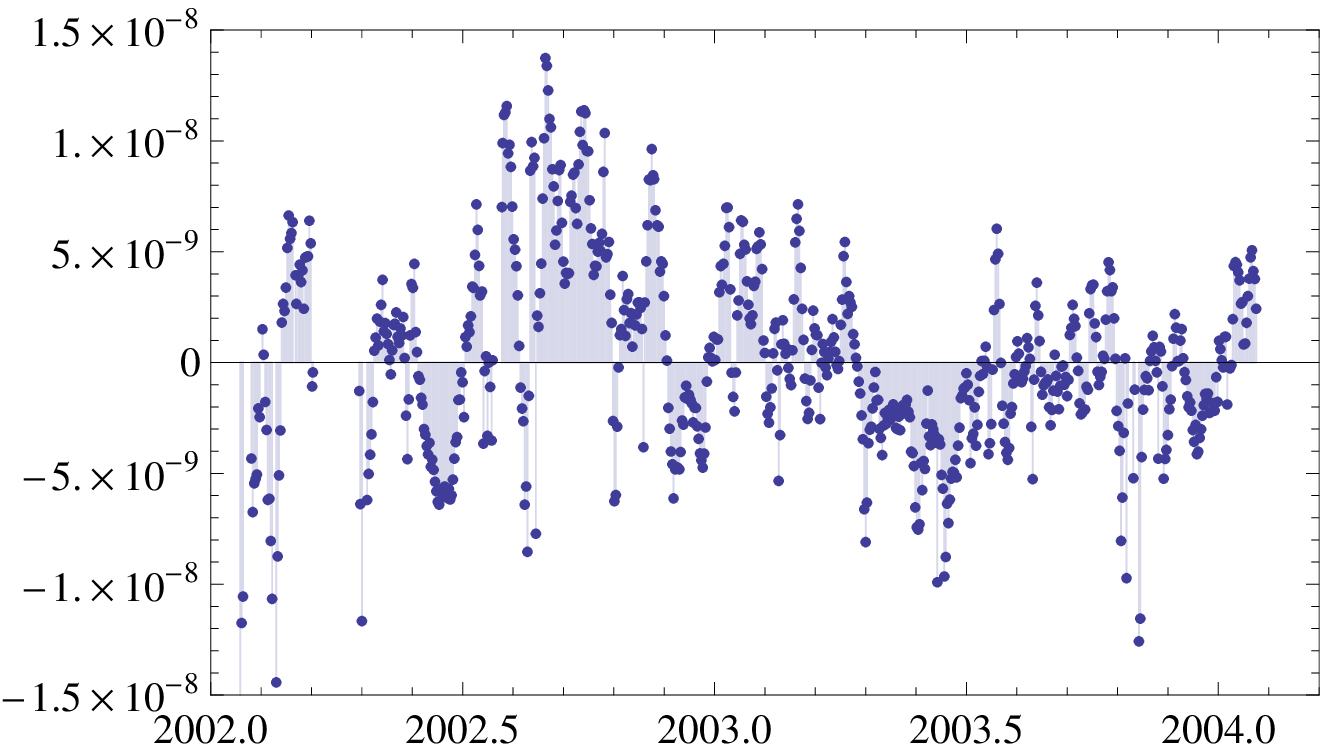}&
			\includegraphics[width=.3\textwidth]{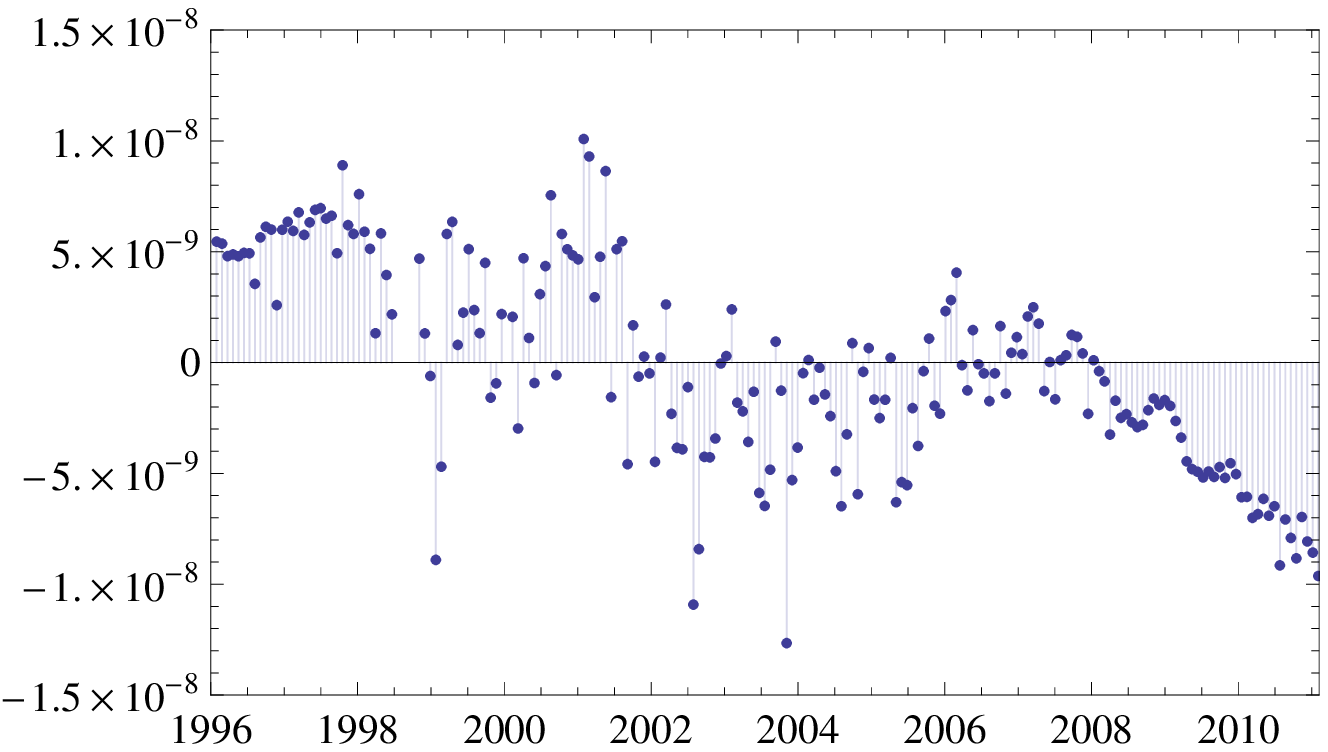}&
			\includegraphics[width=.3\textwidth]{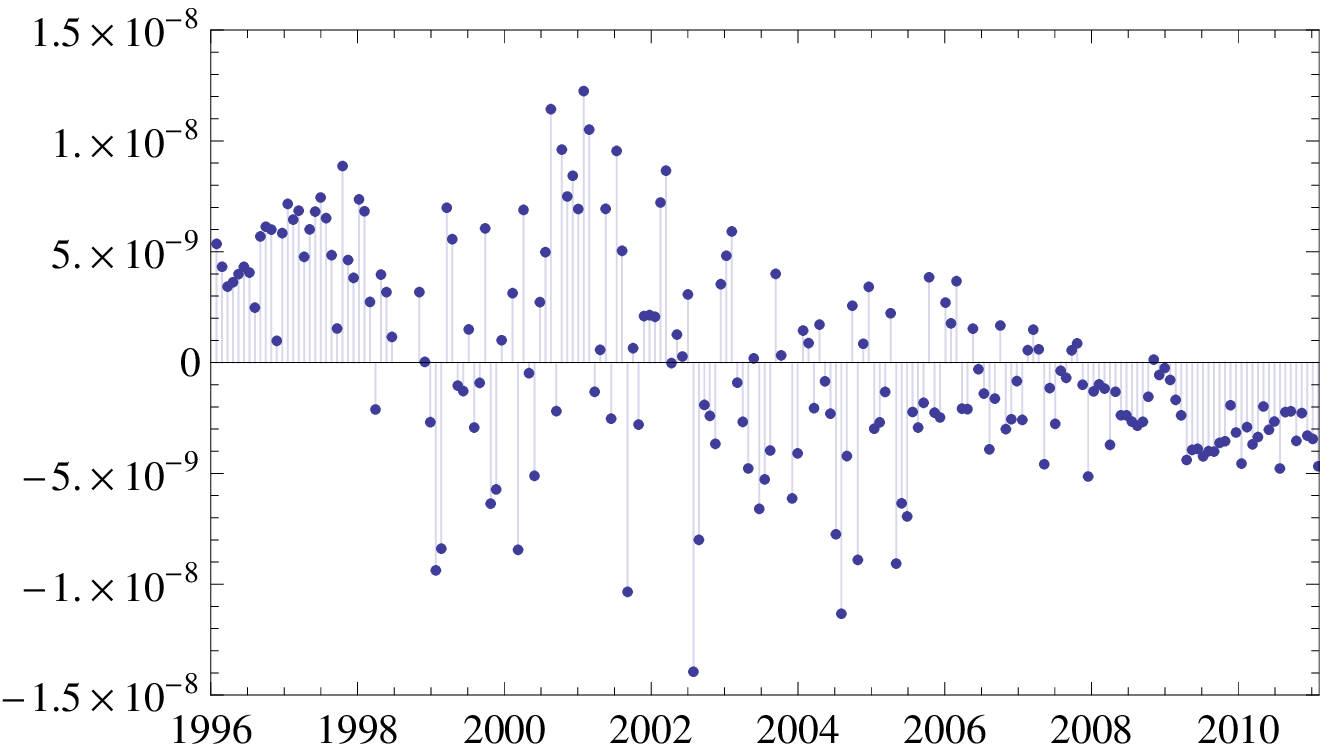}\\ \hline
			\includegraphics[width=.3\textwidth]{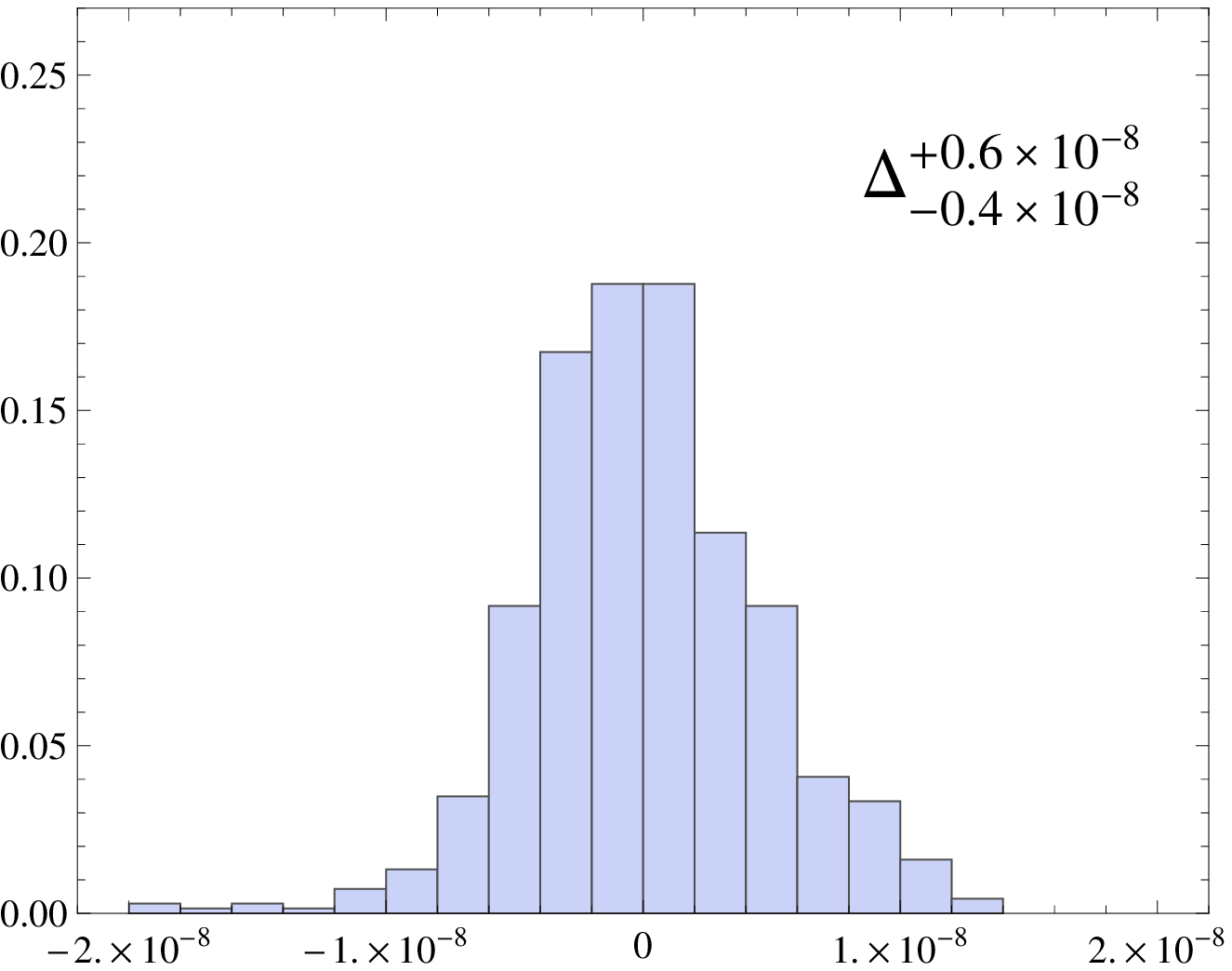}&
			\includegraphics[width=.3\textwidth]{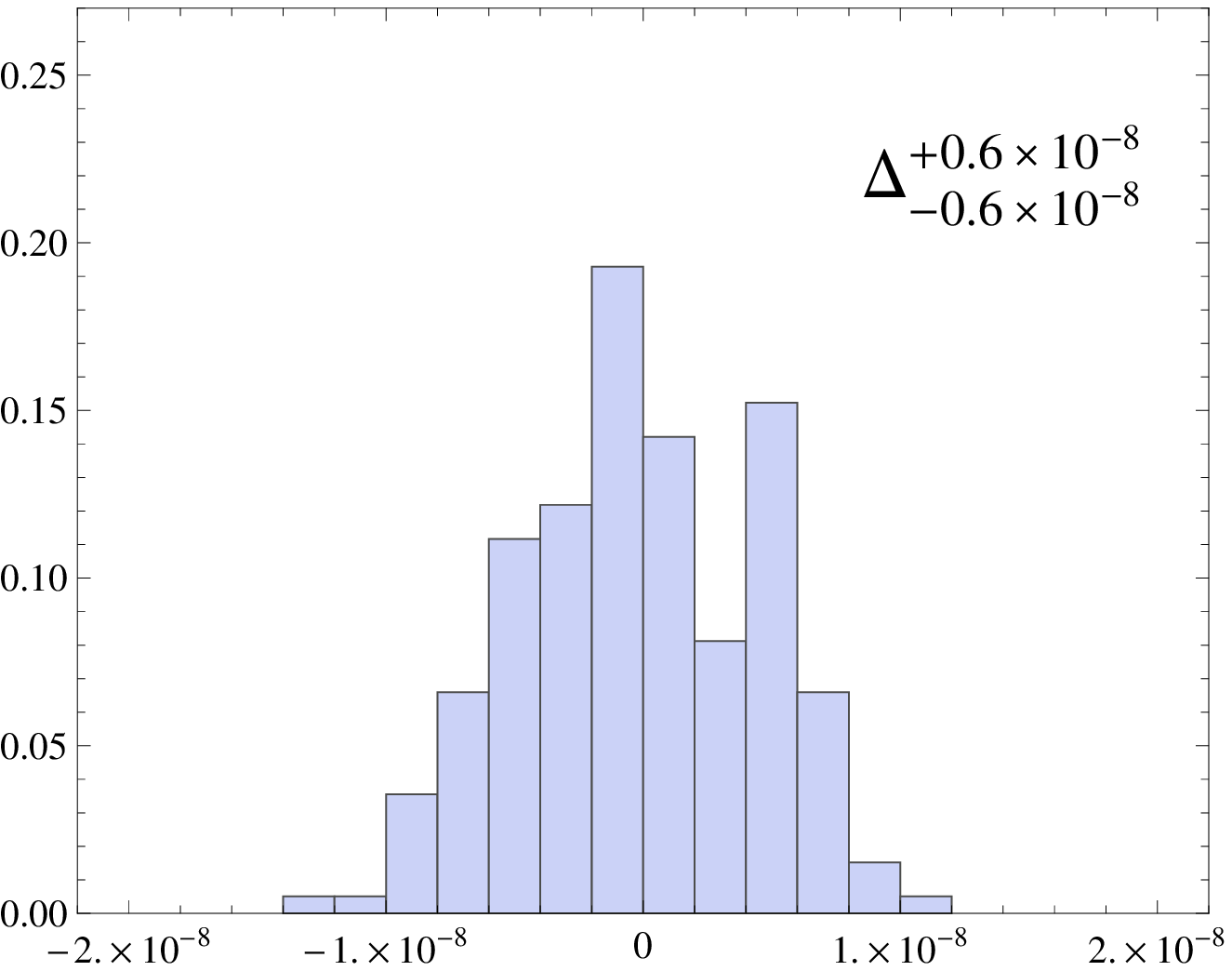}&
			\includegraphics[width=.3\textwidth]{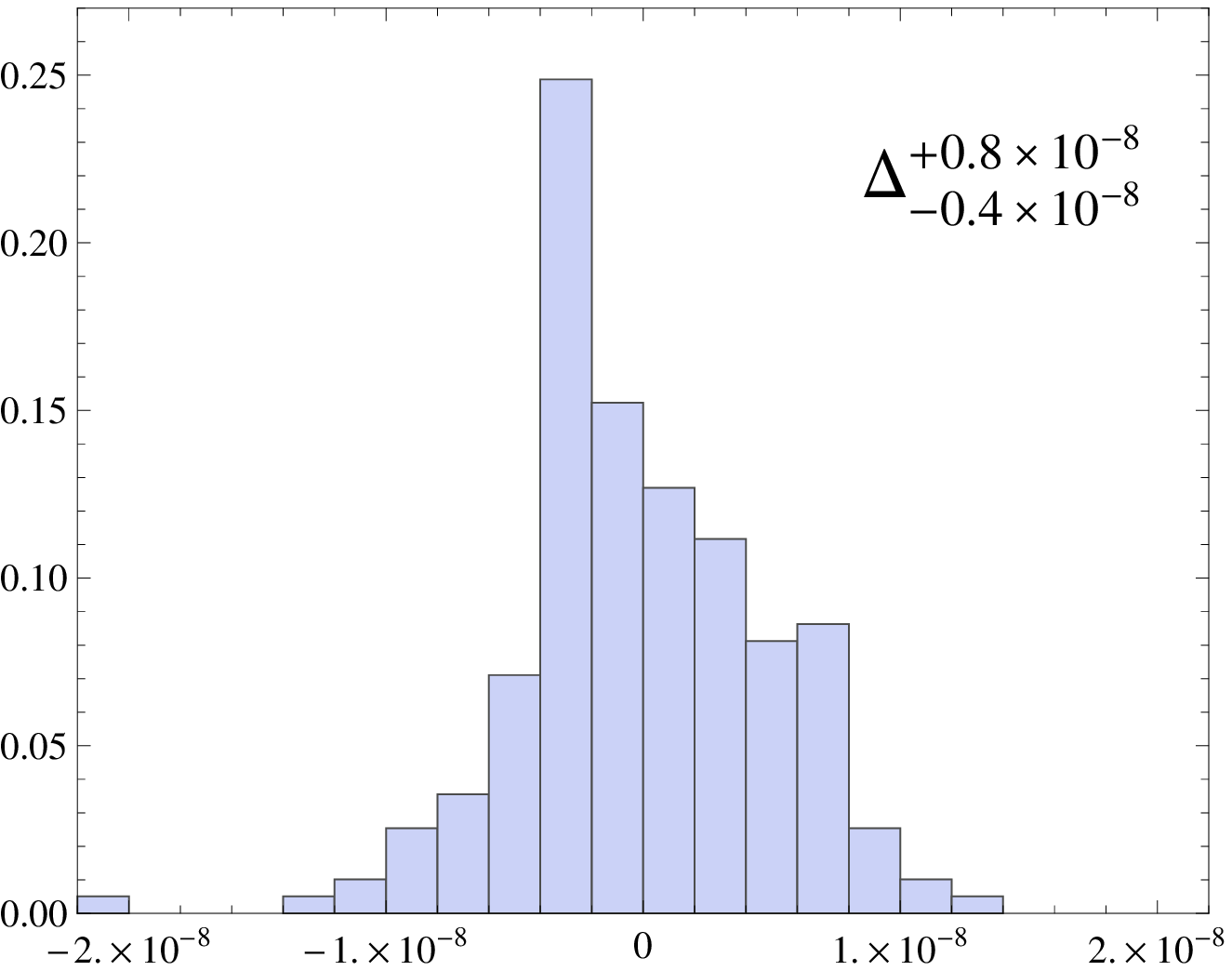}\\ \hline
		\end{tabular}
		\caption{Scatter plot of  model vs data for photoionization models for He defined in Eq.~(\ref{eq:betaSEM}) (left column), Eq.~(\ref{eq:betaMg2F107}) (middle column) and Eq.~(\ref{eq:betaF107}) (right column), fit residuals (middle row), and histograms of the residuals (lower row). The red $y=x$ lines are eye guides. The statistical uncertainties of the models are indicated in insets in the  histogram panels.}
		\label{fig:PhionRatesModelHe}
	\end{figure*}

\begin{figure*}
		\centering
		\begin{tabular}{|c|c|c|}
		\hline
			\includegraphics[width=.3\textwidth]{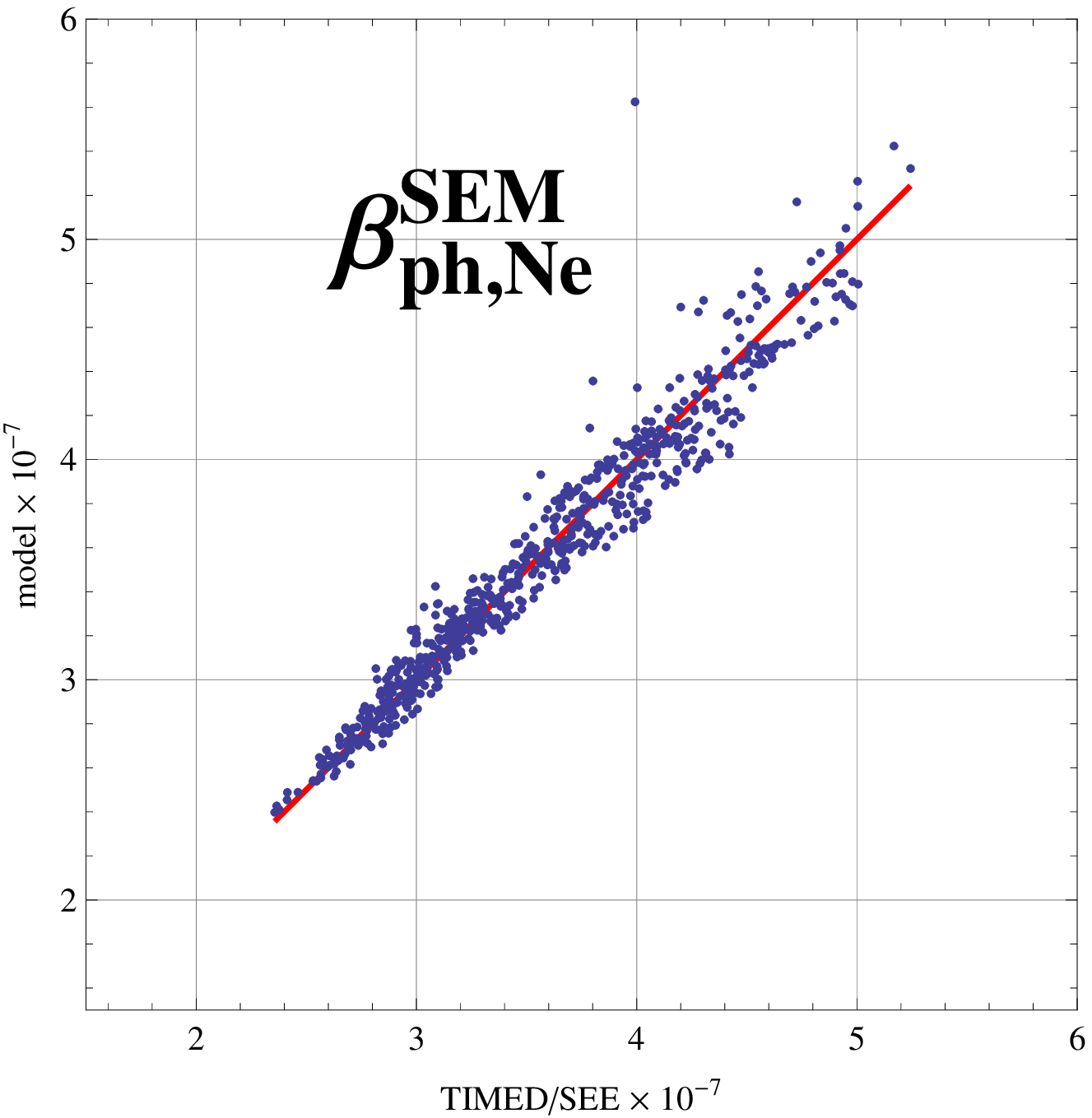}&
			\includegraphics[width=.3\textwidth]{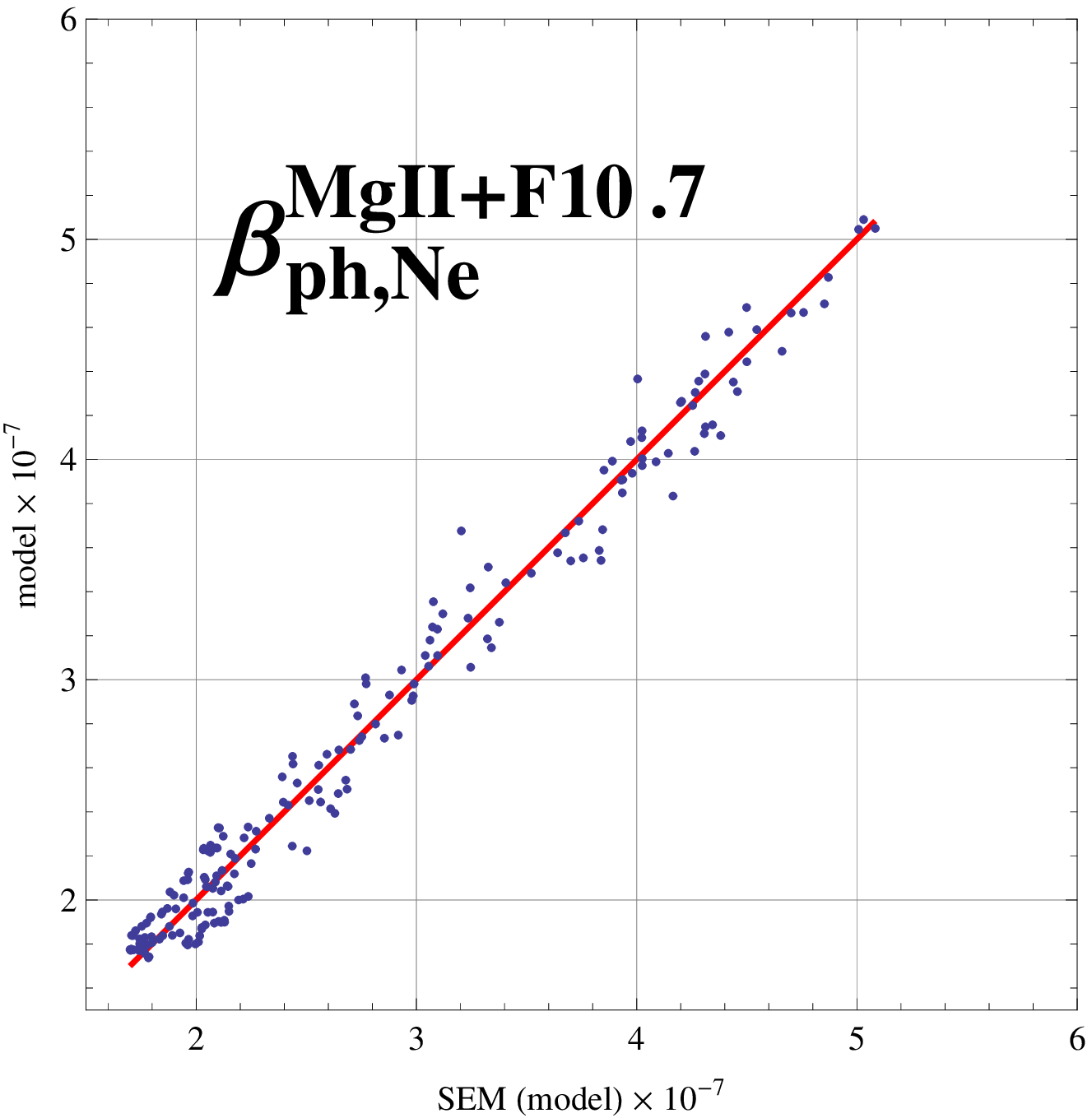}&
			\includegraphics[width=.3\textwidth]{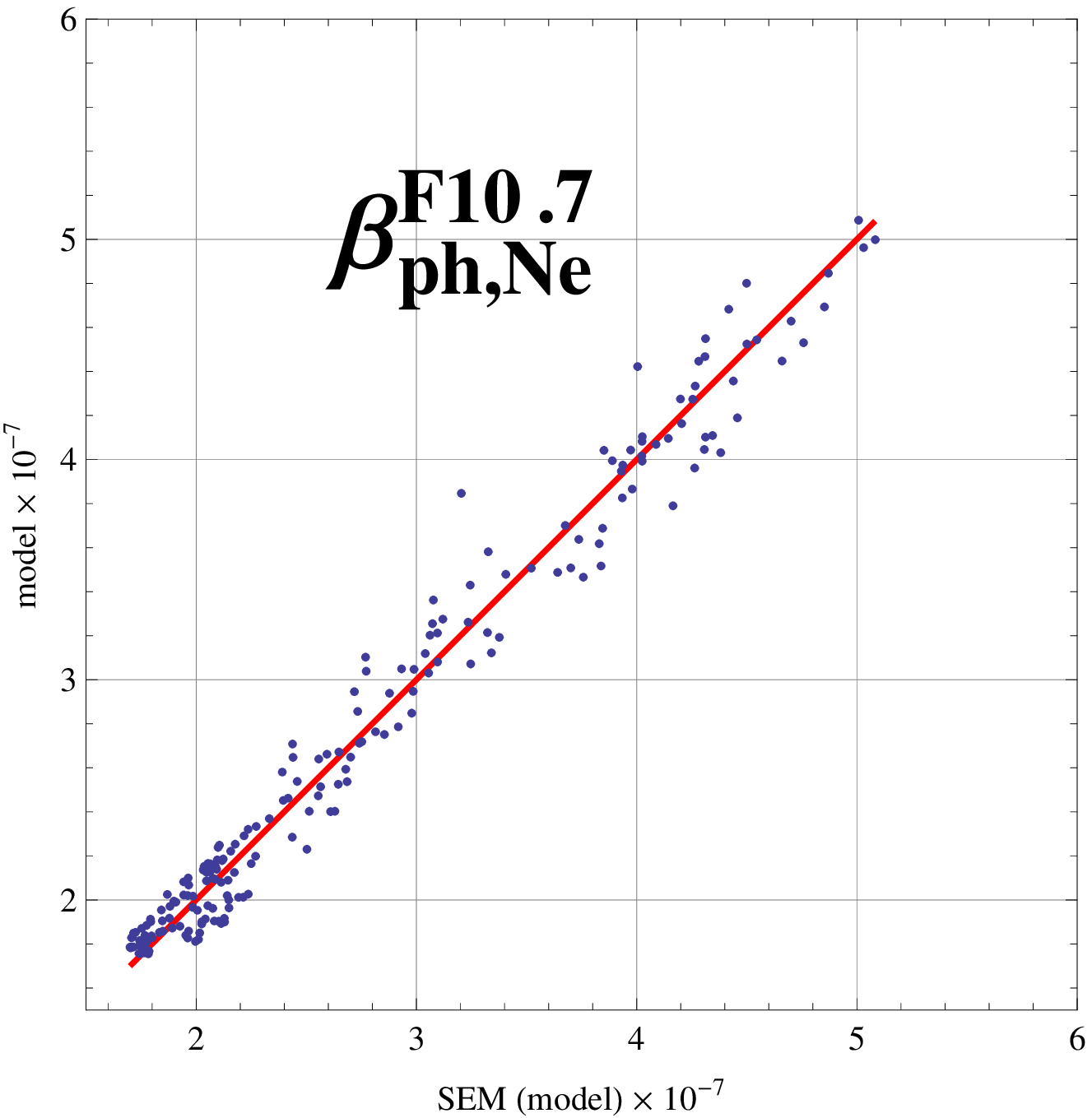}\\ \hline
			\includegraphics[width=.3\textwidth]{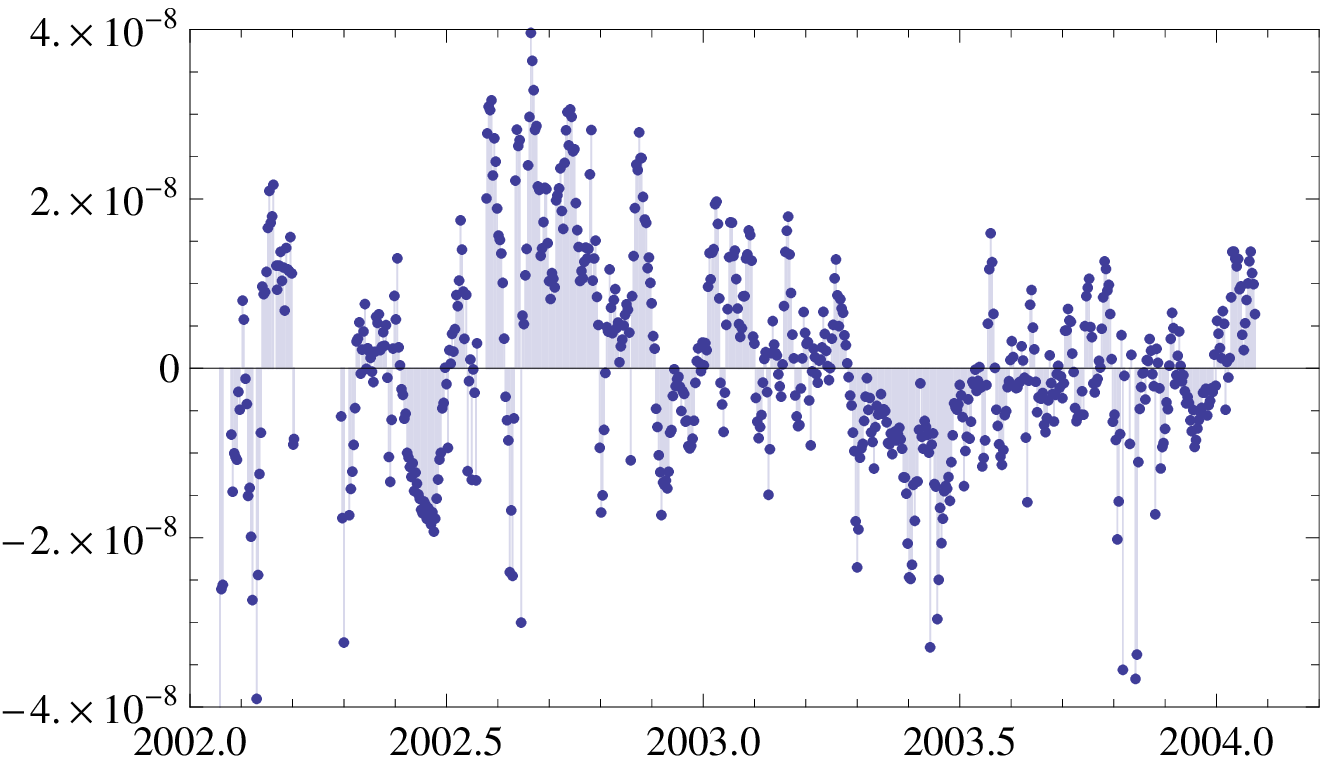}&
			\includegraphics[width=.3\textwidth]{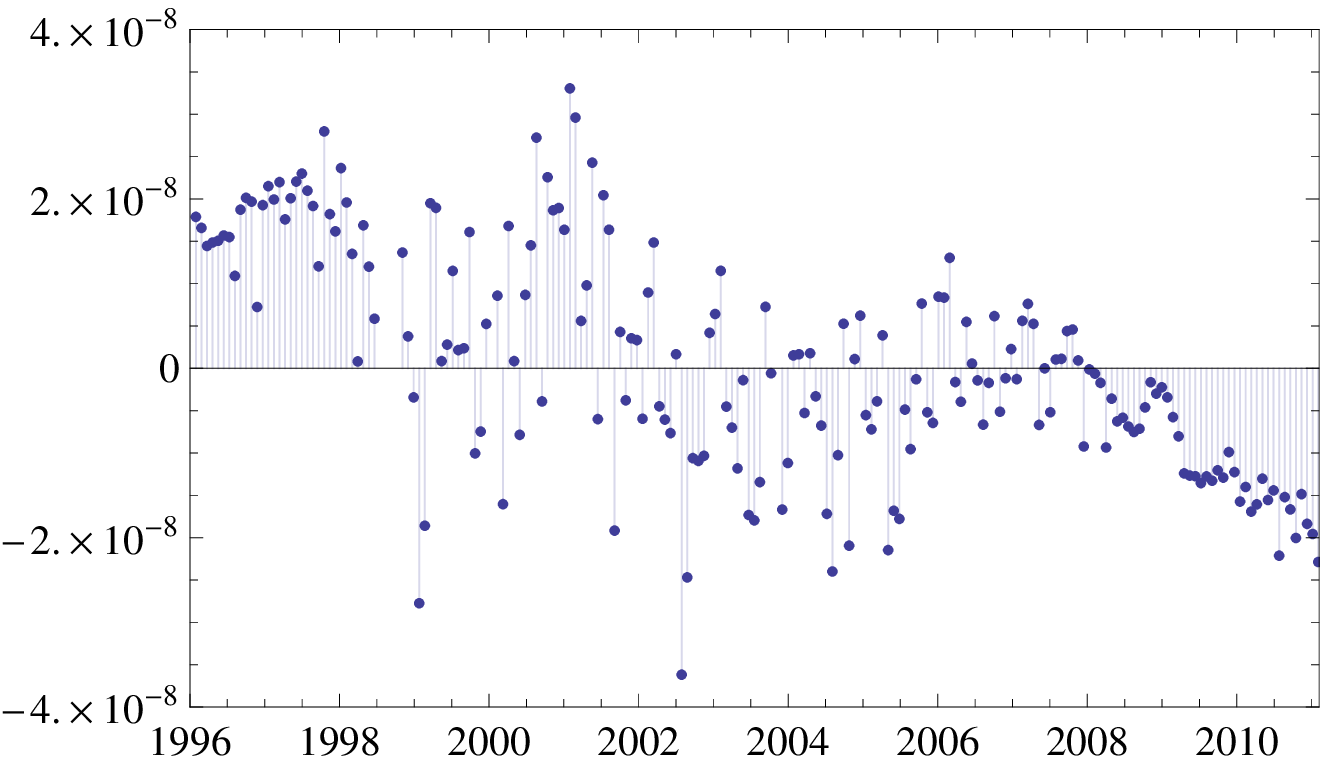}&
			\includegraphics[width=.3\textwidth]{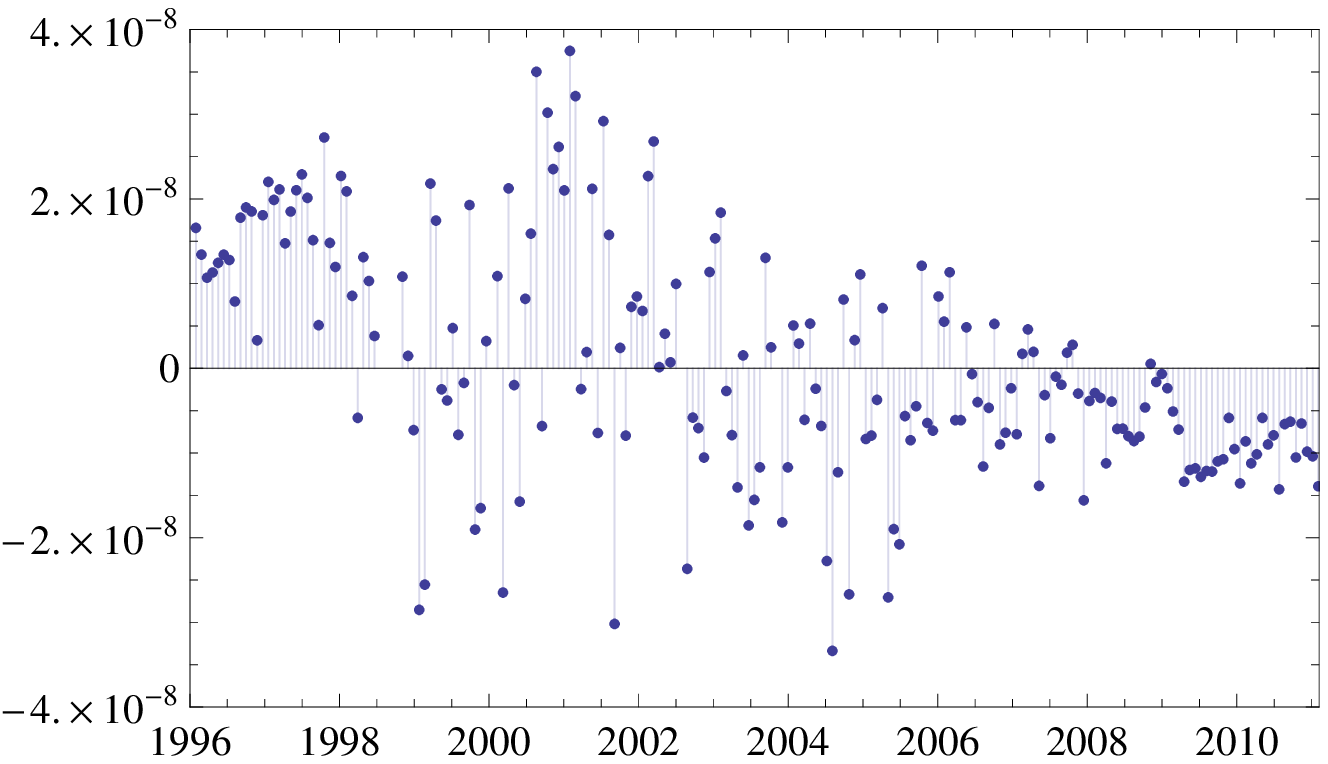}\\ \hline
			\includegraphics[width=.3\textwidth]{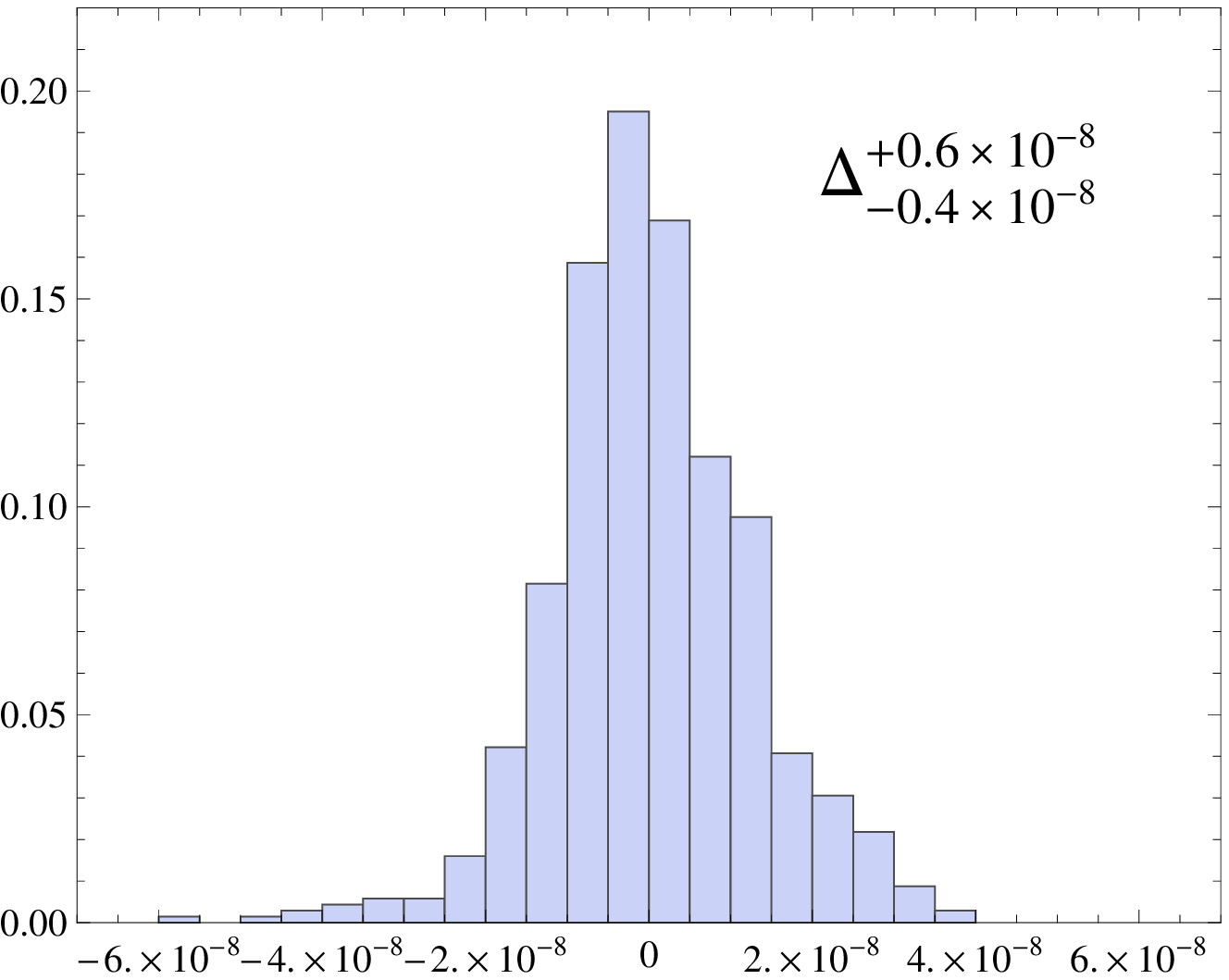}&
			\includegraphics[width=.3\textwidth]{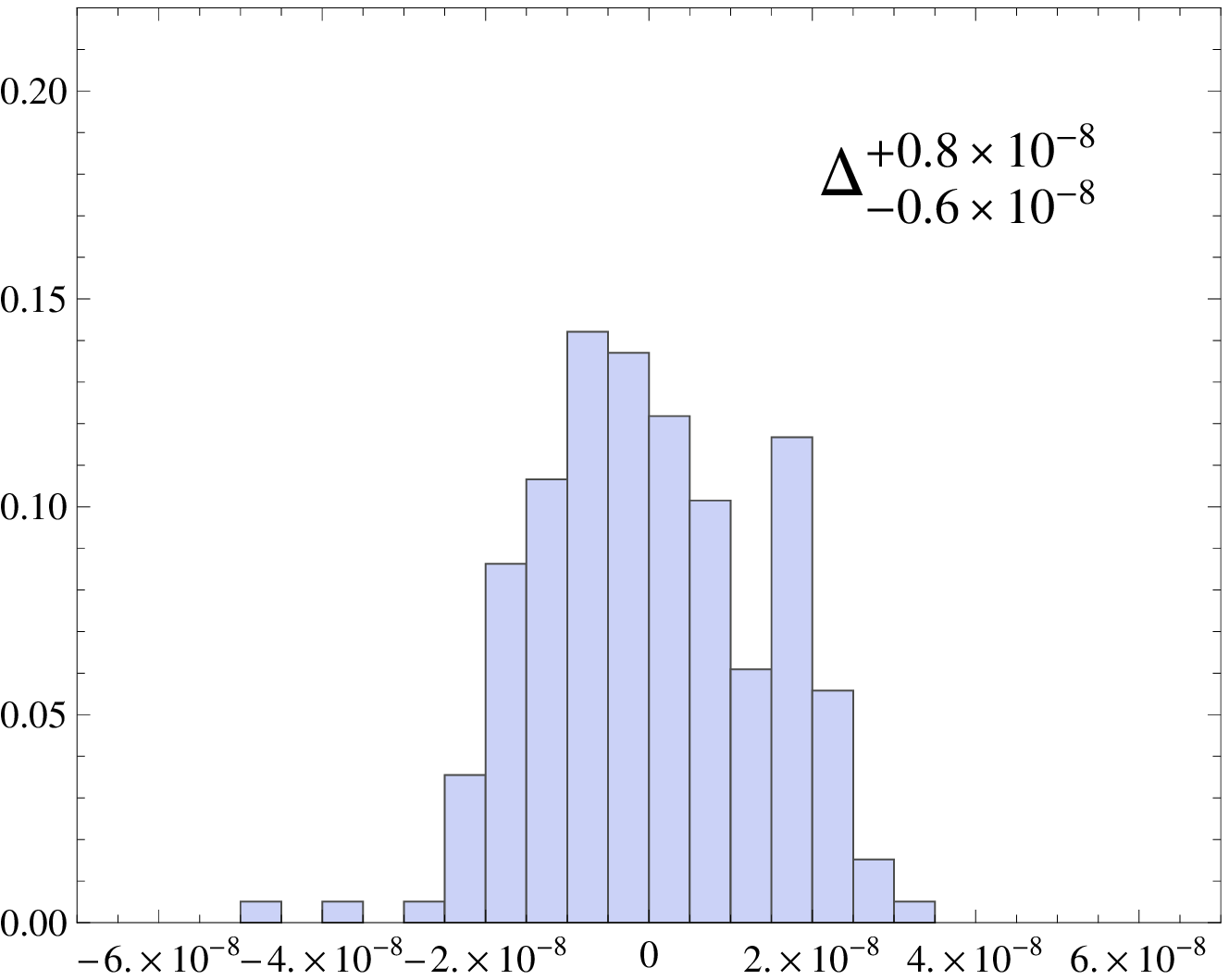}&
			\includegraphics[width=.3\textwidth]{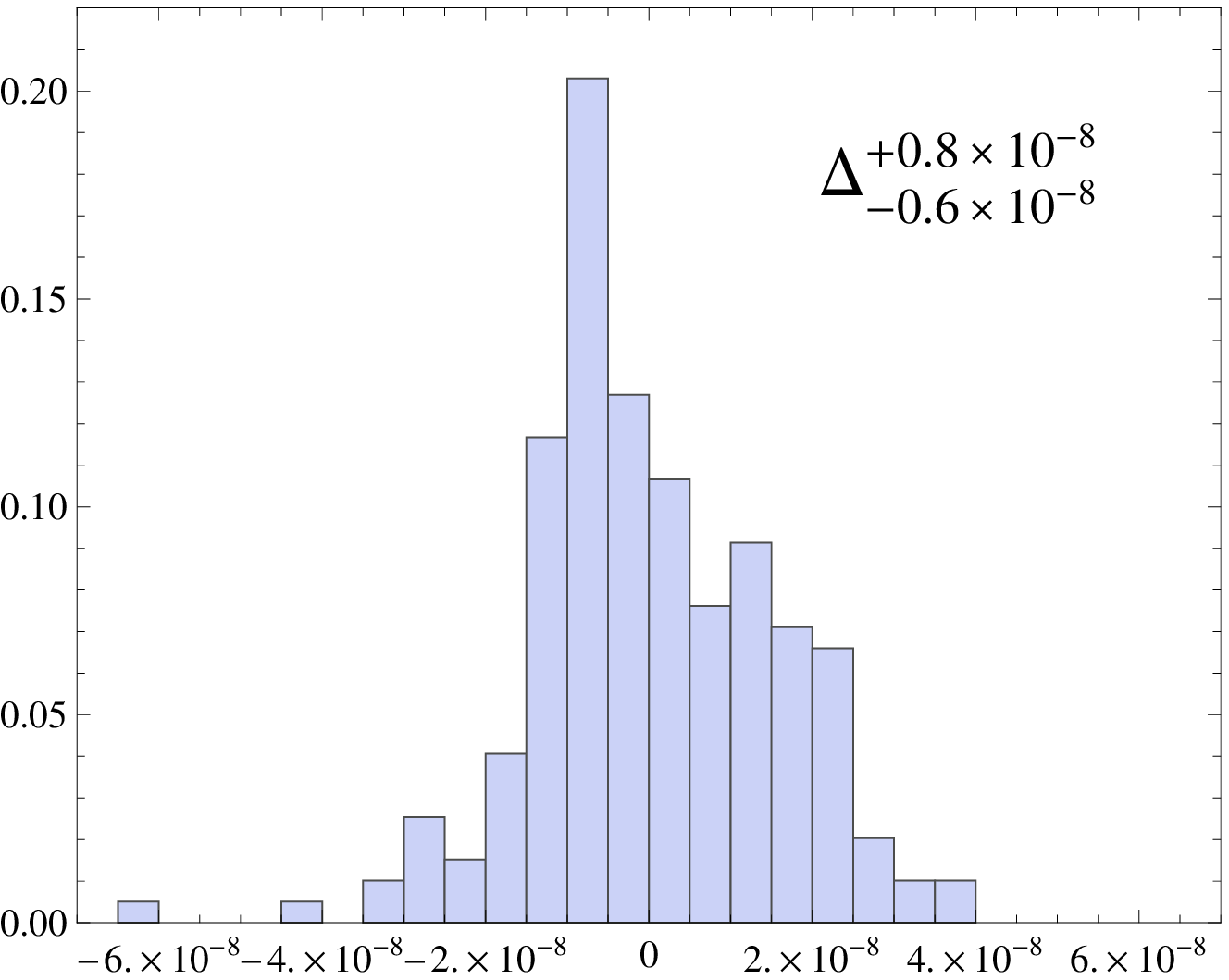}\\ \hline
		\end{tabular}
		\caption{Scatter plot of  model vs data for photoionization models for Ne defined in Eq.~(\ref{eq:betaSEM}) (left column), Eq.~(\ref{eq:betaMg2F107}) (middle column) and Eq.~(\ref{eq:betaF107}) (right column), fit residuals (middle row), and histograms of the residuals (lower row). The red $y=x$ lines are  eye guides. The statistical uncertainties of the models are indicated in the insets in the histogram panels.}
		\label{fig:PhionRatesModelNe}
	\end{figure*}

\begin{figure*}
		\centering
		\begin{tabular}{|c|c|c|}
		\hline
			\includegraphics[width=.3\textwidth]{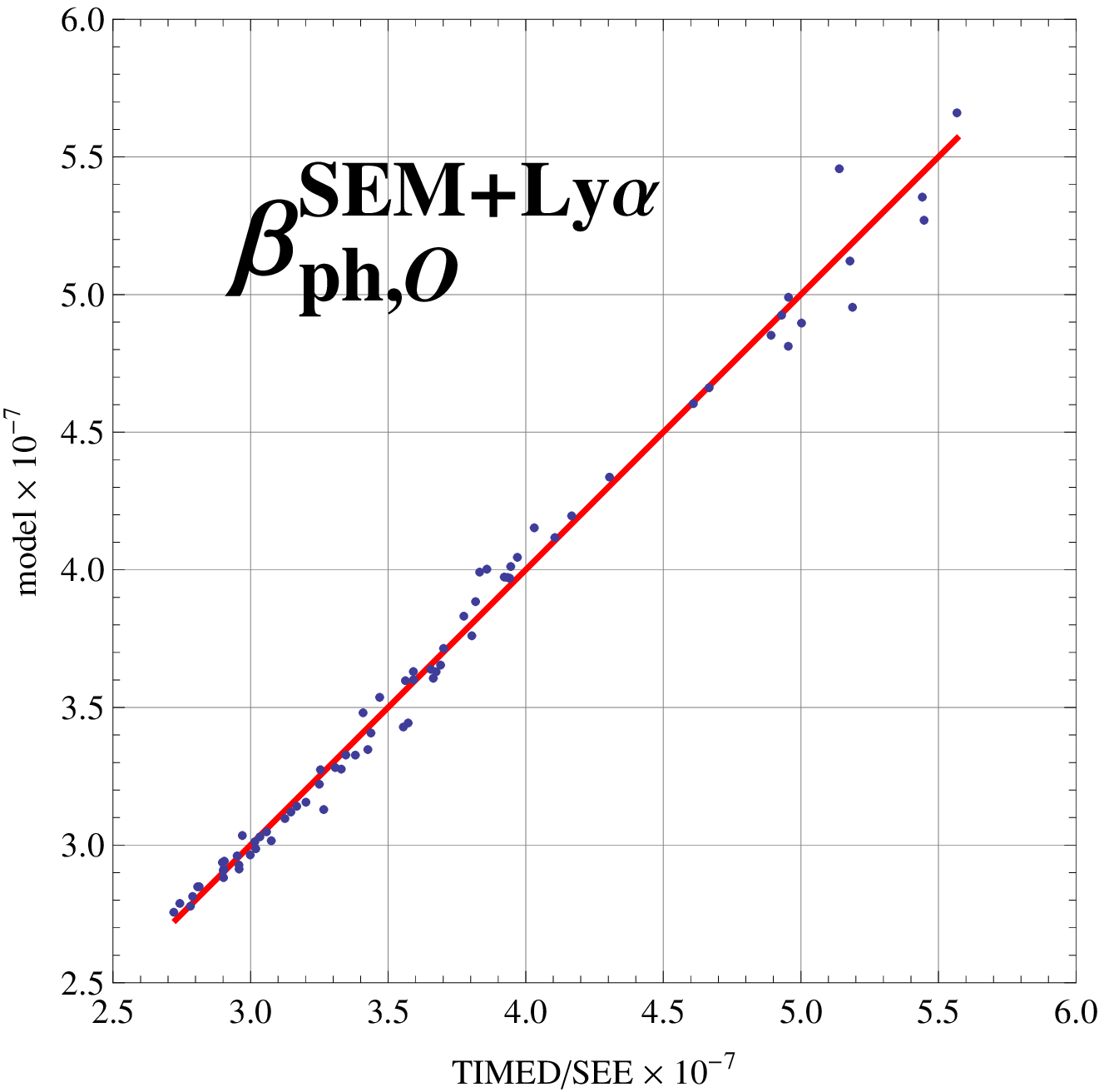}&
			\includegraphics[width=.3\textwidth]{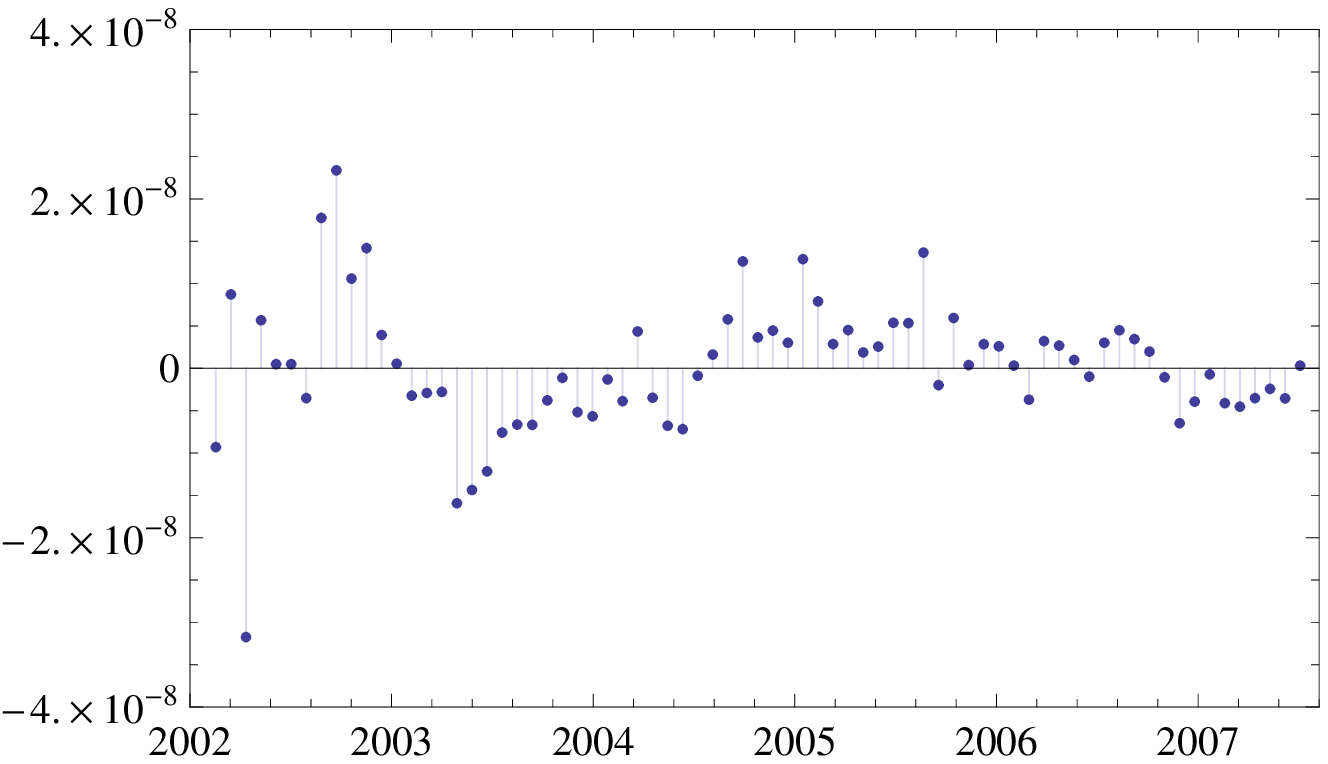}&
			\includegraphics[width=.3\textwidth]{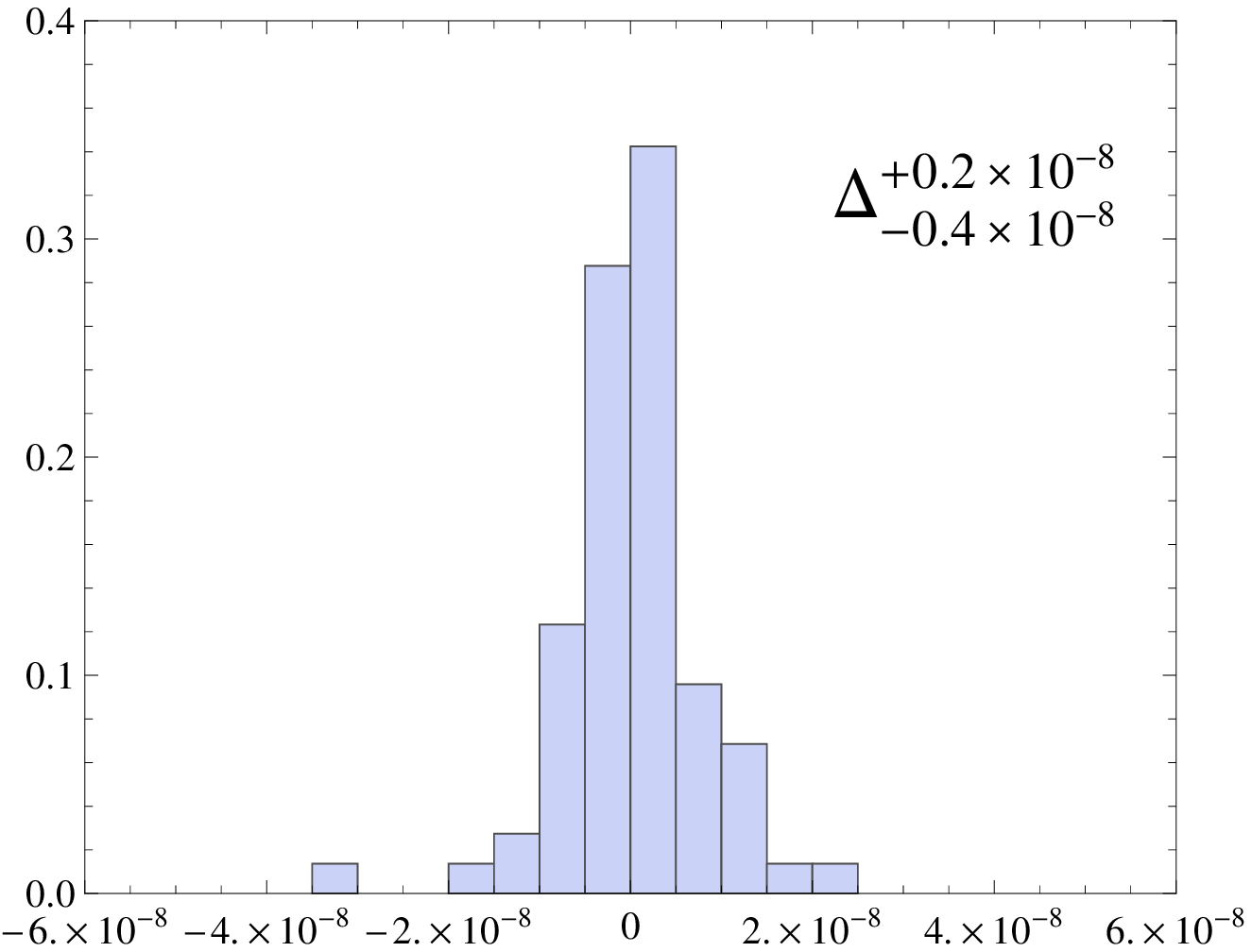}\\ \hline
			\includegraphics[width=.3\textwidth]{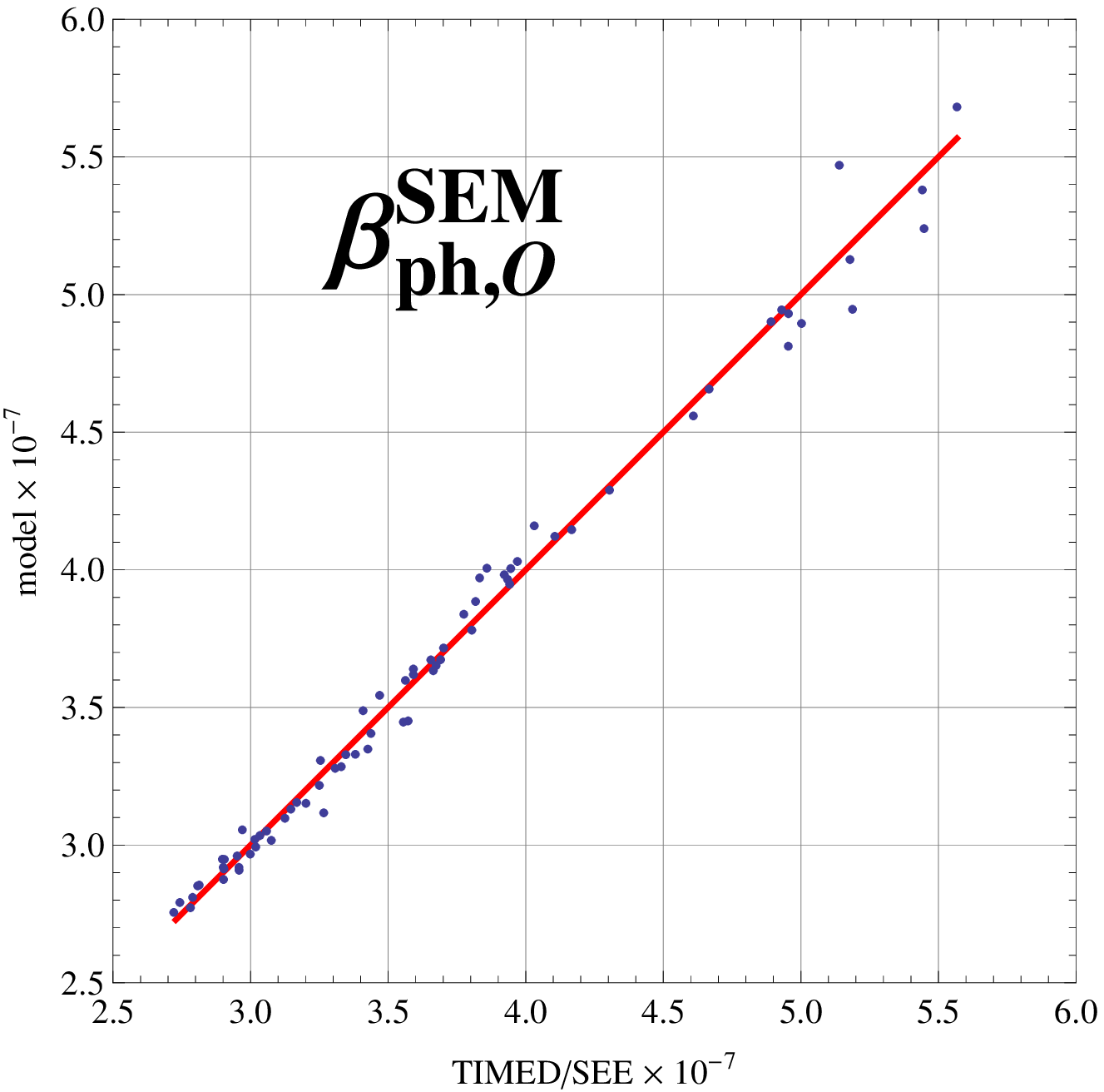}&
			\includegraphics[width=.3\textwidth]{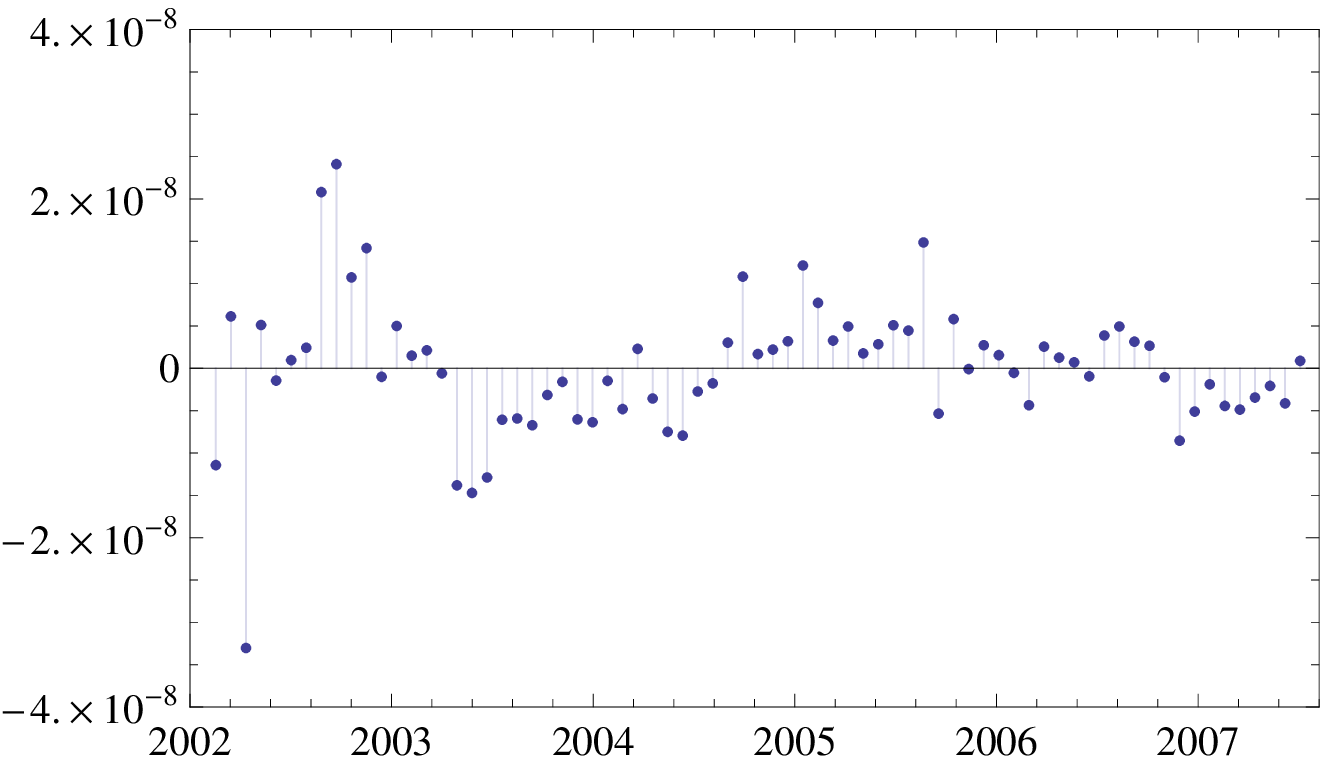}&
			\includegraphics[width=.3\textwidth]{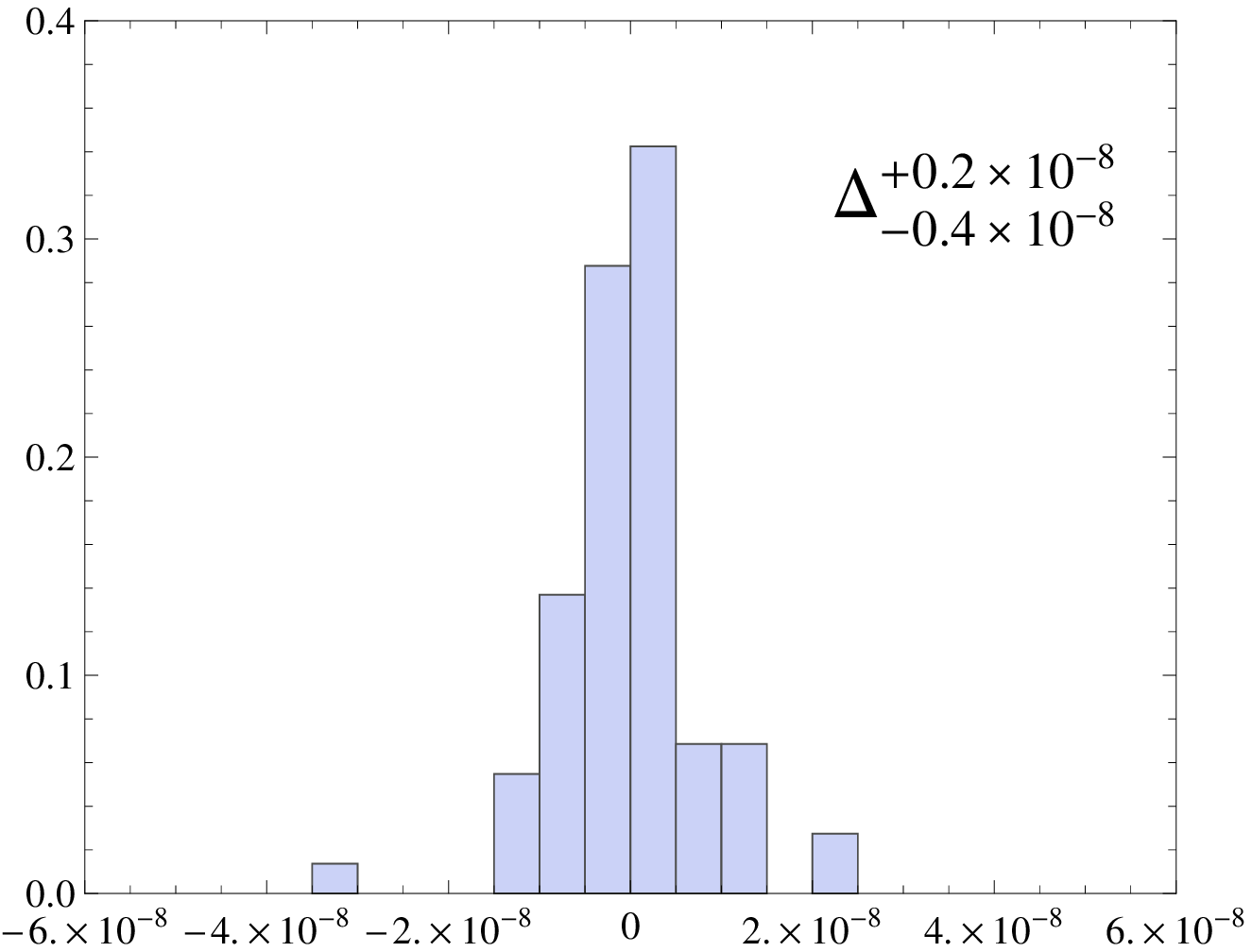}\\ \hline
			\includegraphics[width=.3\textwidth]{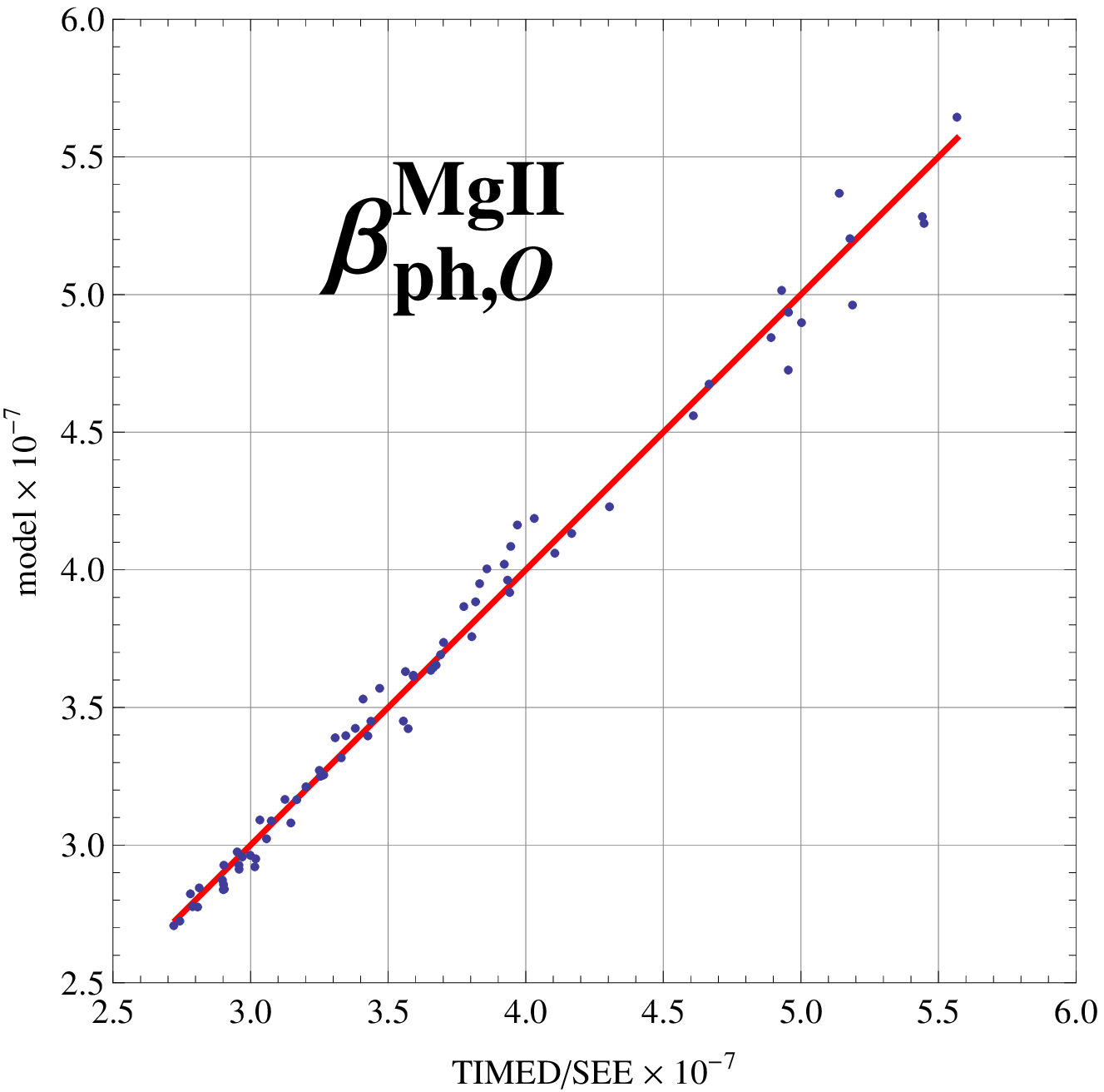}&
			\includegraphics[width=.3\textwidth]{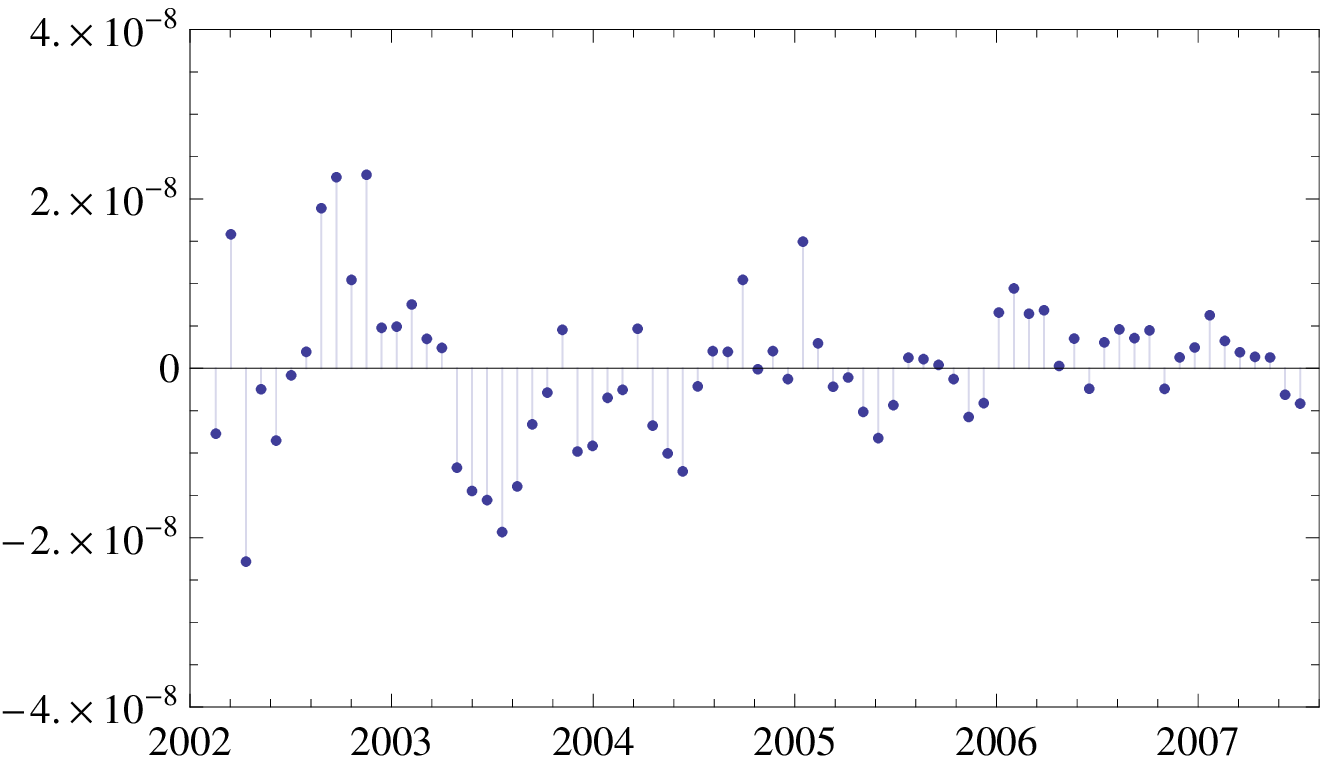}&
			\includegraphics[width=.3\textwidth]{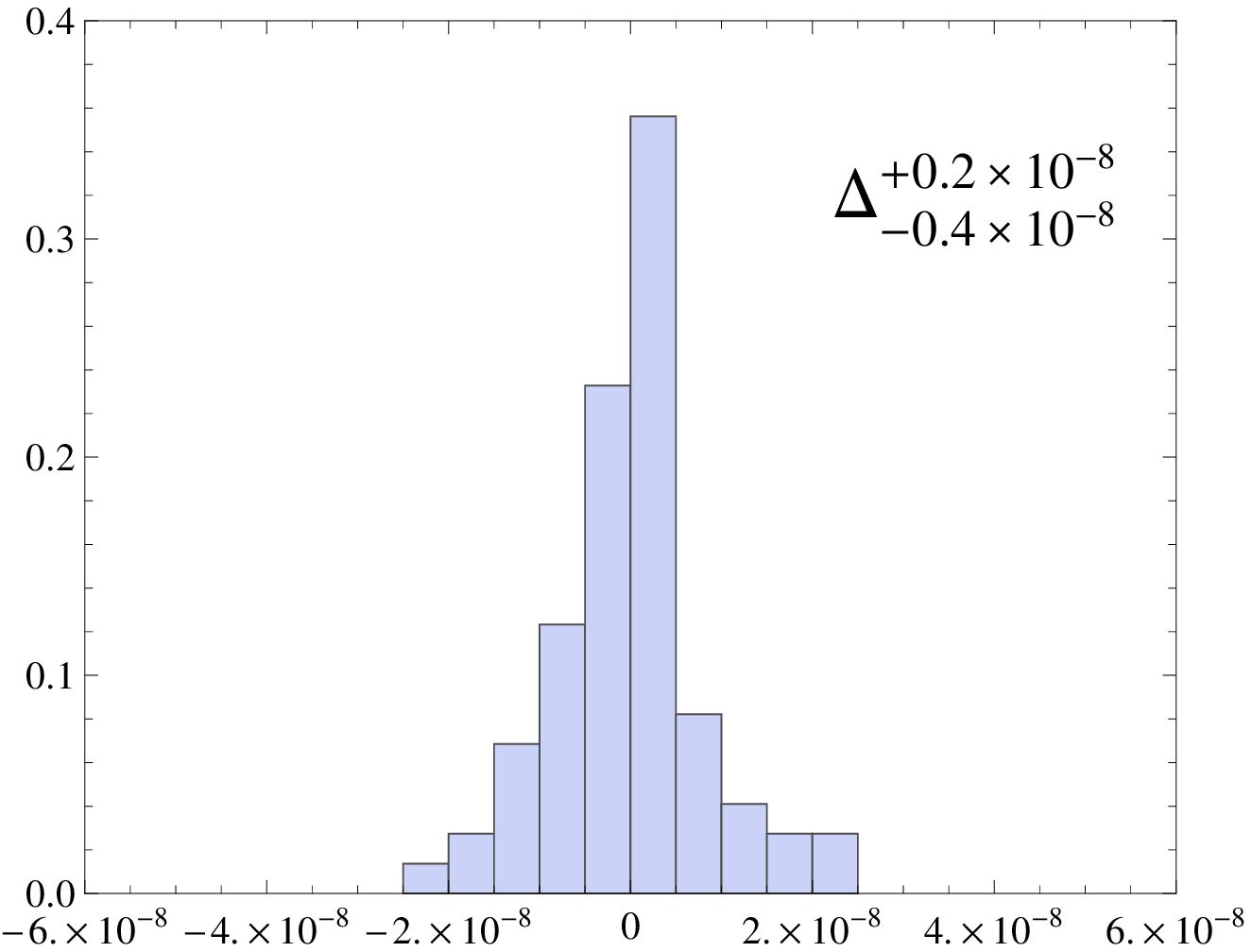}\\ \hline
			\includegraphics[width=.3\textwidth]{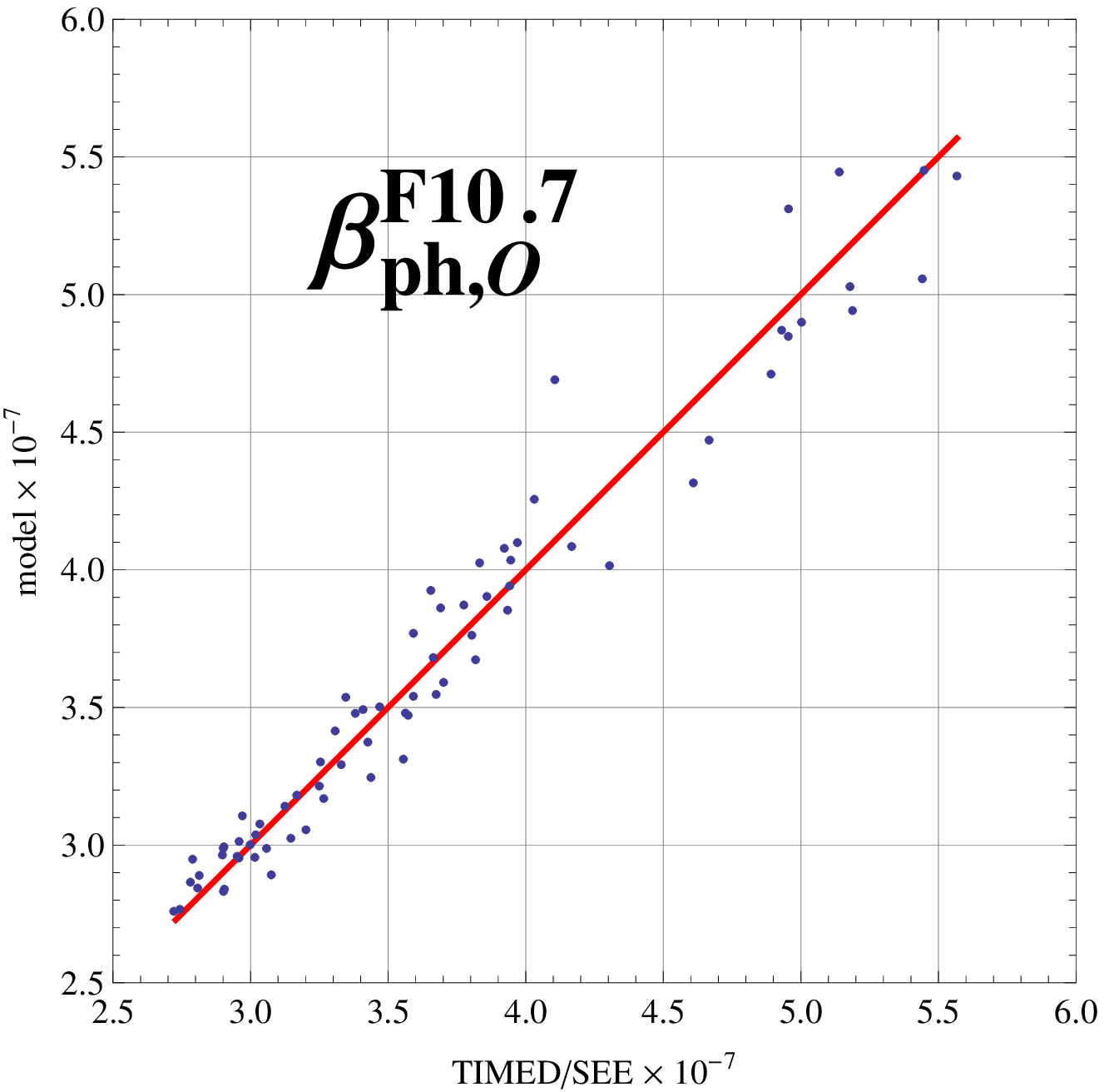}&
			\includegraphics[width=.3\textwidth]{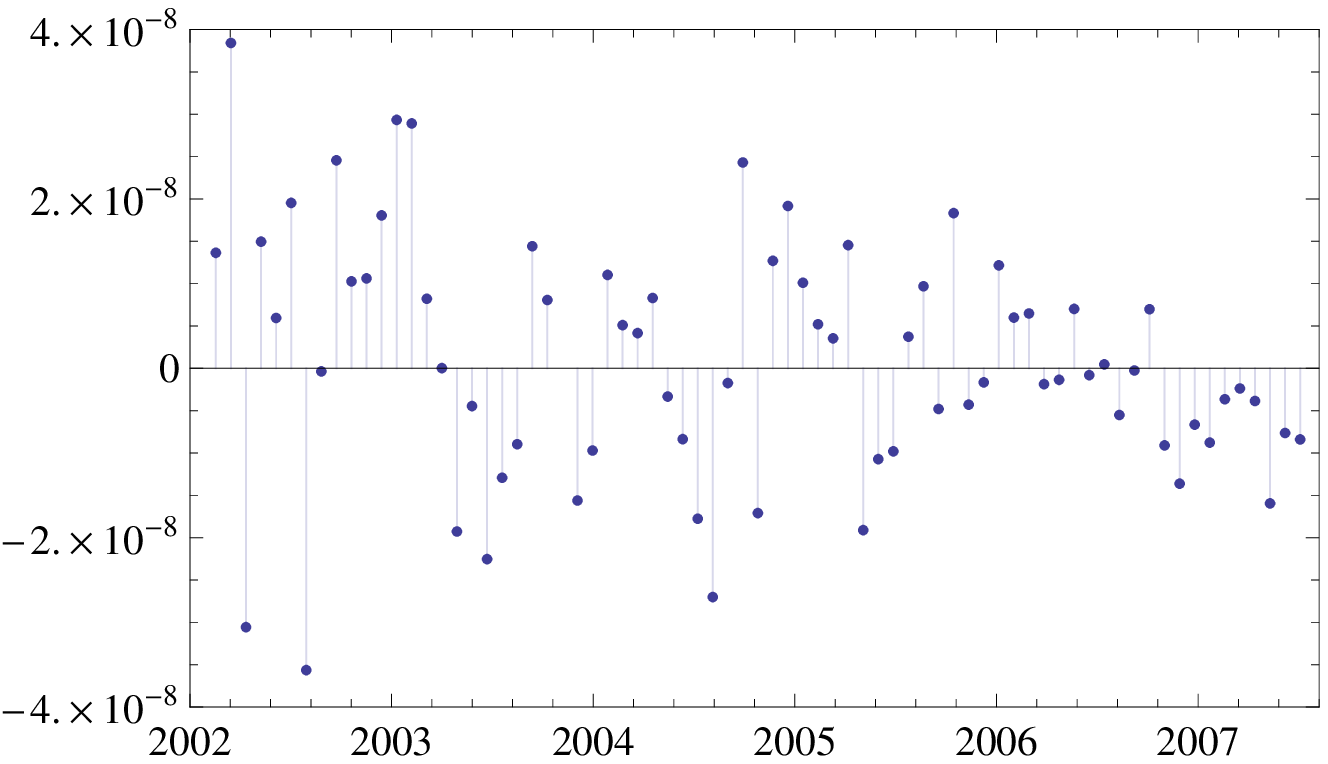}&
			\includegraphics[width=.3\textwidth]{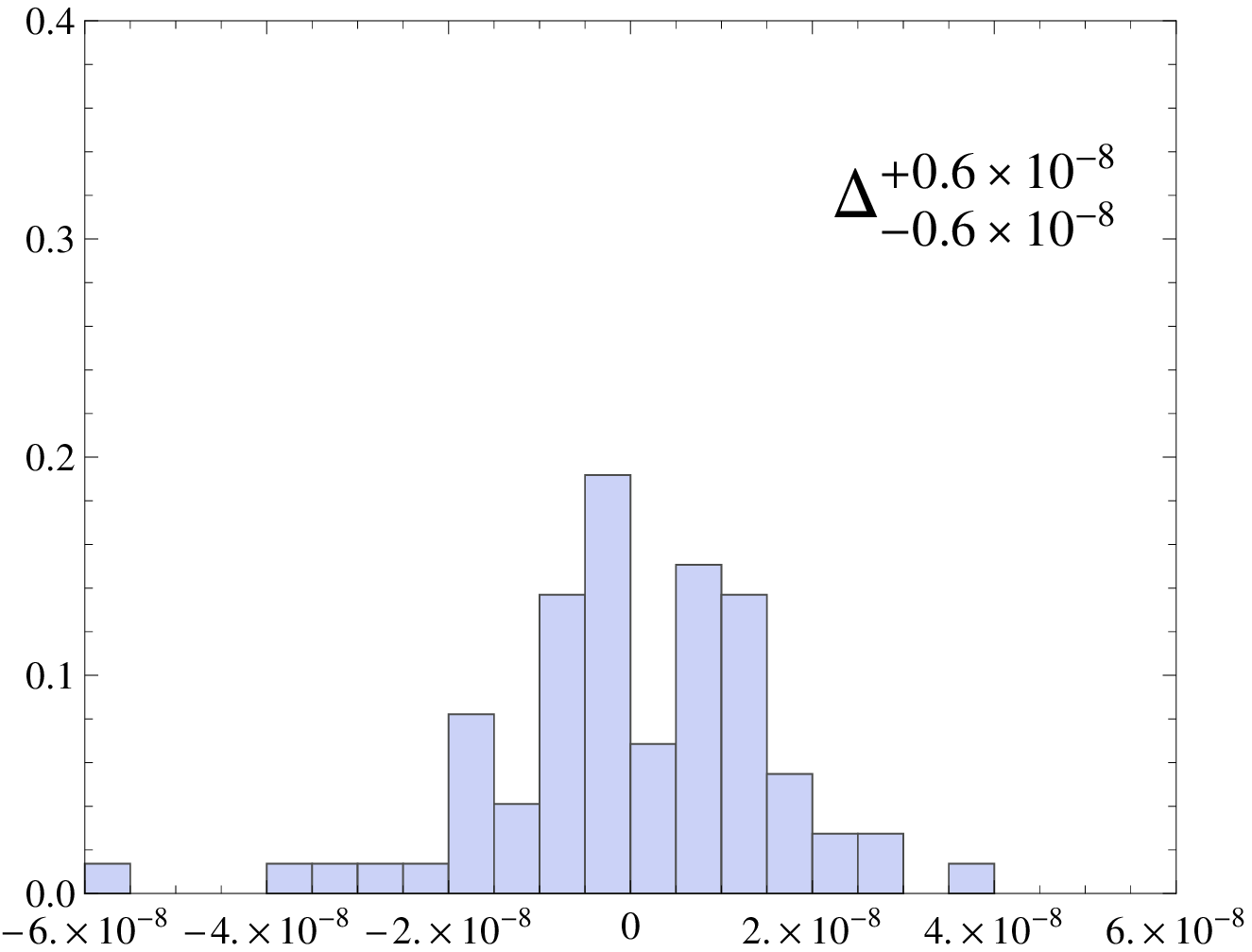}\\ \hline
		\end{tabular}
		\caption{Scatter plot of  model vs data for photoionization models for O defined in Eq.~(\ref{eq:betaOxSEMLya}) (top row), Eq.~(\ref{eq:betaOxSEM}) (second row from the top), Eq.~(\ref{eq:betaOxMg2}) (third row from the top), and Eq.~(\ref{eq:betaOxF107}) (bottom row); fit residuals (middle column); histograms of the residuals (right column). The red $y=x$  lines are eye guides. The statistical uncertainties of the models are indicated in the insets in the histogram panels.}
		\label{fig:PhionRatesModelOx}
	\end{figure*}

To extend backwards the photoionization time series we used the mentioned hierarchy of solar EUV radiation proxies (MgII$_{\mathrm{c/w}}$, Lyman-$\alpha$, and F10.7). We have found the correlation between individual proxies or their combinations and Carrington rotation averaged photoionization rates obtained using the aforementioned EUV measurements. We used them to reproduce the historical record of photoionization rates of He, Ne, and O dating backward to 1948. The following formulae yield the photoionization rates with the highest correlation coefficients. The results of the goodness of the found correlations are gathered in Fig.~\ref{fig:PhionRatesModelHe} and Fig.~\ref{fig:PhionRatesModelNe} for He and Ne, respectively, and in Fig.~\ref{fig:PhionRatesModelOx} for O.

The formula to calculate the photoionization rate from the MgII$_{\mathrm{c/w}}$ index and adjusted F10.7 index expressed in the sfu units (i.e.,$10^{-22}$~W~m$^{-2}$~Hz$^{-1}$; darker green line in Fig \ref{fig:photoRates}) is the following, with the coefficients listed in Table \ref{tab:phModelMg2F107HeNe}:
	\begin{equation}\
		\beta_{\mathrm{ph}}^{\mathrm{MgII_{\mathrm{c/w}}+F10.7}} = a + b \, \mathrm{MgII_{\mathrm{c/w}}} + c \, \mathrm{F10.7}^{p}
		\label{eq:betaMg2F107}
	\end{equation}

	\begin{table*}
		\caption{Parameter values for He and Ne, Eq.~(\ref{eq:betaMg2F107})}
		\label{tab:phModelMg2F107HeNe}
		\centering
		\begin{tabular}{lllll}
		\hline
		Species & $a$ & $b$ & $c$ & $p$\\ \hline \hline
		He & $-6.67462 \times 10^{-7}$ & $2.49477 \times 10^{-6}$ & $7.25706 \times 10^{-9}$ & $0.515151$ \\ 
		Ne & $-1.37371$ & $4.82899 \times 10^{-6}$ & $3.28689 \times 10^{-8}$ & $0.504814$\\ \hline
		\end{tabular}
	\end{table*}

The formula to calculate the photoionization rate using only the F10.7 proxy is the following:
	\begin{equation}\
		\beta_{\mathrm{ph}}^{\mathrm{F10.7}} = a + b \, \mathrm{F10.7}^{p}
		\label{eq:betaF107}
	\end{equation}
with the coefficients available in Table \ref{tab:phModelF107HeNe}. It is drawn with the gray line in Fig.~\ref{fig:photoRates}. Note that the exponent in the F10.7 flux is close to 1/2, not 1, as was frequently adopted in the past.

	\begin{table*}
		\caption{Parameter values for He and Ne, Eq.~(\ref{eq:betaF107})}
		\label{tab:phModelF107HeNe}
		\centering
		\begin{tabular}{llll}
		\hline
		Species & $a$ & $b$ & $p$\\ \hline \hline
		He & $-8.17560 \times 10^{-8}$ & $1.91040 \times 10^{-8}$ & $0.466408$ \\
		Ne & $-2.20271 \times 10^{-7}2$ & $4.99114 \times 10^{-8}$ & $0.492619$ \\ \hline
		\end{tabular}
	\end{table*} 

The procedure for oxygen is very similar to the procedure used for hydrogen, described in \citet{bzowski_etal:12b}. The SEM data are correlated with the TIMED/SEE observations and the Lyman-$\alpha$ (Ly$\alpha$) composite flux according the formula using only Carrington rotation averaged time series (lighter blue line in Fig.~\ref{fig:photoRates}):
	\begin{eqnarray}\
		\beta_{\mathrm{ph,O}}^{\mathrm{SEM+Ly\alpha}} & = & 8.09995 \times 10^{-8} + 2.61569 \times 10^{-19} \, \mathrm{Ly}\alpha + \\ \nonumber
		& - & 2.20543 \times 10^{-18} \, \mathrm{Ch1}^{1.039977} + \\ 
		& + & 3.08725 \times 10^{-18} \, \mathrm{Ch2}^{1.039977} \nonumber
		\label{eq:betaOxSEMLya}
	\end{eqnarray}
When the composite Lyman-$\alpha$ flux is not available, a formula with only the Carrington rotation averaged SEM data is used:
	\begin{eqnarray}\
		\beta_{\mathrm{ph,O}}^{\mathrm{SEM}} & = & 1.27120 \times 10^{-7} - 1.59355 \times 10^{-17} \, \mathrm{Ch1}^{0.947690} \\ \nonumber
		& + & 3.42828 \times 10^{-17} \, \mathrm{Ch2}^{0.947690} 
		\label{eq:betaOxSEM}
	\end{eqnarray}
	
When SEM data are not available, we use a correlation of the Carrington rotation averaged TIMED/SEE flux and MgII$_{\mathrm{c/w}}$ index (lighter green line in Fig.\ref{fig:photoRates}):
	\begin{equation}\
		\beta_{\mathrm{ph,O}}^{\mathrm{MgII_{\mathrm{c/w}}}} = -3.75892 \times 10^{-6} + 1.52594 \times 10^{-5} \, \mathrm{MgII_{\mathrm{c/w}}}  
		\label{eq:betaOxMg2}
	\end{equation}
and from 1948 until the beginning of the MgII$_{\mathrm{c/w}}$ time series in 1978 the photoionization rates for O come from the correlation formula for the F10.7 flux (gray line in Fig.~\ref{fig:photoRates}) based on the comparison with Carrington rotation averaged TIMED/SEE data:
	\begin{equation}\
		\beta_{\mathrm{ph,O}}^{\mathrm{F10.7}} = 2.11313 \times 10^{-8} + 1.27132 \times 10^{-8} \, \mathrm{F10.7}^{0.703322}
		\label{eq:betaOxF107}
	\end{equation}
	
The model described 
calculates Carrington rotation averages of the ionization rates. Extending the proxies into sub-Carrington time scale requires further careful studies related to possible limb darkening/brightening effects that may be present in some proxies but absent in the EUV flux contributing to photoionization. Since in the present study we need only Carrington rotation averaged time series, we leave this aspect  for future analysis. 

It is interesting to note that the photoionization rate for H and He are almost identical during solar maximum, but differ up to $40\%$ during solar minimum. Also worth pointing out is the difference between the photoionization rates of H and O despite their almost identical ionization threshold. This is because of differences in the photoionization cross sections between these species. 


\subsection{Electron impact ionization of Ne, He, and O}
	
The physical aspects of ionization of neutral heliospheric gas by impact of solar wind electrons were originally proposed by \citet{rucinski_fahr:89, rucinski_fahr:91} for He. \citet{bzowski:08a} developed a model of electron ionization rate for H using measurements of electron distribution function from Ulysses \citep{salem_etal:01a, issautier_etal:01a, salem_etal:01a, issautier_etal:08}, carried out between $\sim1.5$ and $\sim5.5$~AU from the Sun, which were recently discussed by \citet{bzowski_etal:12b}. The distribution function is approximated by a bi-Maxwellian model including the dominating cool core and the secondary hot halo populations, with the abundance of the latter increasing towards increasing heliocentric distances. Under these assumptions, and additionally the assumption of quasi-neutrality of solar wind, one can calculate a radial model of electron ionization rate for all species in question (\citet{scime_etal:94}, Fig.~\ref{fig:electronRateRadial}). 

The model values are proportional to solar wind density plus twice the alpha particle content, which approximate the solar wind electron densities and which are taken from the solar wind proton density model used to calculate the charge exchange rates. An extensive discussion of the background solar wind electron temperature model is presented in \citet{bzowski:08a} and \citet{bzowski_etal:12b}.

In our calculation we follow the approach presented in these papers: we approximate the solar wind electron distribution function as bi-Maxwellian, with the dominating cool core and hot halo. The temperatures and relative densities of the core and halo populations evolve with the heliocentric distance, as measured by Ulysses \citep{scime_etal:94}. Calculating the electron ionization rate requires folding the local electron distribution function with the reaction cross section and integrating from the threshold ionization energy to infinity. In this paper we use only a model for the slow solar wind because the NIS atoms in question travel close to the ecliptic plane at all times. The reader is referred to equations (A.9) in \citet{bzowski:08a} for the adopted radial profiles of the core and halo temperatures and (A.11) for the evolution of the abundance of the halo population relative to the core. 

	\begin{figure}
		\resizebox{\hsize}{!}{\includegraphics{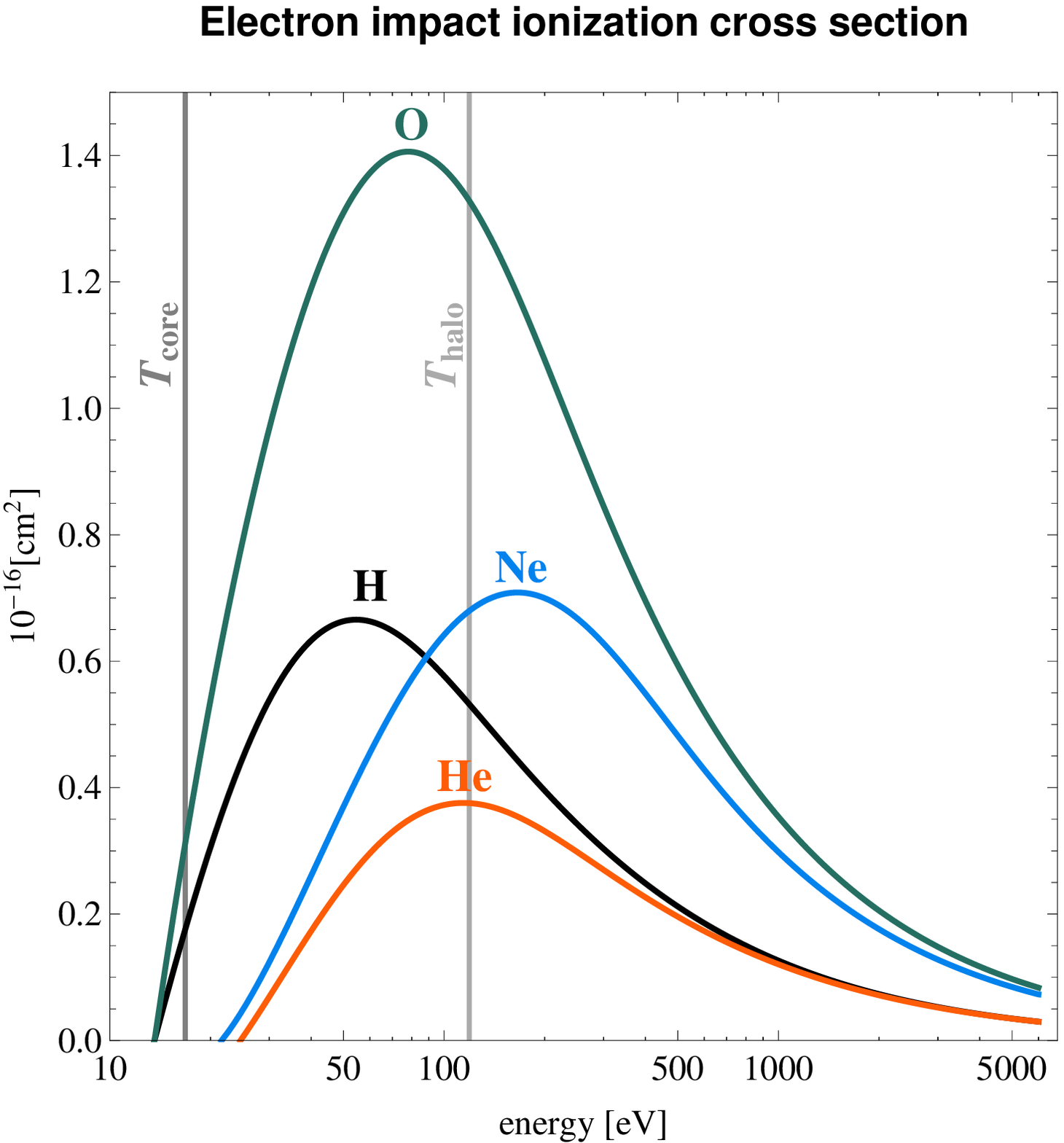}}
		\caption{Cross sections of electron impact ionization for H, He, Ne, and O \citep{lotz:67}. The accuracies for H, He, and O are $10\%$, for Ne $15\%$. The vertical bars mark the energies corresponding to the temperatures of the core and halo solar wind electron populations in the slow solar wind at 1~AU from the Sun. Note that the noble gases can only be ionized by a high-energy wing of the core population and mostly by the halo population.}
		\label{fig:crossSectionsElectrons}
	\end{figure}

The cross sections were taken from \citet{lotz:67} and are presented in Fig.~\ref{fig:crossSectionsElectrons}. They seem sufficiently accurate and reliable in the entire range of energies needed for our purposes and are consistent with more recent studies \citep{godunov_ivanov:99a, bartlett_stelbovics:04a, mattioli_etal:07a}. The two vertical bars in Fig.~\ref{fig:crossSectionsElectrons} represent the energy equal to $kT_e$ for the temperatures of the core and halo populations at 1~AU from the Sun. They show that He and Ne are ionized predominantly by the halo population, with only the fastest wing of the core population being energetic enough to ionize, while H and O are ionized by both core and halo populations. The temperatures of both populations decrease with solar distance and the farther from the Sun, the smaller portion of the total electron population is capable of ionizing the heliospheric species. Electron impact ionization contributes only $\sim10\%$ to the total ionization rate at 1~AU from the Sun and beyond $\sim 5$~AU from the Sun it becomes practically negligible. The electron impact rate is a linear function of the local electron density. 
	
The ionization rate for the core and halo populations was tabulated as a function of solar distance assuming $n_{\mathrm{sw}} = 1$~cm$^{-3}$ and then an analytic approximation formula was fitted separately for each species. Here we present the resulting formulae to calculate the electron impact ionization in the slow and fast solar wind regime. In the formulae below, $n_{\mathrm{sw}}\left( r, t\right)$ is solar wind proton density at heliocentric distance $r$ in AU, in time $t$, the logarithms are natural:
	\begin{equation}\
		\beta_{\mathrm{el}}^{\mathrm{slow}}\left( r, t \right) = n_{\mathrm{sw}}\left( r, t\right) \, \mathrm{exp} \left[ \frac{a+b\, \mathrm{ln}{r}+c\,\mathrm{ln}^2{r}}{d+e\,\mathrm{ln}{r}+f\,\mathrm{ln}^2{r}+g\,\mathrm{ln}^3{r}} \right] 
		\label{eq:betaElSlow}
	\end{equation}
		\begin{equation}\
		\beta_{\mathrm{el,He,Ne}}^{\mathrm{fast}} \left( r, t \right) = n_{\mathrm{sw}}\left( r, t\right) \, \mathrm{exp} \left[ \frac{a+b\, \mathrm{ln}{r}+c\,\mathrm{ln}^2{r}+d\,\mathrm{ln}^3{r}}{e+f\,\mathrm{ln}{r}+g\,\mathrm{ln}^2{r}^2+h\,\mathrm{ln}^3{r}} \right] 
		\label{eq:betaElFastHeNe}
	\end{equation}
	\begin{equation}\
	\beta_{\mathrm{el,O}}^{\mathrm{fast}} \left( r, t \right) =  n_{\mathrm{sw}}\left( r, t\right) \,\mathrm{exp} \left[ \frac{a+b\, \mathrm{ln}{r}+c\,\mathrm{ln}^2{r}}{d+e\,\mathrm{ln}{r}+f\,\mathrm{ln}^2{r}+g\,\mathrm{ln}^3{r}+h\,\mathrm{ln}^5{r}} \right] 
		\label{eq:betaElFastO}
	\end{equation}

The coefficients are gathered in Tables~\ref{tab:betaElSlowCoef} and \ref{tab:betaElFastCoef}. Throughout the paper we assume that the solar wind proton density decreases with heliocentric distance as $1/r^2$. The $\beta_{\mathrm{el}}$ rate in our model is slaved to local solar wind density $n_{\mathrm{sw}}\left(r, t\right)$ and consequently it is calculated as a function of time similarly to $n_{\mathrm{sw}}\left(r, t\right)$.

The ionization rate obtained using the presented model are consistent with the rates published by \citet{arnaud_rothenflug:85}, which for He$^0$, Ne$^0$, and O$^0$ are based on the cross sections and approximation formulae from \citet{lotz:67}.

	\begin{table*}
		\caption{Coefficients for electron impact ionization rate in the slow solar wind (Eq.~(\ref{eq:betaElSlow}))}
		\label{tab:betaElSlowCoef}
		\centering
		\begin{tabular}{llllllll}
		\hline
	 species & $a$ & $b$ & $c$ & $d$ & $e$ & $f$ & $g$ \\ \hline \hline
		He & $-1.7691 \times 10^{4}$ & $5.5601 \times 10^{3}$ & $-9.9697 \times 10^{3}$ & $8.8827 \times 10^{2}$ & $-3.4268 \times 10^{2}$ & $5.2256 \times 10^{2}$ & $-1.9169 \times 10^{1}$ \\ \hline
		Ne & $-8.3072 \times 10^{1}$ & $3.3497 \times 10^{1}$ & $-4.9956 \times 10^{1}$ & $4.2941 \times 10^{0}$ & $-2.0157 \times 10^{0}$ & $2.6833 \times 10^{0}$ & $-9.9570 \times 10^{-2}$\\ \hline
		O & $-8.0999 \times 10^{4}$ & $3.8910 \times 10^{4}$ & $-2.3981 \times 10^{4}$ & $4.4617 \times 10^{3}$ & $-2.4442 \times 10^{3}$ & $1.4332 \times 10^{3}$ & $-5.1963 \times 10^{1}$ \\ \hline
		\end{tabular}
	\end{table*}
	
	\begin{table*}
		\caption{Coefficients for electron impact ionization rate in the fast solar wind (Eqs.~(\ref{eq:betaElFastHeNe}) and (\ref{eq:betaElFastO}))}
		\label{tab:betaElFastCoef}
		\centering
		\begin{tabular}{lllllllll}
		\hline
	   species & $a$ & $b$ & $c$ & $d$ & $e$ & $f$ & $g$ & $h$ \\ \hline \hline
		He & $-7.9357 \times 10^{1}$ & $-1.5044 \times 10^{1}$ & $-2.8540 \times 10^{1}$ & $3.2146 \times 10^{0}$ & $3.7781 \times 10^{0}$ & $5.5937 \times 10^{-1}$ & $1.3935 \times 10^{0}$ & $-2.0605 \times 10^{-1}$ \\ \hline
		Ne & $-1.6140 \times 10^{1}$ & $-2.8934 \times 10^{0}$ & $-1.1702 \times 10^{1}$ & $1.2639 \times 10^{0}$ & $7.9316 \times 10^{-1}$ & $1.0748 \times 10^{-1}$ & $5.7670 \times 10^{-1}$ & $-8.0280 \times 10^{-2}$  \\ \hline
		O & $-3.6108 \times 10^{3}$ & $9.8440 \times 10^{2}$ & $-8.5183 \times 10^{2}$ & $1.8844 \times 10^{2}$ & $-6.2083 \times 10^{1}$ & $4.8152 \times 10^{1}$ & $-1.7787 \times 10^{0}$ & $-2.1580 \times 10^{-2}$ \\ \hline
		\end{tabular}
	\end{table*}

\subsection{Charge exchange rates for He, Ne, and O}

In addition to the photoionization and electron impact rates discussed in the former section, the ionization processes relevant for neutral heliospheric species include charge exchange. The charge exchange reaction was recently discussed by \citet{bzowski_etal:12b} for heliospheric H and most of the theory applies for O, Ne, and He. 

\begin{figure}
		\resizebox{\hsize}{!}{\includegraphics{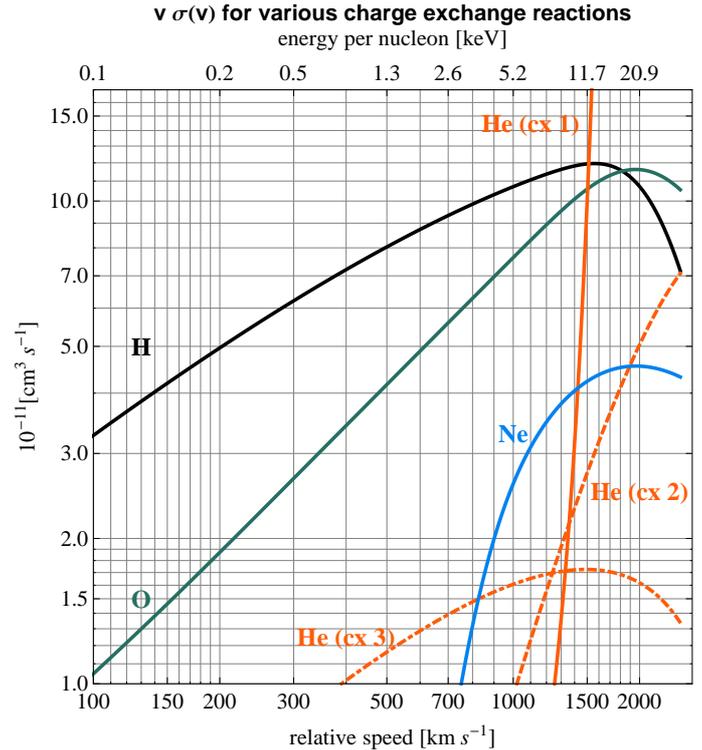}}
		\caption{Products of relative speeds of reaction components and charge exchange cross sections for these speeds for the charge exchange reactions listed in the text (see Section~\ref{sec:ChargeExchange}). To obtain the actual charge-exchange rate, one has to multiply a quantity read off from the figure for a given speed by the local density of the ionized reaction component (solar wind proton, alpha particle). The lower horizontal axis is scaled in~$\mathrm{km}~\mathrm{s}^{-1}$, the upper axis in energy per nucleon in keV.}
		\label{fig:crossSectionsCX}
	\end{figure}

The charge exchange cross sections for He and O used in the present paper come from \citet{lindsay_stebbings:05a} and for Ne from \citet{nakai_etal:87a}. Fig.~\ref{fig:crossSectionsCX} shows the product of relative speed of the component of the reaction and cross section for charge exchange for the NIS species observed by IBEX. To obtain an instantaneous charge exchange rate the value taken from the figure for a given speed must be multiplied by the local ion density, which is assumed to fall off with $r^2$ and is a function of heliolatitude and time \citep{sokol_etal:12a}. The figure explains the huge differences between the charge exchange rates for H and O on one hand and for the noble gases on the other hand, and illustrates that charge exchange rates are not linear functions of the collision speed, as frequently approximated.

The calculations were performed using Carrington rotation averaged solar wind proton densities and speeds from the formula Eq.~(\ref{eq:cxDef}). The values thus obtained are approximate because the atom speed is assumed~$0$. In the simulations shown in the main text we use the full formula with $\mid \vec{v_{\mathrm{rel}}} \mid = \mid \vec{v_{\mathrm{sw}}} - \vec{v_{\mathrm{atom}}} \mid$
	\begin{equation}
	\beta_{\mathrm{cx}}=n_{\mathrm{sw}} v_{\mathrm{rel}} \sigma_{\mathrm{cx}}\left( v_{\mathrm{rel}} \right).
	\label{eq:cxDef}
	\end{equation}
$\vec{v_{\mathrm{sw}}}$ and $n_{\mathrm{sw}}$ were calculated for a given instant of time $t$ and heliolatitude $\phi$ from bi-linear interpolation between the grid nodes fixed at halves of Carrington rotations and full tens of degrees heliolatitude (see \citet{sokol_etal:12a}). For the dependence on heliocentric distance $r$, proton density $n_{\mathrm{sw}}$ was assumed to conform to the $1/r^2$ dependence and $v_{\mathrm{sw}}$ was assumed distance independent. 

\subsection{Total ionization rates for He, Ne, and O}

The total ionization rates for NIS species are algebraic sums of rates of all relevant ionization processes. Fig.~\ref{fig:ionRates} presents the total ionization rates of the contributing reactions for the NIS species observed by IBEX. For the noble gases the main reaction is photoionization, the next is electron impact ionization, which gives $\sim10\%$ input to the total rate. The rate of charge exchange with solar wind protons (and alphas in the case of helium) is very low, most of the time much less than $1\%$ of the total rate. The case of oxygen is more complex. During the previous minimum in 1996, the photoionization and charge exchange rates were comparable ($\sim45\%$ input each to the total rate), but since 1998 the input from charge exchange decreased (to $\sim35\%$) and photoionization became dominant (more than $50\%$). 

The total ionization rates for He and Ne show a clear modulation with solar activity because they mainly depend on photoionization, which varies quasi-periodically with the solar activity cycle. The ionization rate for O does not show a clear modulation with the solar activity cycle because it is a sum of a quasi-periodic component from photoionization and a time-dependent, but not periodic, component from charge exchange and electron impact. Due to the similarity of the first ionization potential for H and O it is usually assumed that the total ionization rates of these species should be very similar. Our analysis shows that it is true most of the times, but not always: the ionization rate for O was higher than for H during a couple of years during the previous solar maximum and it seems that this behavior is going to repeat during the forthcoming maximum as well. The ionization rate of H shown in Fig.~\ref{fig:ionRates} is taken from \citet{bzowski_etal:12b} and show here for comparison.

The uncertainties of the ionization rates used in this work will be a subject of separate paper (Sok{\'o}{\l} et al., 2013, in preparation). They arise from uncertainties in the reaction cross sections, from uncertainties of the measured quantities (the solar wind and solar EUV output), from the statistical uncertainties of the proxies used, and from simplifications used in the models. We estimate that the overall uncertainty in the total ionization rates is 15--20$\%$. The statistical uncertainties of the proxy models are indicated in the figures of histograms (see Figures \ref{fig:PhionRatesModelHe}, \ref{fig:PhionRatesModelNe}, \ref{fig:PhionRatesModelOx}). The other uncertainties are listed in Table~\ref{tab:auxErrors}.

\bibliographystyle{aa}
\bibliography{iplbib}

\end{document}